\documentclass[iop,numberedappendix,twocolappendix,appendixfloats]{emulateapj}
\usepackage{graphicx,subfigure,amsmath, amsfonts, amssymb,aas_macros,times,natbib,mathtools,lipsum}
\usepackage[export]{adjustbox}
\usepackage[usenames,dvipsnames]{color}
\usepackage[toc,page]{appendix}
\usepackage{mathrsfs}

\newcommand{\kms}{km~s$^{-1}$}
\newcommand{\cplus}{\hbox{[C\,\sc{ii}]}}
\newcommand{\hi}{\hbox{H\,{\sc i}}}
\newcommand{\oi}{\hbox{[O\,{\sc i}]}}

\newcommand{\atlas}{\hbox{ATLAS$^{\rm 3D}$}}
\newcommand{\supp}{$\mathscr{S}$}
\shorttitle{CO(1--0) Imaging of Hickson Compact Groups and Suppression of Star Formation}
\shortauthors{K. Alatalo et al.}

\begin{document}

%%%%%%%%%%%%%%%%%%%%%%%%%%%%%%%%%%%%%%%%%%%%%%%
% TITLE AND AUTHOR

\title{Star formation suppression in Compact Group Galaxies: a new path to quenching?}

\author{K. Alatalo,$^{1,2}$\altaffilmark{$\dagger$} P.~N. Appleton,$^{1,3}$ U. Lisenfeld,$^{4}$ T. Bitsakis,$^{3,5}$ L. Lanz,$^{1}$ M. Lacy,$^6$ V. Charmandaris,$^{7,8,9}$ M. Cluver,$^{10}$ M.~A. Dopita,$^{11,12,13}$ P. Guillard,$^{14,15}$ T. Jarrett,$^{16}$ L.~J. Kewley,$^{11}$ K. Nyland,$^{17}$ P.~M. Ogle,$^{1}$ J. Rasmussen,$^{18,19}$ J.~A. Rich,$^{1,2}$ L. Verdes-Montenegro,$^{20}$ C.~K. Xu,$^{1,3}$ M. Yun$^{21}$
}

\affil{$^{1}$Infrared Processing \& Analysis Center,  California Institute of Technology, Pasadena, CA 91125, USA\\
$^{2}$Observatories of the Carnegie Institution of Washington, 813 Santa Barbara Street, Pasadena, CA 91101, USA \\
$^{3}$NASA Herschel Science Center, IPAC,  California Institute of Technology, Pasadena, CA 91125, USA\\
$^{4}$Departamento de F\'isica Te\'orica y del Cosmos, Universidad de Granada, Granada, Spain\\
$^{5}$Instituto de Astronom\'ia, Universidad Nacional Aut\'onoma de M\'exico, Aptdo. Postal 70-264, 04510, M\'exico, D.F., Mexico\\
$^{6}$National Radio Astronomy Observatory, 520 Edgemont Road, Charlottesville, VA 22903, USA\\
$^{7}$Institute for Astronomy, Astrophysics, Space Applications \& Remote Sensing, National Observatory of Athens, GR-15236, Penteli, Greece\\
$^{8}$Department of Physics, University of Crete, GR-71003, Heraklion, Greece\\
$^{9}$Chercheur Associ\'e, Observatoire de Paris, F-75014, Paris, France\\
$^{10}$ Department of Physics, University of the Western Cape, Robert Sobukwe Road, Bellville, 7535, South Africa\\
$^{11}$Research School of Astronomy and Astrophysics, Australian National University, Cotter Rd., Weston ACT 2611, Australia\\
$^{12}$Astronomy Department, King Abdulaziz University, P.O. Box 80203, Jeddah, Saudi Arabia\\
$^{13}$Institute for Astronomy, University of Hawaii, 2680 Woodlawn Drive, Honolulu, HI 96822, USA\\
$^{14}$Sorbonne Universit\'es, UPMC Univ Paris 6 et CNRS, UMR 7095,  Institut d'Astrophysique de Paris, 98 bis bd Arago, 75014 Paris, France\\
$^{15}$Institut d'Astrophysique Spatiale, Universit\'e Paris-Sud XI, 91405 Orsay Cedex\\
$^{16}$Astrophysics Cosmology and Gravity Centre, Dept of Astronomy, University of Cape Town, Private Bag X3, Rondebosch, 7701, Republic of South Africa\\
$^{17}$Netherlands Institute for Radio Astronomy (ASTRON), Postbus 2, 7990 AA Dwingeloo, The Netherlands\\
$^{18}$Dark Cosmology Centre, Niels Bohr Institute, University of Copenhagen, Juliane Maries Vej 30, DK-2100 Copenhagen, Denmark\\
$^{19}$Technical University of Denmark, Department of Physics, Building 309, DK-2800 Kgs. Lyngby, Denmark\\
$^{20}$Departamento Astronom\'ia Extragal\'actica, Instituto Astrof\'isica Andaluc\'ia (CSIC), Glorieta de la Astronom\'ia s/n 18008 Granada, Spain\\
$^{21}$University of Massachusetts, Astronomy Department, Amherst, MA 01003, USA
} 
\altaffiltext{$\dagger$}{Hubble fellow}
\slugcomment{Accepted by the Astrophysical Journal - 18 Sept 2015}
\email{email:kalatalo@carnegiescience.edu}
%%%%%%%%%%%%%%%%%%%%%%%%%%%%%%%%%%%%%%%%%%%%%%%
% ABSTRACT

\begin{abstract}
We present CO(1--0) maps of 12 warm H$_2$-selected Hickson Compact Groups (HCGs), covering 14 individually imaged warm H$_2$ bright galaxies, with the Combined Array for Research in Millimeter Astronomy.  We found a variety of molecular gas distributions within the HCGs, including regularly rotating disks, bars, rings, tidal tails, and possibly nuclear outflows, though the molecular gas morphologies are more consistent with spirals and early-type galaxies than mergers and interacting systems. Our CO-imaged HCG galaxies, when plotted on the Kennicutt-Schmidt relation, shows star formation suppression of $\langle$\supp$\rangle$\,=\,10$\pm$5, distributed bimodally, with five objects exhibiting suppressions of \supp\,$\gtrsim$\,10 and depletion timescales $\gtrsim$\,10\,Gyr. This star formation inefficiency is also seen in the efficiency per freefall time of \citet{krumholz+12}. We investigate the gas-to-dust ratios of these galaxies to determine if an incorrect L$_{\rm CO}$--$M({\rm H_2})$ conversion caused the apparent suppression and find that HCGs have normal gas-to-dust ratios. It is likely that the cause of the apparent suppression in these objects is associated with shocks injecting turbulence into the molecular gas, supported by the fact that the required turbulent injection luminosity is consistent with the bright H$_2$ luminosity reported by \citet{cluver+13}.  Galaxies with high star formation suppression (\supp\,$\gtrsim$10) also appear to be those in the most advanced stages of transition across both optical and infrared color space.  This supports the idea that at least some galaxies in HCGs are transitioning objects, where a disruption of the existing molecular gas in the system suppresses star formation by inhibiting the molecular gas from collapsing and forming stars efficiently. These observations, combined with recent work on poststarburst galaxies with molecular reservoirs, indicates that galaxies do not need to expel their molecular reservoirs prior to quenching star formation and transitioning from blue spirals to red early-type galaxies.  This may imply that star formation quenching can occur without the need to starve a galaxy of cold gas first.
\end{abstract}

%%%%%%%%%%%%%%%%%%%%%%%%%%%%%%%%%%%%%%%%%%%%%%%
% SUBJECT HEADINGS


%%%%%%%%%%%%%%%%%%%%%%%%%%%%%%%%%%%%%%%%%%%%%%%
% ACTUAL PAPER

%%%%%%%%%%%%%%%%% INTRODUCTION %%%%%%%%%%%%%%%%%%%%%

\section{Introduction}
The present-day galaxy population has a bimodal distribution, comprised of blue spiral galaxies and red elliptical and lenticular galaxies \citep{tinsley78,strateva+01,baldry+04}, with a dearth of galaxies at intermediate optical colors \citep{bell+03}.  Their rarity suggests that galaxies transition rapidly in colors.  However, \citet{schawinski+14} showed that selecting transitioning objects based on color leads to an overestimation of morphologically transforming objects, as the majority of green valley objects are not those undergoing the rapid morphological transformation from spiral to elliptical, but were mainly spiral galaxies that were undergoing secular evolution.  More recently, \citet{smethurst+15} supported this picture of spiral galaxies transitioning at intermediate rates in the green valley, but also showed that early type galaxies transition more rapidly. 

%25b & $\checkmark$ & red & $\checkmark$ & $\checkmark$ & $\checkmark$ & $\checkmark$ & D\\
%40c & & green & & & $\checkmark$ & $\checkmark$ & B+R \\
%47a & & blue & & & $\checkmark$ & R,S\\
%55c & & green & & & & D \\
%57a & $\checkmark$ & red & $\checkmark$ & $\checkmark$ & $\checkmark$ & M\\
%57d &  & blue & & & & R\\
%68a & $\checkmark$ & green & & & $\checkmark$ & D\\
%79a & & red & & $\checkmark$ & $\checkmark$ & D\\
%82b & $\checkmark$ & red & $\checkmark$ & $\checkmark$ & $\checkmark$ & M\\
%91a & & blue & & & $\checkmark$ & S\\
%95c & $\checkmark$ & green & $\checkmark$ & & $\checkmark$ &  M\\
%96a & & blue & & & $\checkmark$ & B+R,S\\
%96c & & green & & & &  D\\
%100a & & blue & & & $\checkmark$ & D\\

%%%%%%%%%%%%%%% Table 1 %%%%%%%%%%%%%%%
\begin{table*}[t!]
\centering
\caption{HCG CO(1--0) galaxy properties} \vspace{1mm}
\begin{tabular}{l l l l l r l c r r c c c c c c r c}
\hline \hline
HCG & Principal & RA & Dec & Dist.$^\ddagger$ & Morph.$^\flat$ & $F_{\rm 1.4GHz}$ & \cplus?
& {\sc moheg}$^\diamond$ & Optical & {\em Spitzer} & {\em WISE} & Optical$^\bigstar$ & CO$^\triangle$\\
Name & Name & (J2000) & (J2000) & (Mpc) & t-type & (mJy) & & & Sequence & gap & IRTZ & AGN & Morph.\\
\hline
HCG\,25b & PGC012539  & 03\,20\,45.41 & -01\,02\,40.9 & 85.8 & 0.9$\pm$1.3 & 5.6$^a$ & $\checkmark$ &
$\checkmark$ & red & $\checkmark$ & $\checkmark$ & $\checkmark$ & D\\
HCG\,40c & PGC027508 & 09\,38\,53.61 & -04\,51\,36.6 & 94.1 & 2.3$\pm$2.1& 10.3$^b$ &  
& & green & & & $\checkmark$ &  B+R\\
HCG\,47a & UGC05644  & 10\,25\,46.26 & +13\,43\,00.7 &141 & 2.4$\pm$1.2 & 13.1$^a$ &  
& & blue & & & $\checkmark$ & R,S\\
HCG\,55c & PGC035573  & 11\,32\,05.69 & +70\,48\,38.7 & 222 & 1.3$\pm$1.3 & -- & $\checkmark$ 
& & green & & & & D\\
HCG\,57a & NGC3753 & 11\,37\,53.90 & +21\,58\,53.0 & 127$^\dagger$ & 2.1$\pm$0.6 & 3.8$^a$ & $\checkmark$ 
& $\checkmark$ & red & $\checkmark$ & $\checkmark$ & $\checkmark$ & M\\
HCG\,57d & NGC3754  & 11\,37\,54.92 & +21\,59\,07.8 & 127$^\dagger$ & 3.4$\pm$1.1 & -- & $\checkmark$ 
&  & blue & & & & R\\
HCG\,68a & NGC5353  & 13\,53\,26.69 & +40\,16\,58.9 & 34.6 & -2.0$\pm$0.7 &40.5$^a$ &  
& $\checkmark$ & green & & & $\checkmark$ & D\\
HCG\,68b & NGC5354  & 13\,53\,26.70 & +40\,18\,09.9 & 38.1 & -2.1$\pm$0.7 & 8.0$^a$ & $\checkmark$ 
& $\checkmark$ & -- & -- & -- & -- & --\\
HCG\,79a & NGC6027A  & 15\,59\,11.14 & +20\,45\,17.5 & 64.5 & 0.3$\pm$2.1 & 9.3$^a$ & $\checkmark$ 
& & red & & $\checkmark$ & $\checkmark$ & D\\
HCG\,82b & NGC6163 & 16\,28\,27.91 & +32\,50\,47.0 & 148 & -1.7$\pm$1.1 & -- &  
& $\checkmark$ & red & $\checkmark$ & $\checkmark$ & $\checkmark$ & M\\
HCG\,91a & NGC7214 & 22\,09\,07.68 & -27\,48\,34.1 & 92.6 & 4.5$\pm$0.7& 29.2$^c$ & $\checkmark$ 
& & blue & & & $\checkmark$ & S\\
HCG\,95c & PGC071077  & 23\,19\,31.09 & +09\,30\,10.7 & 158 & 9.0$\pm$2.0 & 4.10$^d$ & $\checkmark$ 
& $\checkmark$ & green & $\checkmark$ & & $\checkmark$ &  M\\ 
HCG\,96a & NGC7674  & 23\,27\,56.72 & +08\,46\,44.5 & 116$^\dagger$ & 3.8$\pm$0.6 & 221.0$^a$ & $\checkmark$ 
& & blue & & & $\checkmark$ & B+R,S\\
HCG\,96c & PGC071505  & 23\,27\,58.78 & +08\,46\,58.1 & 116$^\dagger$ & 5.5$\pm$4.7 & 0.85$^d$ & $\checkmark$ 
& & green & & & &  D\\
HCG\,100a & NGC7803  & 00\,01\,19.97 & +13\,06\,40.5 & 69.5 & 0.1$\pm$1.0 & 12.3$^a$ & $\checkmark$ 
& & blue & & & $\checkmark$ & D\\
%\hline
%HCG\,16c & NGC\,838 & 02~18~38.53 & -10~08~48.1 & 49.8\\
%HCG\,16d & NGC\,839 & 02~18~42.93 & -10~44~02.7 & 50.1\\
%HCG\,37b & NGC\,2783b & 09~18~33.15 & +30~00~00.5 & 97.8\\ 
%HCG\,44a & NGC\,3190 & 10~18~05.63 & +21~49~56.3 & 21.8\\
\hline \hline
\end{tabular} \\
\label{tab:gal_params}
\raggedright {\footnotesize
$^\ddagger$Luminosity distance determined using the Nearby Extragalactic Database (NED)\\
$^\dagger$Distance determined by the more massive of the group members\\
$^\flat$ Morphological t-type determined by HyperLEDA \citep{hyperleda}\\
$F_{\rm 1.4\,GHz}$ continuum from ($a$) \citet{nvss}, ($b$) \citet{menon+85}, ($c$) \citet{brown+11}, and ($d$) \citet{first}.\\
$\diamond$ MOHEG definition based on H$_2$/7.7$\mu$m\,$\geq$\,0.04 \citep{ogle+06,cluver+13}, all except 57d are detected in H$_2$.\\
$^\bigstar$ Optical AGN classification from \citet{martinez+10}, counting both transition objects (TO), LINERs, and Seyferts\\
$^\triangle$Morphological class of the molecular gas: D = disk, R = ring, M = mildly disrupted, B+R = bar/ring, and S = spiral, based on the classification scheme from \citet{alatalo+13}
}
\end{table*}

Many early transitioning scenarios posited that the majority of molecular gas in galaxies is depleted prior to the quenching of star formation \citep{sanders+mirabel96,hopkins+06}, through both supernova and active galactic nucleus (AGN) feedback mechanisms \citep{springel+05}.  However, recent observations have started to question whether the two-stage scenario consisting of (1) the expulsion of star-forming gas followed by (2) the cessation of star formation (SF) is the exclusive evolutionary picture. Studies now show that many poststarburst galaxies contain substantial reservoirs of molecular gas \citep{french+15,rowlands+15} and are dustier than normal galaxies \citep{yesuf+14}.  This shows that removing all star-forming material before transitioning is not a requirement in a galaxy's transformation from blue to red.

With the advent of the {\em Wide-field Infrared Survey Explorer} ({\em WISE}; \citealt{wise}) mission, evidence has accumulated that mid-infrared (IR) colors can also be used to identify phases of transitioning galaxies \citep{ko+13}. \citet{a14_irtz} showed the existence of a prominent bifurcation between star-forming spiral galaxies and quiescent early-type galaxies in the {\em WISE} [4.6]--[12]$\mu$m colors, deeming this the ``Infrared Transition Zone'' (IRTZ).  Objects within the IRTZ have red optical colors (also described in \citealt{ko+13}), suggesting that galaxies traverse the optical green valley before the IRTZ. Assuming that [4.6]--[12]$\mu$m {\em WISE} colors trace the interstellar medium (ISM) within the galaxy and optical colors trace the mean stellar population age \citep{donoso+12}, which further supports the idea that a non-negligible number of galaxies quench SF before shedding their ISMs.

Many plausible mechanisms have been introduced to explain this transition, including major mergers \citep{toomre72,springel+05}; experiencing ram pressure stripping, and strangulation when falling into a galaxy cluster (\citealt{bekki+02,blanton+moustakas09}; and references therein); morphological quenching \citep{martig+09,martig+13}; minor mergers \citep{qu+10,eliche-moral+12,a14_stelpop,a15_sfsupp}; AGN feedback \citep{hopkins+06,fischer+10,feruglio+10,sturm+11,alatalo+11,aalto_1377,cicone+12,cicone+14,alatalo15}, and tidal disruption and harassment through group interactions \citep{hickson+92,zabludoff+98,walker+10,bitsakis+10,bitsakis+14}. Given their low galaxy velocity dispersion and high density, as well as their short-lived nature \citep{hickson82}, compact groups serve as an ideal environment in which to study galaxy transformation.

Hickson compact groups (HCGs) are defined as ``small, relatively isolated systems of typically four or five galaxies in close proximity to one another'' \citep{hickson82,hickson97}.  They tend to have a high fraction of early-type galaxies (E/S0), evidence of tidal interactions, and high density structure with low velocity dispersion \citep{hickson82,hickson97} and deficiencies in \hi\ compared with isolated galaxies \citep{verdes-montenegro+01,borthakur+10}.  Compact groups appear to go through an evolutionary phase that can be traced by neutral gas depletion \citep{verdes-montenegro+01}. In the later stages of depletion, H\,{\sc i} is found less in the galaxies and more in the intragalactic medium (IGM); \citep{borthakur+10}, with a rise in the fraction of groups containing extended group-wide X-ray emission \citep{ponman+96}. However, the origin of the extended X-ray emission is still unclear for HCGs in general (see \citealt{rasmussen+08}), with strong starburst winds being the partial cause in at least one system \citep{osullivan+14b,osullivan+14}.  The fraction of galaxy types also evolves in concert with the neutral gas depletion, with spiral-rich groups at early times and elliptical-rich groups later in the sequence \citep{bitsakis+10,bitsakis+11,bitsakis+14}.

%%%%%%%%%%%%%%% Figure 1 %%%%%%%%%%%%%%%
\begin{figure}[t]
\includegraphics[width=0.49\textwidth]{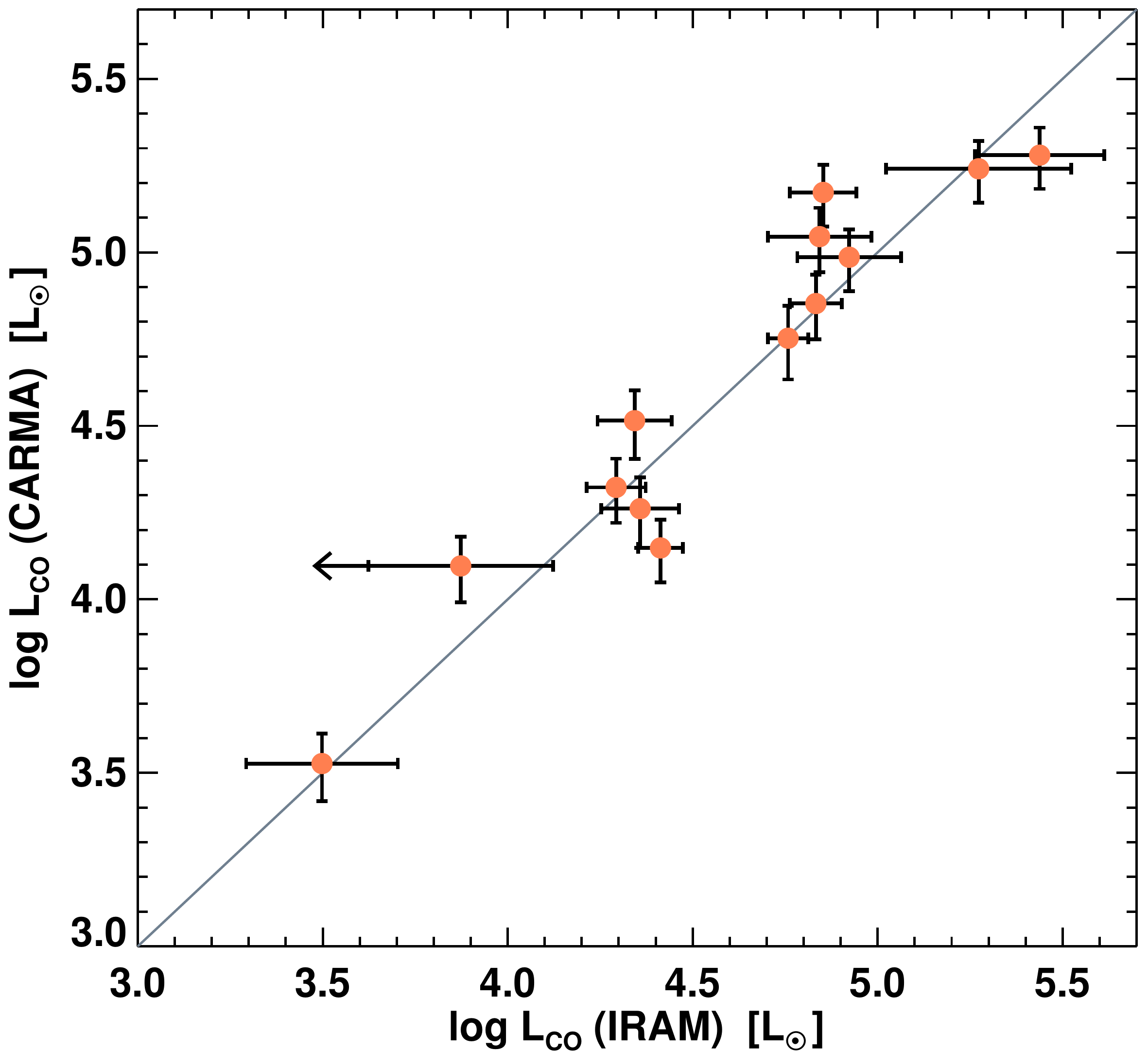}
\caption{A comparison between the total CO luminosity measured by CARMA and IRAM.  Errors for CARMA include both the RMS and the 20\% millimeter flux calibration uncertainty.  For IRAM, the range of values consists of the luminosity measured in the central pointing as a lower limit and the extrapolated total luminosity as the higher estimate.  Overall, the CO luminosities derived from CARMA and IRAM are in good agreement.}
\label{fig:comparison}
\end{figure}

It also appears that the galaxies within the HCG transform rapidly.  No {\em Spitzer} color bimodality was reported by \citet{lacy+04,lacy+07} between early-type and late-type galaxies in a sample of field galaxies, but \citet{johnson+07} documented a marked {\em Spitzer} IR ``gap'' in compact group galaxies, with very few galaxies observed between the star-forming cloud and the quiescent cloud.  \citet{walker+10} suggested that compact group galaxies within this ``gap'' were likely in the midst of a rapid morphological transformation.  \citet{cluver+13} showed that the ``gap'' galaxies in compact groups tend to have warm hydrogen emission traced by the {\em Spitzer} Infrared Spectrograph (IRS, \citealt{spitzerIRS}) that is enhanced beyond the level that photon-dominated regions powered by star formation alone could explain, termed Molecular Hydrogen Emission Galaxies (MOHEGs; \citealt{ogle+07}).  \citet{cluver+13} suggested that this enhanced H$_2$ emission might be energized by shocks caused by collisions with the clumpy intragroup medium.  MOHEG ``gap'' HCGs therefore represent ideal laboratories to test the interplay between rapid galaxy transition, SF quenching, and the disruption of the star-forming fuel and ISM.

Evidence that star formation might be taking place inefficiently in transitioning galaxies has started to mount, including in radio galaxies \citep{nesvadba+10,guillard+15}, AGN-driven molecular outflow host NGC\,1266 \citep{a15_sfsupp}, as well as in HCGs. Both the shock in Stephan's Quintet \citep{appleton+06,guillard+09,guillard+12,konstantopoulos+14} and the HCG galaxy HCG\,57a \citep{a14_hcg57} seem to exhibit suppressed star formation. In these cases, the authors suggested that the injection of turbulence might be causing this star formation inhibition, either from collisions with the radio jets or AGN outflows, or the collisions and interactions within the group environment.

We present new Combined Array from Research in Millimeter Astronomy (CARMA)\footnote{http://www.mmarray.org} CO(1--0) observations of 12 HCGs, including multiple galaxies considered ``gap'' galaxies and MOHEGs (shown in Table \ref{tab:gal_params}).  In \S\ref{sec:obs}, we describe the sample selection and observations from CARMA, including reduction and analysis methods.  In \S\ref{sec:results}, we present the molecular properties of the sample, including their position on the Kennicutt-Schmidt (K-S) relation \citep{ken98}.  In \S\ref{sec:discussion}, we discuss these results in the context of transitioning galaxies.  In \S\ref{sec:summary}, we summarize our results.  We comment and present the maps of individual galaxies in the Appendix. The cosmological parameters $H_0 = 70~$km~s$^{-1}$, $\Omega_{\rm m} = 0.3$ and $\Omega_\Lambda = 0.7$ \citep{wmap} are used throughout.

%%%%%%%%%%%%%%%%% OBSERVATIONS %%%%%%%%%%%%%%%%%%%%%
\section{Observations and Analysis}
\label{sec:obs}
\subsection{Sample Selection}
While single dish observations are able to provide information about the CO luminosity, and possibly some information about the molecular gas kinematics (such as line-width and line-shape), it is unable to provide spatial information about the molecular gas.  Interferometry is able to provide information about the extent and distribution of the molecular gas, allowing for direct comparisons between the CO and stellar mass, or star formation. For this reason, we chose to use CARMA to follow up a subset of the CO-bright, warm H$_2$-bright HCG galaxies presented in \citet{lisenfeld+14}.

The galaxies chosen for the new CARMA imaging were drawn from the HCGs that were detected in warm H$_2$ using the {\em Spitzer} IRS\footnote{One galaxy, HCG\,57d, was not included in the original {\em Spitzer} IRS footprint.} (usually with boosted $L_{\rm H_2,warm}/L_{7.7\mu{\rm m PAH}}$ ratios; \citealt{ogle+07,cluver+13}) as well as detected in CO(1--0) with the Institut de Radioastronomie Millim\'etrique (IRAM) 30m \citep{lisenfeld+14}. Only 10\% of HCG galaxies are MOHEGs \citep{cluver+13}. From this subset of CO-bright, H$_2$-strong galaxies, we chose a subset that maximized the overlap with {\em Herschel} \cplus-detected objects (Appleton et al., in preparation).  These objects also tended to show early-type morphologies based on their t-types from HyperLEDA\footnote{http://leda.univ-lyon1.fr/} \citep{hyperleda}. 11 HCGs altogether were chosen for the CARMA sample, including 8 HCGs with corresponding \cplus\ and \oi\ {\em Herschel} observations.  HCG\,96, although not observed through the {\em Spitzer} HCG program, had data in both the CARMA and {\em Herschel} archives, and was therefore added to our sample.  All sample galaxies and their properties are listed in Table~\ref{tab:gal_params}.

%%%%%%%%%%%%%%% Table 2 %%%%%%%%%%%%%%%
\begin{table*}[t]
\centering
\caption{CARMA CO(1--0) Observing Parameters} \vspace{1mm}
\begin{tabular}{l c c c c c c c}
\hline \hline
{\bf Name} & Semester & Total & Gain &  $\theta_{\rm maj}\times\theta_{\rm min}$ & Kelvin & $\Delta$V$^\ddagger$\\
&& Hours$^\flat$ & calibrator & ($''$) & per Jy  & (km\,/\,s)  \\
\hline
HCG\,25 & 2014a & 12.14 & 0339-017 & $6.1\times3.6$ & 4.2734 & 21\\
HCG\,40 & 2014a & 17.67 & 0825+031 & $5.9\times3.9$ & 4.0316 & 21\\
HCG\,47 & 2014a & 12.41 & 0956+252 & $4.9\times3.4$ & 5.4678 & 32\\
HCG\,55 & 2013b & 7.64 & 1048+717 & $7.4\times7.1$ & 1.7767 & 33\\
HCG\,57 & 2013a & 8.61 & 1224+213 & $4.6\times3.3$ & 6.0303 & 31\\
HCG\,68 & 2013b & 16.24 & 1310+323 & $5.1\times3.7$ & 4.9006 & 30\\
HCG\,79 & 2013b & 9.75 & 1613+342 & $7.4\times6.4$ & 1.9520 & 31\\
HCG\,82 & 2014a & 10.83 & 1635+381 & $4.3\times3.6$ & 5.9389 & 21\\
HCG\,91 & 2013b & 8.06 & 2258-279 & $8.5\times4.1$ & 2.6671 & 21\\
HCG\,95 & 2013a & 13.26 & 3C454.3 & $3.9\times3.0$ & 7.9045 & 21\\
HCG\,96$^\dagger$ & 2010b & 7.33 & -0010+109 &$2.5\times2.4$ & 15.0203 & 10\\
HCG\,100 & 2013b & 9.75 & 3C454.3 & $4.6\times3.6$ & 5.5428 & 21\\
\hline \hline
\end{tabular} \\
\label{tab:co_params}
\raggedright {\footnotesize
$^\ddagger$Channel width\\
$^\flat$Total time on-source time\\
$^\dagger$Archival CARMA data\\
%$^\diamond$OVRO data used the mean of the moment1 map for the $v_{\rm sys}$
}
\end{table*}

%%%%%%%%%%%%%%% Table 3 %%%%%%%%%%%%%%%
\begin{table*}
\centering
\caption{CARMA Derived Properties} \vspace{1mm}
\begin{tabular}{l r r c c c c c c r r c}
\hline \hline
Name & $v_{\rm sys}$~~~~ & {\em z}$_{\rm co}$~~~~  & Vel. Range & $F_{\rm 100\,GHz}^\ddagger$ & RMS$^\flat$ & Flux$^\ddagger$ & $L_{\rm CO}$ & $M$(H$_2$)$^{\ddagger,\natural}$ & \multicolumn{2}{c}{Area} & $\Sigma($H$_2$)$^{\ddagger,\natural}$ \\
& (km\,/\,s) & & (km\,/\,s) & (mJy) & (mJy\,/\,bm) & (Jy~km\,/\,s) & ($10^4 L_\odot$) & ($10^9 M_\odot$) & ($\Box''$) & (kpc$^2$) & ($M_\odot$\,/\,pc$^{2}$) \\
\hline
25b & 6371 & 0.02125 & 6037--6684 & $<0.525$ & 4.83 & $21.16\pm 2.37$ & $1.83\pm 0.20$ & $1.60\pm 0.18$ & 191 & 30.4 & $52.7\pm 5.9$ \\
40c & 6419 & 0.02141 & 6085--6753 & $<0.634$ & 6.69 & $93.32\pm 2.97$ & $9.70\pm 0.31$ & $8.49\pm 0.27$ & 563 & 107.7 & $78.9\pm 2.5$ \\
47a & 9637 & 0.03215 & 9429--9844 & $<0.307$ & 3.25 & $64.56\pm 1.89$ & $14.91\pm 0.44$ & $13.06\pm 0.38$ & 992 & 408.4 & $32.0\pm 0.9$ \\
55c & 15760 & 0.05257 & 15394--16125 & $<0.595$ & 3.97 & $10.09\pm 1.33$ & $5.67\pm 0.74$ & $4.96\pm 0.65$ & 247 & 233.1 & $21.3\pm 2.8$ \\
57a & 8723 & 0.02910 & 8214--9231 & $<0.391$ & 3.85 & $59.16\pm 3.94$ & $11.12\pm 0.74$ & $9.74\pm 0.65$ & 487 & 164.6 & $59.1\pm 3.9$ \\
57d & 8944 & 0.02983 & 8753--9134 & $<0.391$ & 3.85 & $7.50\pm 0.33$ & $1.41\pm 0.06$ & $1.23\pm 0.05$ & 176 & 59.3 & $20.8\pm 0.9$ \\
68a & 2249 & 0.00750 & 1457--3040 & $6.687\pm 0.198$ & 5.07 & $23.65\pm 2.16$ & $0.34\pm 0.03$ & $0.30\pm 0.03$ & 152 & 4.2 & $71.1\pm 6.5$ \\
68b$^\dagger$ & $-$ & $-$ &  --  & $1.597\pm 0.243$ & -- & -- &  --  &  --  & -- & -- & $-$ \\
79a & 4149 & 0.01384 & 3841--4457 & $0.756\pm 0.172$ & 5.02 & $25.42\pm 1.91$ & $1.25\pm 0.09$ & $1.10\pm 0.08$ & 382 & 35.4 & $31.0\pm 2.3$ \\
82b & 10415 & 0.03474 & 10181--10652 & $0.483\pm 0.126$ & 3.69 & $12.92\pm 1.28$ & $3.28\pm 0.32$ & $2.87\pm 0.28$ & 99 & 44.5 & $64.6\pm 6.4$ \\
91a & 6857 & 0.02287 & 6543--7171 & $<0.991$ & 9.53 & $173.54\pm 4.80$ & $17.44\pm 0.48$ & $15.28\pm 0.42$ & 2050 & 377.4 & $40.5\pm 1.1$ \\
95c & 11540 & 0.03849 & 11270--11831 & $0.710\pm 0.169$ & 4.15 & $24.74\pm 1.66$ & $7.13\pm 0.48$ & $6.25\pm 0.42$ & 173 & 87.3 & $71.6\pm 4.8$ \\
96a & 8638 & 0.02881 & 8482--8793 & $3.911\pm 0.383$ & 7.74 & $121.75\pm 1.59$ & $19.09\pm 0.25$ & $16.72\pm 0.22$ & 533 & 150.5 & $111.1\pm 1.5$ \\
96c & 8809 & 0.02938 & 8592--9025 & $<1.150$ & 7.74 & $9.61\pm 0.13$ & $1.59\pm 0.02$ & $1.39\pm 0.02$ & 18 & 5.3 & $260.3\pm 3.5$ \\
100a & 5220 & 0.01741 & 4888--5551 & $0.429\pm 0.090$ & 4.32 & $36.99\pm 2.34$ & $2.11\pm 0.13$ & $1.84\pm 0.12$ & 343 & 36.3 & $50.7\pm 3.2$ \\

\hline \hline
\end{tabular} \\
\label{tab:derived_params}
\raggedright {\footnotesize
$^\ddagger$Does not include 20\% absolute flux calibration uncertainty\\
$^\flat$RMS noise per channel \\
$^\natural$Does not include 30\% conversion uncertainty \citep{bolatto+13} in $\alpha_{\rm CO}$, assuming $X_{\rm CO}$\,=\,2$\times$10$^{20}$\,cm$^{-2}$\,(K~km~s$^{-1}$)$^{-1}$\\
$^\dagger$Detected only in 3mm continuum; in the CARMA observations, see \citet{lisenfeld+14} for its detected CO properties based on IRAM data\\
}
\end{table*}

\subsection{CARMA}
The HCGs were observed with CARMA, an interferometric array of 15 radio dishes (6$\times$10.4m and 9$\times$6.1m) located in the Eastern Sierras in California \citep{carma}. Observations were taken over the course of three semesters between 12 Mar 2013 -- 16 Jun 2014. One, HCG\,96, was taken from the archive.  Thus, we observed a total of 12 HCGs (and 15 individual galaxies, including HCG\,68b, detected only in 3mm continuum)\footnote{HCG\,40e was also tentatively detected, but below a signal-to-noise ratio of 3, and only a small subset of the velocity structure was recovered.}.  The observing strategy and data reduction were performed in a manner identical to that of the ATLAS$^{\rm 3D}$ galaxies in \citet{alatalo+13}.  Table \ref{tab:co_params} presents the semester, gain calibrator, bandpass calibrator, total hours on source, and beam full-width at half-maximum (FWHM) for each of the HCGs observed with CARMA in the D-array (with baselines between 11--150m).  Galaxies with new CARMA observations had a correlator configuration of 8$\times$500\,MHz window in each sideband,\footnote{The observations of HCG\,96 were taken with a correlator configuration of 5$\times$250 and 3$\times$500\,MHz per sideband, totalling 5500\,MHz of bandwidth.} with the CO(1--0) line utilizing the highest resolution 500\,MHz mode (with channel resolution of $\approx 15$\,km~s$^{-1}$).  This meant there was a sufficiently large bandwidth to measure 3mm continuum, which was successfully detected in 7/15 sources (listed in Table \ref{tab:derived_params}).  Continuum contributions were subtracted in $uv$-space using the {\sc miriad} task {\tt uvlin} \citep{miriad}, as detailed in \citet{alatalo+13}.

The resulting channel maps and moment maps were also constructed in identical fashion to \citet{alatalo+13}.  Figures \ref{fig:hcg25}--\ref{fig:hcg100} showcase the CARMA data for each individual HCG, including the channel maps, integrated intensity (moment0), and mean velocity (moment1) maps, as well as integrated spectra and position-velocity diagrams (PVD).  The PVD was constructed by creating a slice in the velocity cube at a certain position angle (shown as a dashed line on the moment0 figure), and integrating across a slice in space.

The integrated spectrum was constructed by using the moment0 map to create a clip-mask and integrating the flux within the moment0-defined (unmasked) aperture.  This was done separately for each galaxy.  The root mean square (RMS) noise was then taken by calculating the standard deviation of all pixels in the cube that were outside of the moment0-aperture per channel and is listed in Table \ref{tab:derived_params}.  An additional noise correction of 30\% was also added in quadrature to the RMS noise to account for the oversampling of the maps (see: \citealt{a15_co13} for details).  The RMS noise per channel for the integrated spectrum was then calculated by multiplying the RMS of the entire data cube by the square root of the total number of beams in the moment0-aperture.

%%%%%%%%%%%%%%% Figure 2 %%%%%%%%%%%%%%%
\begin{figure*}[t]
\subfigure{\includegraphics[width=0.49\textwidth]{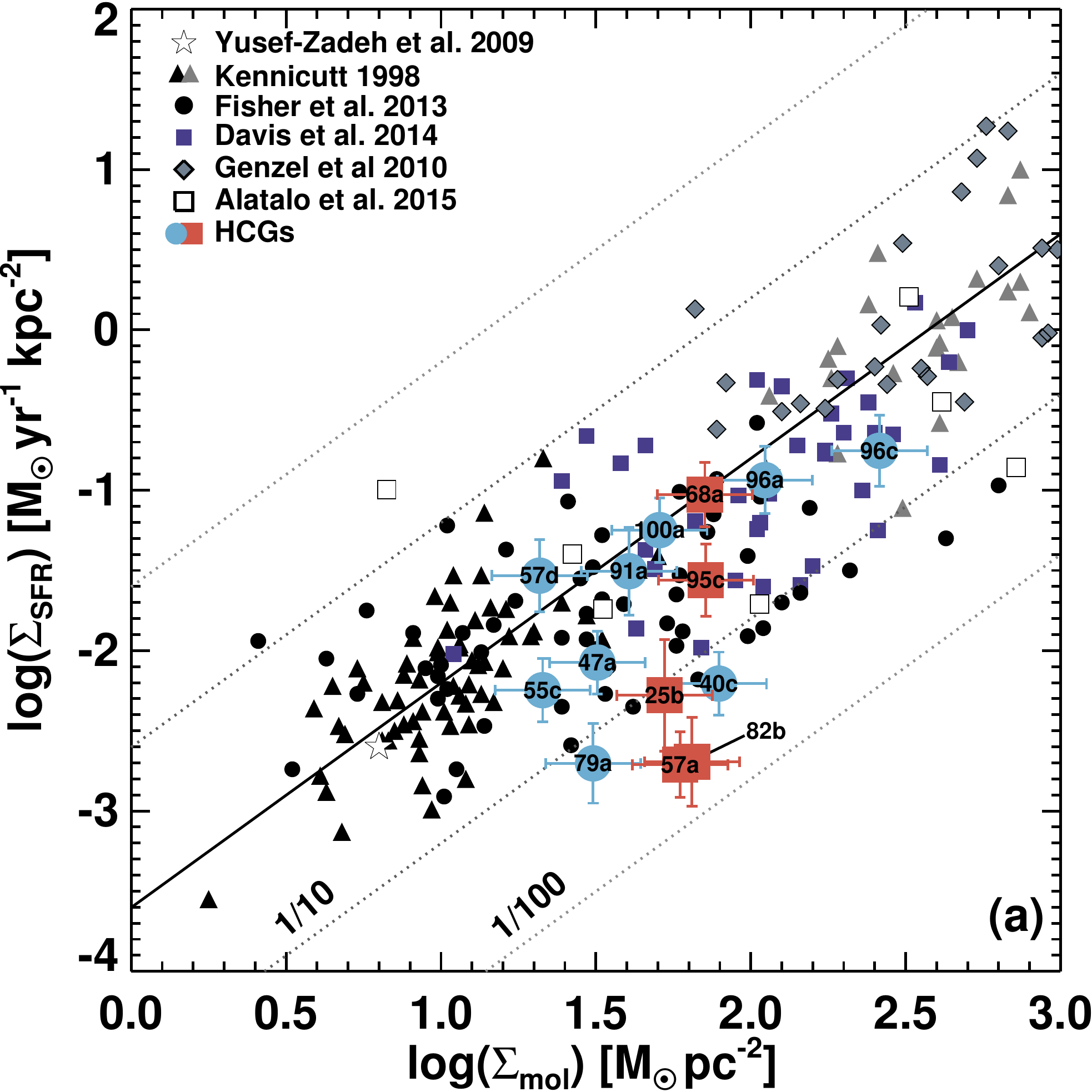}}
\subfigure{\includegraphics[width=0.49\textwidth]{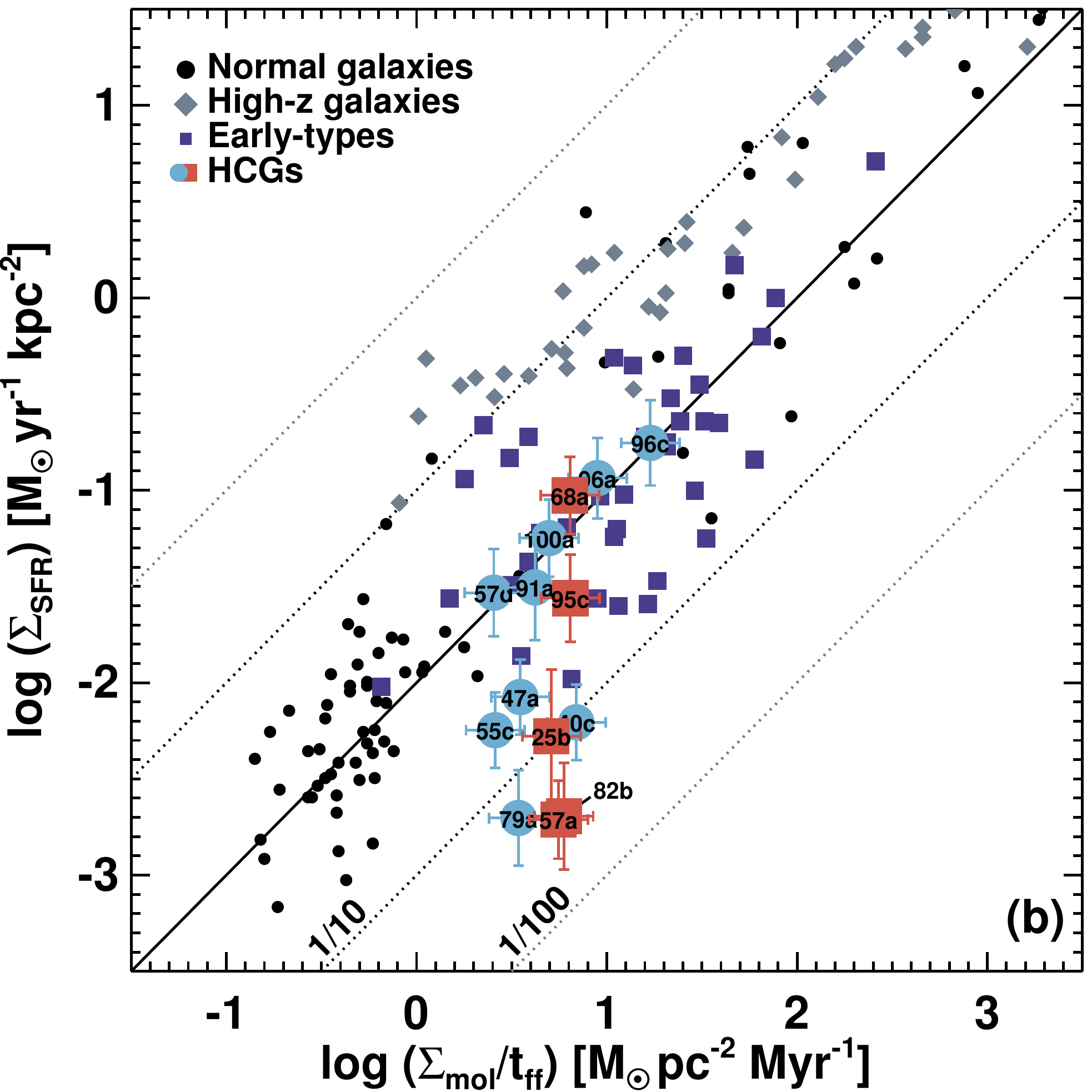}}
%\subfigure{\includegraphics[width=0.49\textwidth]{figures/supp_hist.pdf}}
\caption{{\bf (Left):} SFR and gas surface density in our HCG sample are shown in comparison on the Kennicutt-Schmidt relation \citep{ken98}.  The SFRs were taken from \citet{bitsakis+14}, normalized to a Salpeter Initial Mass Function (IMF; \citealt{salpeter55}).  The HCGs are compared to the Milky Way \citep{yusef-zadeh+09}, normal galaxies or LIRGs \citep{ken98,fisher+13}, CO-imaged early-type galaxies \citep{davis+14}, high redshift objects \citep{genzel+10}, and radio galaxies \citep{ogle+10,a15_sfsupp}, all renormalized to a Salpeter IMF. The solid black line represents the Kennicutt-Schmidt relation, and dashed lines represent levels of 1/10 and 1/100 suppression (below) and enhancement (above).  The shapes and colors of the HCG points are based on whether they meet the MOHEG criterion from \citet{ogle+07}, with red squares for HCGs with $L_{\rm H_{2,warm}}/L_{\rm 7.7\mu m PAH}$\,$>$\,0.04, considered confirmed MOHEGs, and blue circles representing those below this threshold. The SFR error bars include the 10\% errors quoted in \citet{bitsakis+14}, and the 0.38\,dex scatter in {\sc magphys}-derived SFRs from \citet{lanz+13}, added in quadrature. {\bf(Right):} The SF laws from \citet{krumholz+12} are plotted for nearby and high-$z$ galaxies (black dots and gray diamonds, respectively; \citealt{krumholz+12} and references therein), and early-type galaxies (indigo squares; \citealt{davis+14}), all normalized to a Salpeter IMF \citep{salpeter55}. The axes of panel b were chosen to match the axes shown in \citet{krumholz+12}/ Dotted lines represent suppressions and enhancements of 10 and 100.  The HCGs found to be suppressed in panel a are also suppressed in panel b.}
%The distribution of SF suppression (\supp; see \S\ref{sec:ks}) in the HCGs compared with all other galaxies represented in the K-S plot.  Although there are fewer HCGs represented, they are far more suppressed on average than the comparison sample.}
\label{fig:KS}
\end{figure*}

To calculate an integrated line flux for each galaxy, we determined the extent of emission within the cube (shaded blue in the integrated spectrum) and summed across them.  The line flux RMS was then calculated by multiplying the RMS per channel of the spectrum by the velocity width (listed in Table \ref{tab:co_params}) and the square root of the total number of channels integrated to derive the line flux.

Table \ref{tab:derived_params} lists the properties derived from the imaging data of the individually detected galaxies, including the RMS noise in the channel maps, the detected continuum levels, and the total detected CO line fluxes. The spatial extent of the molecular gas was determined by summing the total number of unmasked pixels (the moment0-aperture) in the moment maps.  CO luminosities (shown in Table \ref{tab:derived_params}) were calculated using the luminosity distance to the source (listed in Table \ref{tab:gal_params}). The H$_2$ mass was determined using the $L_{\rm CO}$--$M$(H$_2$) relation: 

\begin{equation}
M({\rm H}_2) = 1.05\times10^4\frac{S_{\rm CO}\Delta v D_{\rm L}^2}{1+v_{\rm sys}/c}~M_\odot,
\end{equation}

\noindent which assumes $X_{\rm CO} = 2\times10^{20}$\,cm$^{-2}$~(K~km~s$^{-1}$)$^{-1}$ (the mean conversion factor presented in \citealt{bolatto+13}). $S_{\rm CO} \Delta v$ is the CO(1--0) flux (in Jy~km~s$^{-1}$), $D_{\rm L}$ is the luminosity distance (in Mpc), $v_{\rm sys}$ is the optically-defined systemic velocity (in km~s$^{-1}$), and $c$ is the speed of light (in km~s$^{-1}$).  The time variability of flux calibrators at 3mm continuum adds an additional $\approx$\,20\% uncertainty to the CO(1--0) flux, and the $L_{\rm CO}$--$M$(H$_2$) conversion also carries a 30\% uncertainty \citep{bolatto+13}, imposing an additional 35\% absolute flux uncertainty to the measured CO masses\footnote{Errors reported in Table~\ref{tab:derived_params} do not include the 35\% uncertainty, though it is included in error bars in all figures.}.

\subsection{Comparison of the CARMA and IRAM data}
Thirteen of the 14 CO-imaged HCG galaxies presented in this paper were also observed by \citet{lisenfeld+14} using the IRAM\,30m (only HCG\,96c was not observed). Figure~\ref{fig:comparison} compares the CO(1--0) luminosities derived from both sets of data.  The measured CO(1--0) luminosities from CARMA and the IRAM 30m are in good agreement, confirming that the CARMA observations do not resolve out substantial flux, and are therefore a reasonable representation of the molecular gas in these systems.  HCG\,57d, for which CARMA detected only half of the flux detected with the IRAM\,30m (possibly due to resolving out flux, and sensitivity), is an outlier \citep{a14_hcg57}.

\section{Results}
\label{sec:results}
\subsection{Molecular gas morphologies of the HCG galaxies}
Following the morphological classification used by \citet{alatalo+13} for ATLAS$^{\rm 3D}$ early-type galaxies, we have morphologically classified the molecular gas in all of the CO-imaged HCGs (listed in Table~\ref{tab:gal_params}; detailed discussion can be found in Appendix~\ref{app:individuals}).  We classify each galaxy as being either a disk (D), spiral (S), bar+ring (B+R), ring (R), mildly disrupted (M), or a combination of multiple distinct classifications.  Details of the morphological classifications can be found in \citet{alatalo+13}.  The morphologies seen in our HCGs tends to be a mix of regular rotation and dynamically excited structures, with a lack of strongly disrupted objects, which are quite prevalent in Ultraluminous Infrared Galaxies (ULIRGs; \citealt{sanders+mirabel96}) and interacting galaxies \citep{wilson+08}.  Overall though, our CO-imaged galaxies have appearances comparable to field galaxies, including both spirals \citep{bimasong} and early-types \citep{alatalo+13}.

\subsection{The Kennicutt-Schmidt relation in warm, H$_2$-bright HCG galaxies}
\label{sec:ks}
Figure~\ref{fig:KS} displays the molecular gas surface density of each HCG compared with its star formation rate (SFR) surface density (calculated assuming the SF and gas are co-spatial), and using the SFR derived in \citet{bitsakis+14}.  SFRs calculated by {\sc magphys} \citep{magphys} seem to have a scatter of 0.38\,dex \citep{lanz+13}, which we have included in our SF rate uncertainty.  To test whether the SF histories assumed in \citet{bitsakis+14} could significantly alter our conclusions, we input varying SF histories (including continuous and truncated models) and derived SF rates using {\sc cigale} \citep{ciesla+15}. The SF rates derived from {\sc cigale} did not vary by more than the assumed scatter reported in \citet{lanz+13}.

%%%%%%%%%%%%%%% Figure 3 %%%%%%%%%%%%%%%
\begin{figure}[t]
\includegraphics[width=0.49\textwidth]{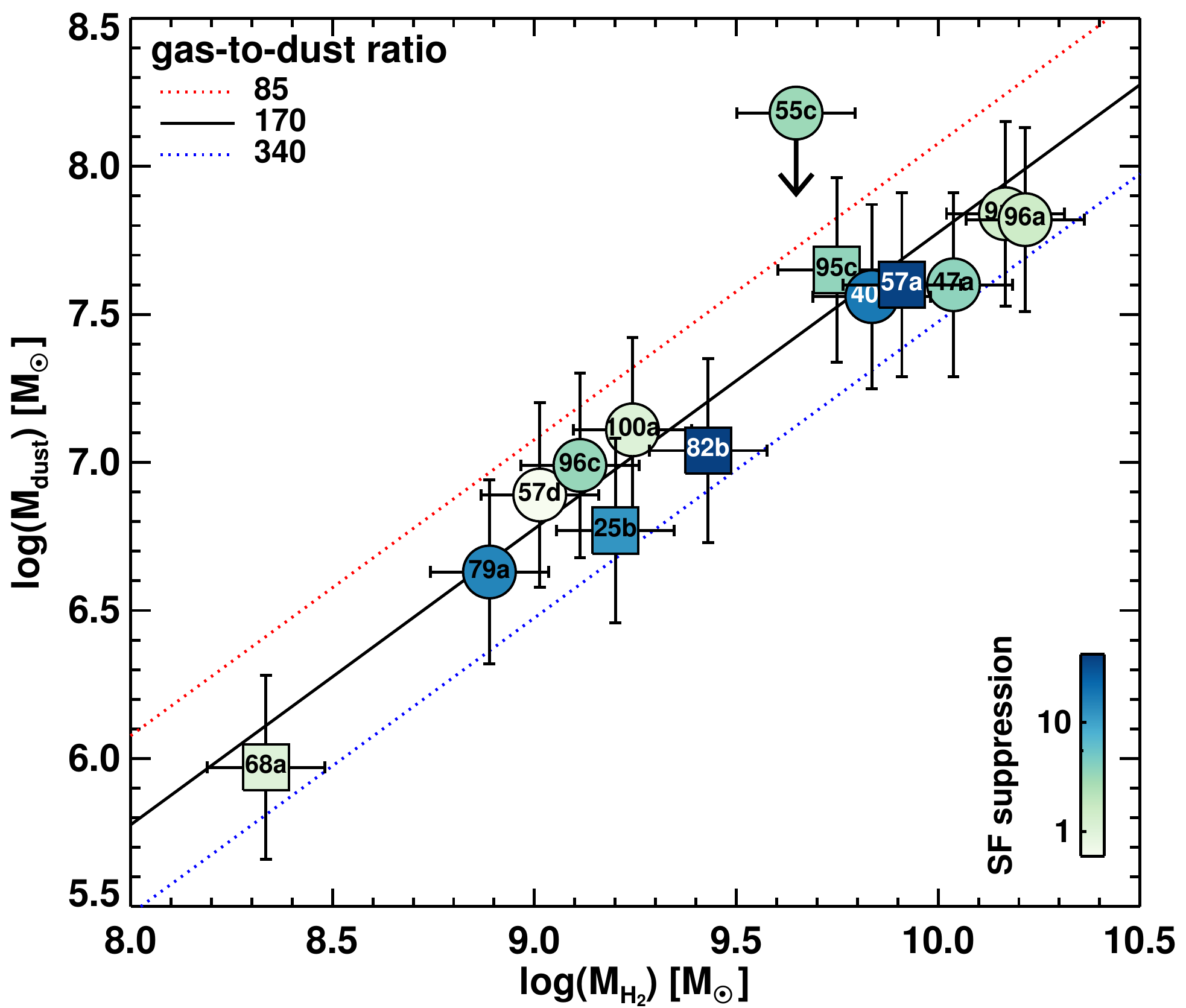}
\caption{The molecular gas-to-dust ratios of the HCGs in this sample, using the molecular gas mass determined in this work, and the dust masses from \citet{bitsakis+14}, with an uncertainty of 0.5\,dex \citep{dacunha+10}.  The dust mass of HCG\,55c is an upper limit.  Red and blue dotted lines represent factors of two of the average gas-to-dust ratio. The colors of the points represent the suppression of SF in the molecular gas in each galaxy.  The gas-to-dust ratios of the HCGs appear consistent across a large range in molecular gas masses, as well as values of \supp.}
\label{fig:gastodust}
\end{figure}

$\Sigma_{\rm SFR}$ for the HCGs is compared with $\Sigma_{\rm SFR}$ of the Milky Way \citep{yusef-zadeh+09}, normal galaxies \citep{ken98, fisher+13}, high redshift galaxies \citep{genzel+10}, luminous infrared galaxies (LIRGs; \citealt{ken98}), radio galaxies \citep{ogle+10,a15_sfsupp}, and CO-imaged early-type galaxies \citep{davis+14}.  The HCGs are color-coded based on their $L_{\rm H_2, warm}/L_{\rm 7.7\mu m PAH}$ ratios from \citet{cluver+13}, with MOHEGs represented by red squares, and non-MOHEGs represented by blue circles.  Overall, most HCG galaxies fall within the scatter of the K-S relation \citep{ken98}, but HCG galaxies as a group are more likely to be found in the lower half of the scatter than the upper half (although a larger sample will be required to determine whether this is statistically significant). This result is consistent with the single-dish results from \citet{lisenfeld+14}, and there are some objects (in particular HCG\,25b, 40c, 57a, 79a, and 82b) that exhibit a much more substantial scatter off of the K-S relation.  We define the degree of SF suppression, \supp, as the ratio between the expected SFR surface density ($\Sigma_{\rm SFR}$) from the measured molecular gas surface density $\Sigma_{\rm mol}$ using the K-S relation \citep{ken98}, and the observed $\Sigma_{\rm SFR}$.  Values of \supp\ for each HCG are listed in Table~\ref{tab:supp}. 

%%%%%%%%%%%%%%% Table 5 %%%%%%%%%%%%%%%
\begin{table*}[t]
\centering
\caption{Star formation suppression values} \vspace{1mm}
\begin{tabular}{l r r r r r r r r r}
\hline \hline
Name & $t_{\rm ff}^a$~ & \supp$_{\rm KS}^b$ & \supp$_{\rm \epsilon_{ff}}^c$ & $t_{\rm dep}^d$ & $\sigma_{\rm norm\,SF}^e$ & $\sigma_{\rm Q=1}^f$ &
 $E_{\rm turb}^g$ & $L_{\rm turb}^h$ & $L_{\rm H_2,warm}/L_{\rm turb}^i$ \\
& (Myr) & & & (Gyr) & (km\,/\,s) & (km\,/\,s) & ($10^{55}$ ergs) & ($10^{39}$ erg\,/\,s) \\
\hline
25b & 10.3 & 12.2 & 9.1 & 10.0 & 91 & 51 & 7.2 & 31 & 0.34\\
40c & 11.4 & 18.2 & 10.4 & 12.7 & 104 & 62 & 57 & 303 & 0.36\\
47a & 9.1 & 3.8 & 3.9 & 3.8 & 39 & 40 & 35 & 120 & 0.50\\
55c & 8.2 & 3.2 & 4.3 & 3.8 & 43 & 32 & 9.0 & 25 & 2.15\\
57a & 10.6 & 38.9 & 26.8 & 30.4 & 268 & 54 & 49 & 226 & 0.82\\ 
57d & 8.2 & 0.6 & 0.8 & 0.7 & 8 & 32 & 2.2 & 6.0 & $-$\\
68a & 11.1 & 1.0 & 0.6 & 0.8 & 6 & 59 & 1.8 & 9.0 & 0.75\\ 
79a & 9.0 & 15.4 & 16.2 & 15.7 & 162 & 39 & 2.9 & 9.6 & 1.32\\
82b  & 10.8 & 42.2 & 27.4 & 31.9 & 274 & 56 & 16 & 76 & 0.34\\
91a & 9.7 & 1.4 & 1.3 & 1.3 & 13 & 44 & 52 & 200 & 0.33\\
95c & 11.1 & 3.6 & 2.2 & 2.6 & 22 & 59 & 37 & 193 & 0.57\\  
96a & 12.4 & 1.6 & 0.7 & 1.0 & 7 & 74 & 158 & 997 & 0.14\\
96c & 15.4 & 3.4 & 0.9 &1.5 & 9 & 113 & 31 & 297 & 0.22\\
100a & 10.2 & 1.1 & 0.8 & 0.9 & 8 & 50 & 7.9 & 34 & 0.95\\
\hline \hline
\end{tabular} \\
\label{tab:supp}
\raggedright {\footnotesize
Derived SF suppression parameters for the molecular gas in HCGs including ($a$) the free-free timescale using Equation\,2, assuming that $\sigma_{\rm gas}$\,=\,10\,km~s$^{-1}$, ($b$) \supp\ (the deviation from the K-S relation, ($c$) the star formation suppression from the \citet{krumholz+12} law using $t_{\rm ff}$ (and assuming $\sigma_{\rm gas}$\,=\,10\,km~s$^{-1}$) and ($d$) the depletion timescale ($M_{\rm mol}$/SFR) for each of the CO-imaged HCG galaxy. ($e$) The necessary molecular velocity dispersion ($\sigma_{\rm gas, norm\,SF}$) of the giant molecular clouds within the galaxies in order for each galaxy to have normal SF efficiency from \citet{krumholz+12}. ($f$) The required {\em global} molecular gas velocity dispersion to stabilize the molecular gas against collapse (Toomre $Q$\,$>$\,1). ($g$) The corresponding total turbulent energy injection required to stabilize the molecular disk. ($h$) The turbulent luminosity (the total turbulent energy injected per rotational period). ($i$) The ratio of the total warm H$_2$ luminosity from {\em Spitzer} IRS \citep{cluver+13} to the required turbulent luminosity.
}
\end{table*}

The molecular gas depletion (the molecular gas mass divided by the star formation rate) timescales associated with the HCG galaxies are also shown in Table~\ref{tab:supp}.  As expected, objects with normal \supp\ appear to have depletion timescales similar to those found in normal, star-forming galaxies ($\sim$\,1\,Gyr; \citealt{bigiel+08,saintonge+11}), and early-type galaxies \citep{davis+14}.  The objects with \supp\,$>$10 tend to have depletion timescales that are of order a Hubble time or longer, with HCG\,57a and HCG\,82b having depletion timescales of $>$\,30\,Gyr. These depletion timescales suggest that these systems will nearly indefinitely contain molecular gas, if the suppressions present are continuous (rather than discrete and sporadic) processes.  Given the timescale for galaxies within Hickson Compact groups to transform ($\sim$\,1Gyr; \citealt{walker+10,bitsakis+10,bitsakis+11}), it is likely that the suppression is also a transient phenomenon.  

%\textcolor{red}{This might be a good place to intro the concept of "depletion time". Depletion times on the main sequence are around 0.7 Gyrs typically I think (it would be good to find a ref for this). in the next section you talk about efficiency, which is just the inverse of thye depletion time, so perhaps we should decide which way to phrase it for consistency with the rest of the paper. }

In order to compare to early-type galaxies, another suppressed set of objects, we have calculated \supp\ for CO(1--0) imaged ATLAS$^{\rm 3D}$ galaxies \citep{ybc08,crocker+11,alatalo+13,davis+14}. We calculated the total SFRs of these objects using the (uncorrected) WISE 22$\mu$m emission,\footnote{The scatter associated with the $K_s$-band based factor used to correct the 22$\mu$m emission for aged stars from \citet{davis+14} is almost as large as the observed inefficiency.} assuming that the majority of the 22$\mu$m emission originates from SF in most CO-bright objects without strong AGNs (\citealt{calzetti+07}; Nyland et al., submitted).  In the CO-imaged subsample ATLAS$^{\rm 3D}$ galaxies from \citet{davis+14}, we find \supp$\approx$2.6, consistent with the predictions of \citet{martig+13} that the bulges in early-type galaxies can stabilize molecular disks, creating suppressions of 2--5.

For the HCGs in this work, we find the mean $\langle$\supp$_{\rm HCG}\rangle$\,$\approx$\,10$\pm$5. There also appears to be a marked bimodality seen between galaxies forming stars at normal efficiency, and a few that are extremely inefficient (with \supp\,$>$10), including HCG\,25b, 40c, 57a, 79a, and 82b. Given the compact nature of the molecular gas and strong X-ray emission in HCG\,68a \citep{liu11}, it is quite possible that a non-negligible fraction of its far-infrared emission originates from the AGN, rather than from SF, similar to NGC\,1266 \citep{a15_sfsupp}.  The contributions of AGNs to the spectral energy distributions of individual galaxies is discussed in Appendix \ref{app:decompir}.
%In HCG galaxies, morphological quenching and gravitationally stabilizing their molecular disks is not able to explain the extreme values of \supp\ that are observed, meaning that an alternative mechanism must be invoked to explain their systematically reduced SF rates.

We have also plotted the SF relation following the procedure in \citet{krumholz+12}. These authors claim that the SF rate in normal (non-LIRG) molecular configurations is dependent on the local freefall time of each individual molecular cloud, converting about 1\% of the molecular mass to stars per freefall time. \citet{krumholz+12} argue that since the majority of stars form in giant molecular clouds (GMCs), their individual properties are the dominant determinant of the star formation efficiency in the molecular gas of a galaxy, and thus the {\em local} GMC freefall time is more important than the {\em global} dynamical time (as is used by \citealt{ken98}). Assuming that the GMC distribution within swathes of galaxies is fairly consistent, \citet{krumholz+12} use this framework to universalize star formation laws, across normal, star-forming galaxies and interactions. \citet{davis+14} have also been able to show that this also applies successfully to early-type galaxies, while \citet{utomo+15} have shown that the GMCs within example early-type galaxy, NGC\,4526, matches the distributions of late-type local galaxies. We use the \citet{krumholz+12} relation to see if our star formation suppression can be reconciled by examining the GMC-scale efficiencies, rather than the global scales.  This framework was able to link Milky Way, local group galaxies, starburst, and high-redshift galaxy scales.  The free-fall time for each HCG was calculated using Equation 4 from \cite{krumholz+12}:

\begin{equation}
t_{\rm ff} = \frac{\pi^{1/4}}{\sqrt{8}} \frac{\sigma}{G(\Sigma_{\rm GMC}^3\Sigma_{\rm gas})^{1/4}},
\end{equation}

\noindent where $G$ is the gravitational constant, $\Sigma_{\rm GMC}$ is the average surface density of the giant molecular clouds within the system (estimated to be 85\,$M_\odot$~pc$^{-2}$; \citealt{bolatto+08}). $\Sigma_{\rm gas}$ for each object is presented in Table \ref{tab:derived_params}, and $\sigma$ is the velocity dispersion of the molecular gas within the individual GMCs, which we have assumed to be 10\,km~s$^{-1}$ (consistent with the 8\,km~s$^{-1}$ assumed in \citealt{krumholz+12}). $\sigma_{\rm gas}$ does not appear to vary by more than a factor of two in disk galaxies \citep{dib+06,walter+08,chung+09}, though $\sigma_{\rm gas}$\,$\approx$\,10~km~s$^{-1}$ might be an underestimate given the disrupted kinematics seen in some HCGs \citep{a14_hcg57}. Figure~\ref{fig:KS}b shows that some HCGs do not appear to fall on the typical efficiency (molecular mass converted into stars per free-fall time), $\epsilon$\,$\approx$\,0.01 relation with the other galaxies, and those that are suppressed in the original K-S plot (Fig~\ref{fig:KS}a) continue to appear suppressed in SF efficiency space as well. The least efficient galaxy, HCG\,82b, is 27 times less efficient than the mean efficiency from \citet{krumholz+12}. A larger sample of objects with suppressions at this level is needed to be able to constrain a duty cycle and determine how long objects are observed with these large depletion times.

\subsection{Gas-to-dust ratios in H$_2$-bright HCG galaxies}
\label{sec:gastodust}
Could the observed suppression in these HCGs be the result of using an incorrect $L_{\rm CO}$--$M$(H$_2$) conversion?  \citet{downes+solomon98} were able to show that ULIRGs did not follow the standard Milky Way relation. The molecular gas in ULIRGs is not distributed in discrete GMCs, instead being a more continuous distribution of molecular gas, and the velocity widths associated with the additional gas motions boosted the CO luminosity per unit gas mass. Anomalous $L_{\rm CO}$--$M$(H$_2$) conversion factors have also been identified in other MOHEGs, including NGC\,4258 \citep{ogle+14} and 3C\,293 \citep{lanz+15}, and thus is an important factor to check in our sample.

In order to determine whether our observations overestimate the molecular gas mass in the suppressed HCG galaxies, we turn to the gas-to-dust ratio as a test.  Figure~\ref{fig:gastodust} shows the gas-to-dust ratios of the CO-imaged HCG sample using dust masses calculated by \citet{bitsakis+14} using full UV--to--sub-mm SED fitting in {\sc magphys} \citep{magphys}.  Data points are color-coded based on the galaxy's \supp\ value.  The mean gas--to--dust ratio seen in our sample is $\approx170$, within the range found for solar metallicity nearby galaxies \citep{sandstrom+13} and the Key Insights on Nearby Galaxies: A Far-Infrared Survey with Herschel sample \citep{remy-ruyer+14}.  The HCG galaxies show a relationship between their dust mass and gas mass that both matches the standard value and has little--to--no dependence on \supp\ or classification of MOHEG.  This seems to indicate that the enhanced \supp\ values are due in part to a physical mechanism within the molecular gas, rather than an issue with converting CO luminosities into molecular gas masses.  Additional observations in $^{12}$CO isotopologues (e.g. $^{13}$CO and C$^{18}$O), as well as dense gas (e.g. HCN, HCO$^+$, and CS), are necessary to confirm whether these systems require a different conversion factor, but the consistent gas-to-dust ratio appears to support a standard conversion factor.

\section{Discussion}
\label{sec:discussion}
\subsection{AGNs in the HCG sample}
Table~\ref{tab:gal_params} lists the properties of each of the CO-imaged galaxies in our sample, including optical and radio signatures of AGNs. The detection of unresolved significant 3mm continuum in 43\% of the CARMA-imaged HCGs indicates the presence of AGNs, as the SFRs in these HCGs would not produce sufficient free-free emission at 3mm continuum to be detectable by CARMA.  Of the 7 HCGs detected in 3mm continuum, 6 have 1.4\,GHz radio detections as well (see Table \ref{tab:gal_params} for fluxes and sources).  HCG\,82b is the only object detected in 3mm continuum but not 1.4\,GHz.  The optical nuclear spectra of the 3mm-detected HCGs all show signatures of AGNs (Table~\ref{tab:gal_params}), either in the form of Seyfert-like spectra, low ionization nuclear emission line-region (LINER), or composite spectra \citep{martinez+10,cluver+13}. However, in many of these cases, slow and fast shocks might be mimicking these nuclear line ratios (\citealt{allen+08,rich+11}; Cales et al. 2015, submitted).  
%Given the interactions that take place in compact groups, molecular gas proximate to the supermassive black hole, exciting an AGN is unsurprising.

%%%%%%%%%%%%%%% Figure 4 %%%%%%%%%%%%%%%
\begin{figure}[t]
\raggedright
\includegraphics[width=0.47\textwidth]{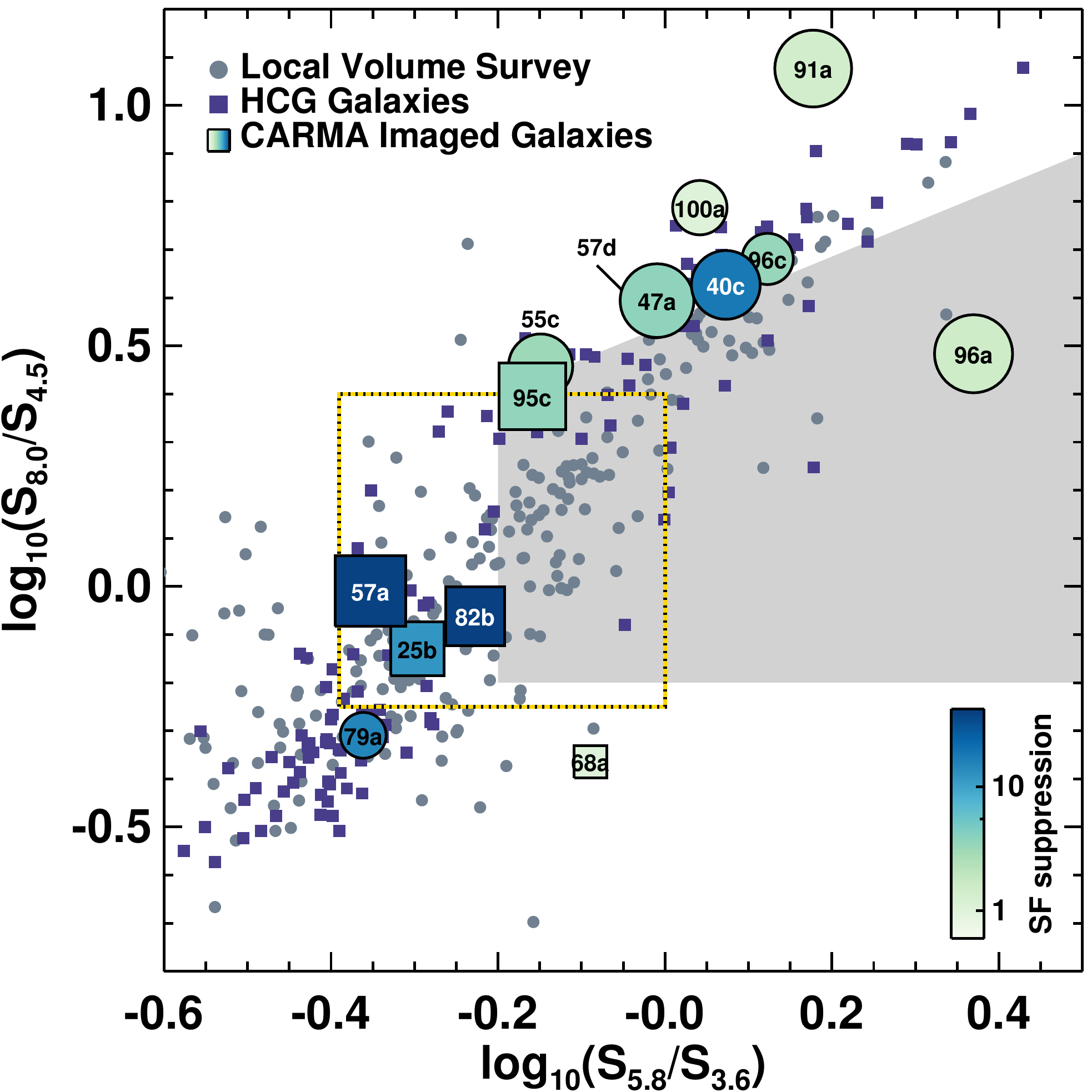}
\caption{The $S_{\rm 5.8\mu m}/S_{\rm 3.6\mu m}$ versus $S_{\rm 8.0\mu m}/S_{\rm 4.5\mu m}$ {\em Spitzer} IRAC colors of a sample of HCGs (blue squares; \citealt{bitsakis+11,cluver+13}) compared to the {\em Spitzer} local volume legacy sample (gray dots; \citealt{dale+09}), with the so-called Lacy wedge (circumscribing the AGN region in the plot) shown in light gray \citep{lacy+04,lacy+07,lacy+13}.  This plot shows the gap seen for HCGs described by \citet{johnson+07} and \citet{walker+10}, highlighted as a yellow and black dashed line. The CO-imaged HCGs are overplotted with the color of the point representing the level of SF suppression, the size of the points representing the mass of the molecular reservoir in each HCG galaxy, and the shapes indicating whether the galaxy is a MOHEG from \citet{cluver+13}. Amongst the labeled HCG galaxies, squares represent MOHEGs ($L_{\rm H_2,warm}/L_{\rm 7.7\mu m PAH}$\,$>$\,0.04) and circles non-MOHEGs.  HCGs with the strongest suppression tend to be those that occupy the infrared gap in {\em Spitzer} color space.}
\label{fig:lacy}
\end{figure}

%%%%%%%%%%%%%%% Figure 5 %%%%%%%%%%%%%%%
\begin{figure}[t]
\raggedright
\includegraphics[width=0.46\textwidth]{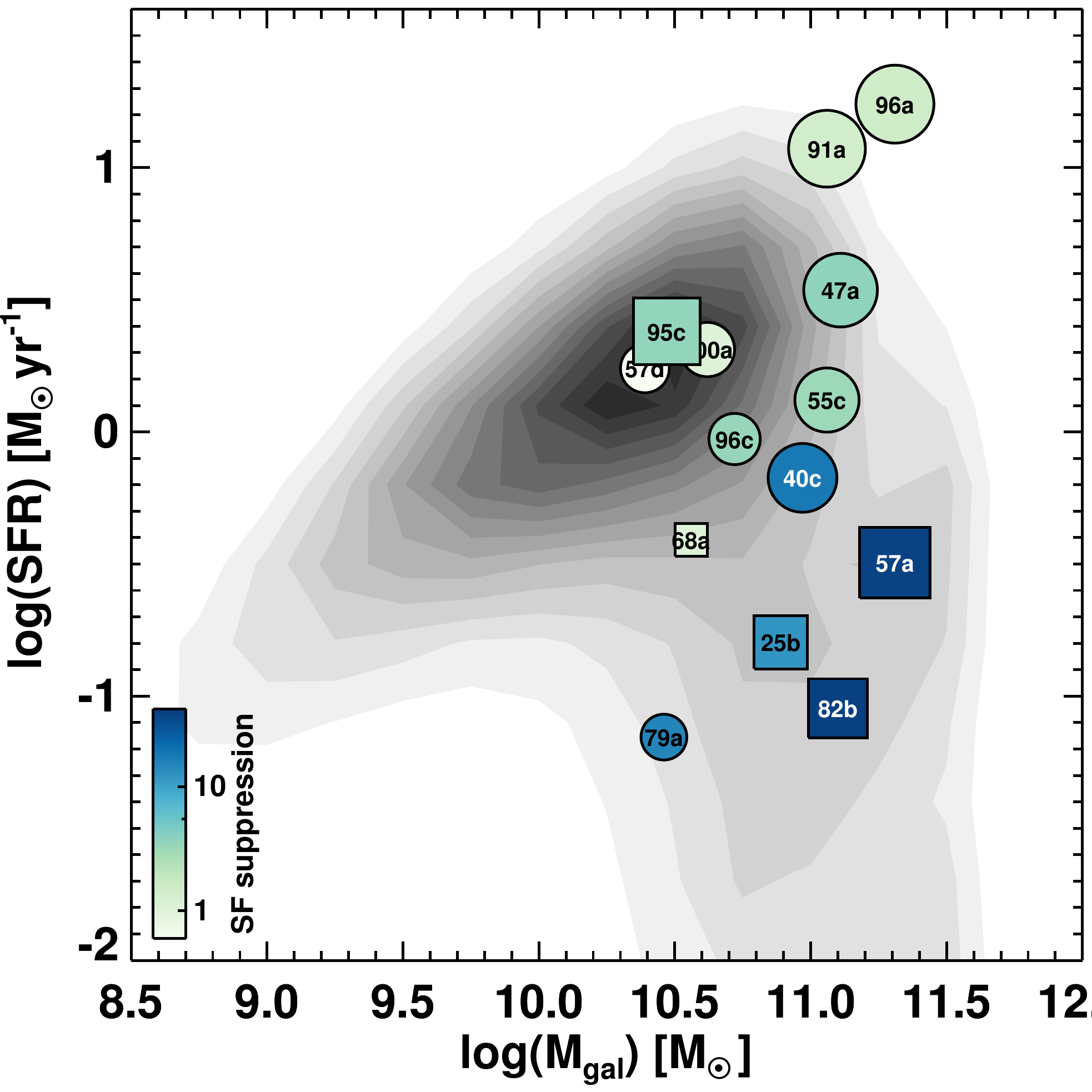}
\caption{The galaxy mass -- star formation relation of a large sample of SDSS galaxies (gray contours), with SFRs and stellar masses derived using {\sc magphys} \citep{magphys,chang+15}, showing the so-called ``star formation main sequence'' \citep{wuyts+11}.  Early-type galaxies are seen as the tail of high mass, low SFR.  The CARMA-imaged HCGs are also overlaid, with colors coded based on their SF suppressions and symbol sizes based on their molecular gas masses.  It is clear from this plot that the HCG galaxies with the most SF suppression are also the ones that are found to be farthest from the star formation main sequence.}
\label{fig:sfrMS}
\end{figure}

There is an abundance of dynamically-excited features amongst our sample of warm H$_2$ bright HCG galaxies, especially amongst galaxies with large values of \supp\ (although HCG\,91a seems to also show signs of a broad blue line wing in its spectrum), which might be a sign that some of the AGNs seen are interacting with the nuclear molecular gas, possibly adding to the suppression (as is seen in NGC\,1266; \citealt{a15_sfsupp}). Deeper analyses of these systems might reveal molecular outflows, as evidence is mounting that outflows are a common feature in objects with molecular reservoirs near an AGN \citep{garcia-burillo+14,garcia-burillo+15}. A detailed analysis to determine whether these line wings and central broad lines are indeed due to molecular outflows would require higher resolution, higher sensitivity observations and is therefore beyond the scope of this paper.

\subsection{\supp\ and its connection to turbulence}
\citet{martig+13} used simulations to predict that in a set of early-type galaxies with resolved molecular gas and SF, it was possible that massive bulges served to stabilize the molecular disks in these systems against gravitational collapse, reducing the SF efficiency of the molecular gas and thus suppressing SF. For our sample, \supp\ does not correlate with the stellar mass of the system, the molecular gas fraction, ($M_{\rm mol.~gas}/M_{\rm star}$), or the visually-derived morphological type of the galaxy, suggesting that an alternative driver of suppression must be at play in these systems.  HCGs that fall off of the K-S relation in Figure~\ref{fig:KS}a show similar behavior in Figure~\ref{fig:KS}b, suggesting that gravitational shears are not the main driver of the observed suppression in our sample (as was the case in early-type galaxies; \citealt{davis+14}). Alternatively, the derived properties for these objects might be inaccurate.  We could be underestimating $t_{\rm ff}$, for instance, because the molecular gas is not distributed in standard (Milky Way-like) molecular clouds, or because we have underestimated the molecular gas velocity dispersion (given the gas disruption that is common in HCGs).  In fact, in these systems, it is possible that both diverge from the norm.

\citet{johnson+07} noted that a fraction of galaxies within HCGs seem to undergo a rapid transition between star-forming and quiescence, with a notable lack of galaxies with intermediate infrared colors.  Figure~\ref{fig:lacy} plots the {\em Spitzer} Infrared Array Camera (IRAC; \citealt{spitzerIRAC}) colors of the CO-imaged HCGs originally plotted in \citet{lacy+04}, overlaid with results from the {\em Spitzer} local volume legacy survey \citep{dale+09}, as well as the HCGs from \citet{bitsakis+11}. Figure~\ref{fig:lacy} shows that those of our galaxies located in the Spitzer IR gap tend to also have a high degree of SF suppression.

\citet{cluver+13} showed that the HCG galaxies that lie within the gap (and thus are rapidly transitioning; \citealt{walker+10}) also tend to be those with prominent warm H$_2$ signatures that required mechanisms in addition to photon-dominated regions. Objects with \supp\,$\lesssim$\,10 might be able to be reconciled with normal SF of \citet{krumholz+12} by revising our estimate of $\sigma_{\rm gas}$ upward in the free-fall time equation, given that observations of high-{\em z} galaxies show $\sigma_{\rm gas}$ with values up to 50\,km~s$^{-1}$ \citep{cresci+09}. Table~\ref{tab:supp} shows what each galaxy requires to be reconciled with a normal SF efficiency. Shocks heat the gas, injecting turbulence into the system, and boosting the molecular gas velocity dispersion by factors of a few to an order of magnitude \citep{guillard+09}.  HCG\,25b, 40c, 57a, 79a, and 82b have \supp\,$>$\,10, too high to reconcile with normal SF by applying a much higher $\sigma_{\rm gas}$.

SF requires gravitational binding energy to be greater than turbulent and radiative energies. If additional energy is introduced into the system, this balance is disrupted, leading to SF becoming inefficient \citep{krumholz+12}.  One example of turbulence-induced SF suppression in the Milky way is the galactic center cloud G0.253+0.016.  This object shows evidence for a recent collision with another cloud \citep{longmore+12} and has a lower SFE compared to similar objects, especially in regions with high velocity dispersion \citep{kauffmann+13}.  The shock in Stephan's Quintet is detected in CO(1--0), but appears to have extremely weak associated SF \citep{konstantopoulos+14}.  \citet{guillard+12} suggest that \supp\ in this region could be a factor of 75 or higher, a result of the turbulence that has been generated by the shocks from the colliding galaxies.  In fact, an increasing number of objects with strong turbulence have been shown to exhibit inefficient SF, including the AGN-driven molecular outflow host NGC\,1266 \citep{a15_sfsupp}, and the radio galaxy 3C\,326N \citep{guillard+15}.  In this sample of objects, which are known to be in collisional systems, large turbulent motions are observed associated with the suppression. The suppressed HCG galaxies in our sample seem to share some similarities with these systems, since many of them also contain gas exhibiting peculiar motions. 

How much energy injection is be required in order to suppress star formation? If we assume that the molecular gas in the HCG galaxies are mainly rotationally supported, we can use the Toomre criterion (Q\,$>$\,1; \citealt{toomreQ}), which describes the balance between rotation and turbulence, and gravitational binding energy, to determine the required energy budget necessary to stabilize the molecular gas against collapse, effectively inhibiting star formation. To derive the necessary energy for stability ($Q$\,=\,1), we use equation (7) in \citet{krumholz+12}:

\[Q = \frac{\sqrt{2(\beta+1)}~\sigma\Omega}{\pi G\Sigma_{\rm mol}} \]

\noindent where $\sigma$ is the global gas velocity dispersion, $\Omega$ is the rotation frequency, $G$ is the gravitational constant, $\Sigma_{\rm mol}$ is the molecular gas surface density, and $\beta$\,=\,0 for the flat part of the rotation curve. We then use force balance to determine the rotation frequency, $\Omega$. If we assume that all disks have equivalent scale heights to the Milky Way (300\,pc; \citealt{burton71,malhotra95}), the rotation frequency, $\Omega$, simplifies to:

\[\Omega = \sqrt{\frac{3}{4\pi}G\frac{\Sigma_{\rm mol}}{H}}\]

\noindent where $H$ is the disk scale height. In order for the disk to be stable against collapse ($Q$\,=\,1), assuming flat rotation, the {\em global} molecular gas velocity dispersion must be:

\begin{equation}
\frac{\sigma}{\rm km\,/\,s} \approx 7.0 \left(\frac{\Sigma_{\rm mol}}{1\,M_\odot\,{\rm pc}^{-2}}\right)^{1/2}\left(\frac{H}{\rm 300\,pc}\right)^{1/2}
\end{equation}

\noindent Assuming that the injected turbulent energy is  \hbox{$E_{\rm turb} = 1/2\pi R_{\rm disk}^2\Sigma_{\rm mol}\sigma^2$}, the required $E_{\rm turb}$ must therefore be:

\begin{equation}
\frac{E_{\rm turb}}{\rm ergs}\approx 8\times10^{51}\left(\frac{R_{\rm disk}}{\rm 3\,kpc}\right)^2\left(\frac{H}{\rm 300\,pc}\right)\left(\frac{\Sigma_{\rm mol}}{1\,M_\odot\,{\rm pc}^{-2}}\right)^2
\end{equation}

\noindent And the turbulent luminosity (energy injection rate per rotational period) is:

\begin{equation}
\frac{L_{\rm turb}}{\rm ergs/s} \approx
4.8\times10^{35} \left(\frac{H}{\rm 300\,pc}\right)^\frac{1}{2}\left(\frac{R_{\rm disk}}{\rm 3\,kpc}\right)^2\left(\frac{\Sigma_{\rm mol}}{1 M_\odot/{\rm pc}^2}\right)^\frac{5}{2}
\end{equation}

Table \ref{tab:supp} lists these values for each of the HCG galaxies in our sample. Overall, the required gas velocity dispersions to stabilize the disk of the suppressed systems are reasonable, with the total injected turbulent energy representing $\sim$1\% of the total molecular kinetic energy (similar to what is seen in radio galaxies; \citealt{guillard+12b}). Comparing the required turbulent injection to stabilize the disk to the extrapolated warm H$_2$ luminosity from \citet{cluver+13}, we find that in most cases, $L_{\rm H_2,warm}$ is within a factor of 3 of the required turbulent injection energy (save for HCG\,96a and 96d, which is a known luminous infrared galaxy and star-bursting system, which likely has a gravitationally unstable disk). Many of the largest $L_{\rm H_2,warm}/L_{\rm turb}$ values being found in galaxies with the largest suppressions, though there is also uncertainty in the extrapolation from the {\em Spitzer} IRS footprint to the galaxy. The fact that the H$_2$ luminosity, which is likely driven by turbulence \citep{appleton+06,guillard+09,guillard+12,guillard+12b,cluver+13} is comparable to the turbulent luminosity seems to indicate that turbulence could indeed be the driver of the star formation suppression that we are seeing in these systems, especially given the large contribution we expect from [C\,{\sc ii}] cooling (up to a factor of 2 larger than $L_{\rm H_2,warm}$; \citealt{guillard+15}), which will be discussed in detail in an upcoming paper (Appleton et al., in preparation).

Other mechanisms than turbulence could be responsible for the suppression of star formation in those groups, like alternative gas heating or tidal disruption processes. The cosmic ray heating of the ISM may be increased in interacting galaxies \citep{scalo+elmegreen04}, but even in systems showing extreme H$_2$ line emission and star formation suppression like Stephan's Quintet and other groups or galaxy interactions, this heating mechanism would require an extremely high cosmic ray flux \citep{guillard+09,guillard+12,peterson+12,cluver+13}. Alternatively, the tidal field induced by the galaxy interaction can be responsible for some expansion of the gas on the external regions of the merger, and lower the average ISM pressure \citep{struck99,palous05,renaud+09}, which could reduce the star formation efficiency locally. However this effect, which depends on the geometrical configuration of the tidal field and the relative position of the galaxy with respect to this field, is difficult to quantify without a proper numerical simulation of the interaction, and is generally thought to globally increase the compressive mode of turbulence (e.g. \citealt{renaud+14}).

%%%%%%%%%%%%%%% Figure 6 %%%%%%%%%%%%%%%
\begin{figure*}[t]
\centering
\includegraphics[width=0.82\textwidth,clip,trim=0cm 0.5cm 0cm 0cm]{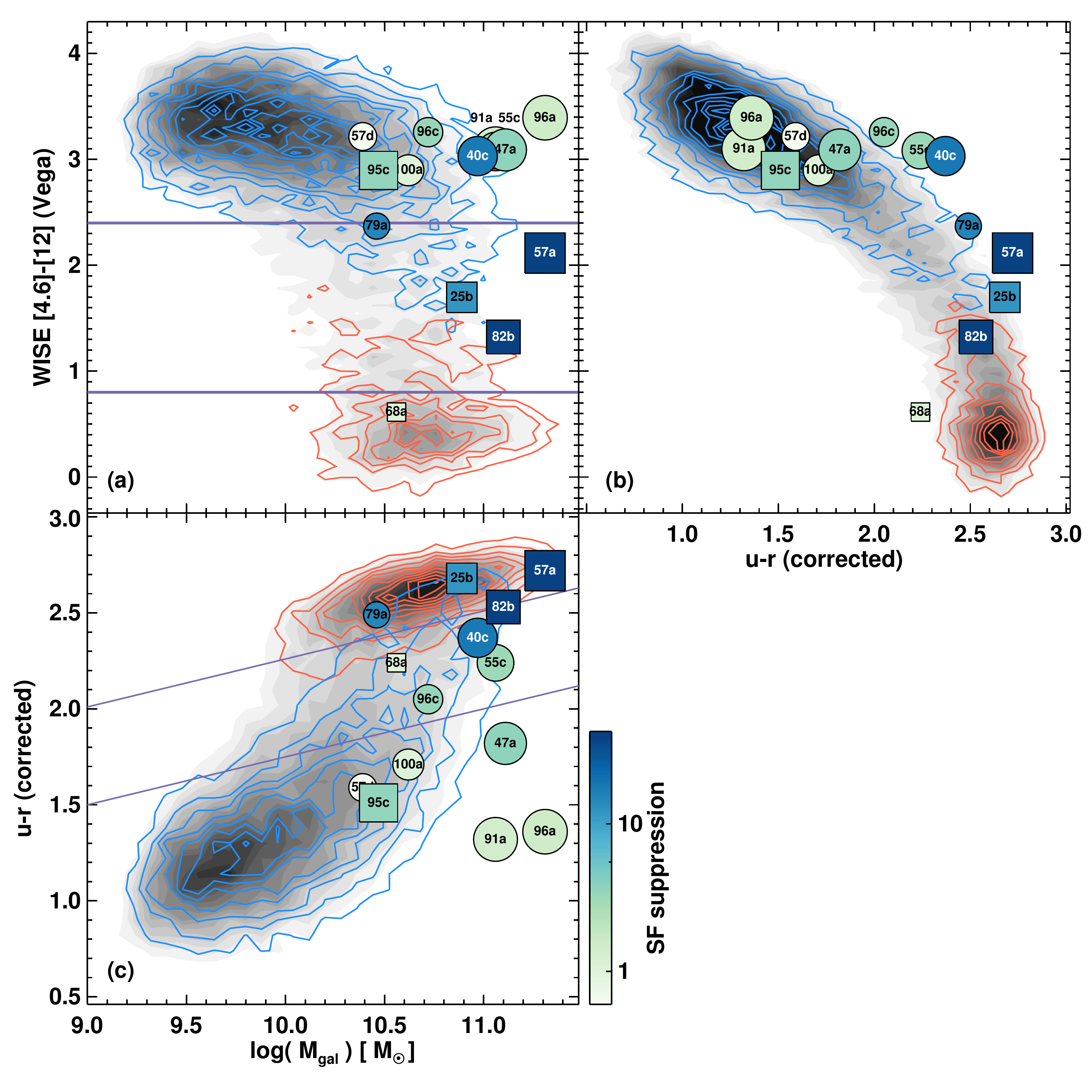}
\caption{Early-type (red contours) and late-type (blue contours) galaxies from the Galaxy Zoo \citep{lintott+08,schawinski+14,a14_irtz} are compared to the HCGs.  The color of the points represent the level of SF suppression, the size of the points correspond to the size of the molecular reservoir, and the shapes indicate whether the galaxy is a MOHEG from \citet{cluver+13}. Squares represent MOHEGs and circles non-MOHEGs.  {\bf(Top left):} The stellar mass vs. {\em WISE} [4.6]--[12]$\mu$m colors, overlaid with indigo lines defining the IRTZ.  {\bf(Top right):} The {\em u--r} vs. [4.6]--[12]$\mu$m sequence identified by \citet{a14_irtz}, with HCGs overplotted.  {\bf(Bottom):} The stellar mass vs. {\em u--r} colors of the Galaxy Zoo galaxies from \citet{schawinski+14}, overlaid with the segment defining the optical green valley. The HCGs that are overplotted had both optical and IR colors derived from \citet{bitsakis+14}. The most suppressed galaxies are the most likely to be in the {\em WISE} IRTZ and on or near the optical red sequence.}
\label{fig:cmd}
\end{figure*}

Major mergers share many properties with HCGs, including the presence of violent interactions.  In particular, the violent interactions that these galaxies are experiencing result in shocks and turbulent gas \citep{cresci+09}. Rather than suppressing SF, most of these objects exhibit super-efficient (but short timescale) SF \citep{sanders+91,sanders+mirabel96,U+12,lanz+13}. The reason HCGs, which share many of these properties, have been found with suppressed SF rather than enhanced SF is possibly the result of the different interaction timescales, larger scale environments, and gas properties.  The CO(1--0) imaged HCG galaxies in our sample are not observed during a major merger, and they are gas-poor relative to ULIRGs.

The observed HCG galaxies have molecular gas masses and surface densities an order of magnitude smaller than what is typically seen in interacting galaxies \citep{downes+solomon98,iono+09}.  The molecular gas in many ULIRGS is more compact compared to HCG galaxies as well \citep{bryant+99,wilson+08,ueda+14}. This likely means that although turbulence can be a disruptive force, it is also transient ($\sim$\,1\,Myr; \citealt{guillard+09}). Without a continuous input of energy, cooling lines are able to dissipate the turbulent energy rapidly, allowing the molecular gas in ULIRGS to restart forming stars quickly, due to the higher gas densities and shorter free-fall times.

The timescale over which the violent interaction is taking place in major mergers is also short ($\approx10^8$\,yr; \citealt{hopkins+08,lanz+14}) compared to the total time in which gravitational forces within the compact group impact the individual galaxies ($\approx$3\,Gyr; \citealt{plauchu-frayn+12}). In major mergers, the gravitational encounters taking place involve only the two merging galaxies \citep{toomre72,privon+13}.  The dynamics within groups are much more complicated. HCG galaxies are interacting with the intragroup medium, and their low density gas is being ram pressure stripped \citep{rasmussen+08}. Unlike the coalescence timescale in major mergers, galaxies in HCGs are undergoing sustained gravitational encounters with the other group members, in which collisions do not necessarily result in coalescence. This extends the timescale of turbulent injection, possibly resulting in the differences observed between major mergers and HCGs.
 
 An in-depth analysis by \citet{a14_hcg57} of one of the most suppressed systems, HCG\,57a, suggested that an ongoing shock from a recent direct collision with HCG\,57d has continuously pumped turbulence into the system. H$_2$ and far-IR cooling lines have been unable to completely dissipate the extra energy, leading to the observed SF suppression.  Once the shock has traversed the system (and thus is no longer pumping energy into the molecular gas), H$_2$, [C\,{\sc ii}], and other far-IR cooling lines should efficiently cool the gas, allowing SF to return to normal efficiency. It is possible that the group environment will extend this timescale as the group members continue to interact, shocking and expelling the interstellar media and ultimately transforming into early-type galaxies. This would be consistent with the compact group evolution picture suggested by \citet{bitsakis+10} that older groups contain a higher fraction of early-type galaxies.

While it appears we have identified a sample of galaxies where turbulent energy has suppressed SF, the exact driver of this turbulence remains undetermined.  {\em Herschel} data of far-IR cooling lines such as [C\,{\sc ii}] and [O\,{\sc i}] will be able to advance our understanding on the interplay between the molecular gas, SF suppression, and cooling mechanisms, potentially providing us with a deeper understanding of how turbulence can impact the way in which galaxies form stars (Appleton et al., in preparation).

%%%%%%%%%%%%%%% Figure 6 %%%%%%%%%%%%%%%
%\begin{figure}[t]
%\includegraphics[width=0.49\textwidth]{figures/color_supp.pdf}
%\caption{We plot \supp\ against many of the HCG properties (with MOHEGs shown as red points and non-MOHEGs shown as blue points), determining what properties \supp\ might correlate against.  Including (a) $u-r$ (corrected) colors for each galaxy and (b) the [4.6]--[12]$\mu$m WISE color.  A Spearman rank test was run on \supp\ and these properties, and only the [4.6]--[12]$\mu$m WISE and $u-r$ colors were found to correlate ($P_0 < 0.05)$, despite an investigation of many other possible correlating properties.}
%\label{fig:fgas_supp}
%\end{figure}

\subsection{SF suppression in warm H$_2$-bright HCGs: a connection to transition?}
%\textcolor{red}{Though if galaxies are shedding their interstellar media as they cross the infrared transition zone (IRTZ), then it is puzzling why these objects still have large gas reservoirs available.  Also, morphology. $f_{\rm gas}$ might be important, but morphology is interesting too.  This might be a chicken or egg problem...}

Figure~\ref{fig:sfrMS} plots the stellar mass -- SF main sequence for normal star-forming galaxies \citep{wuyts+11}, compiled from the {\sc magphys}-derived SFRs from SDSS DR7 \citep{sdssdr7,chang+15}, with our CO-imaged objects overplotted and color-coded by \supp. In this phase space, the HCGs with the most SF suppression fall the farthest from the main sequence, suggesting they are transitioning.  Many of the other (non-suppressed) galaxies sit on the relation, including MOHEGs.  In fact, the location of objects on the main sequence seems to mirror the situation for {\em Spitzer} infrared colors (Fig~\ref{fig:lacy}), strongly suggesting that the most suppressed objects are also the ones that are currently transitioning between the star-forming spiral and quiescent early-type populations.

Figure~\ref{fig:cmd} plots the CARMA-imaged HCG galaxies on the optical and IR color-magnitude diagrams presented in \citet{a14_irtz}.  The HCGs we have studied span a large range of optical and IR colors, although they are generally more massive than the underlying Galaxy Zoo distributions \citep{lintott+08,schawinski+14}.  One would expect that the total molecular mass or molecular gas fraction might be the determining factor of the position of the HCGs on the color-magnitude diagrams, but that is not the case.  HCG\,40c and 57a both contain large reservoirs of molecular gas, while appearing in the optical green valley or red sequence (compared to HCG\,91a and 96a with equivalent molecular mass, which are quite blue).  In fact, Figure~\ref{fig:cmd} shows that the mass of the molecular reservoir does not determine the color of the galaxy. Spearman rank tests were used to search for correlations between \supp\ and other galaxy properties.The only galaxy property significantly correlated with \supp\ was color. Including HCG\,68a, a Spearman correlation of -0.71 with a {\em p}-value\,=\,0.0044 was found when investigating \supp\ and {\em u--r} color.  If HCG\,68a is removed, we find a \supp--({\em u--r}) Spearman correlation of -0.71 (equal to the test with HCG\,68a included) with a {\em p}-value\,=\,0.0011. Without HCG\,68a, \supp--([4.6]--[12]$\mu$m) also shows a Spearman correlation of 0.80 with a {\em p}-value\,=\,0.0067.  With or without HCG\,68a, the optical and IR colors were the only galaxy properties shown to significantly correlate with \supp. The extreme IR colors of HCG\,68a might be due to a buried AGN (Alatalo et al. 2015c, in prep).

These results seem to suggest that a galaxy does not need to shed its ISM to then quench SF (the standard galaxy transition picture; \citealt{hopkins+06}) to morphologically transform. Instead, changing the state of the molecular gas can act to quench SF {\em before} the ISM has been completely shed. It is thus possible that in many of our CO-imaged galaxies, the molecular gas has been rendered infertile due to shocks pumping turbulence into the system. The quenching of SF, and the beginning of the transition across the green valley, occurs before the galaxy loses the majority of its molecular ISM. This is consistent with the findings of \citet{leon+98} and \citet{martinez-badenes+12} that HCG galaxies contain comparable molecular gas reservoirs to isolated galaxies \citep{lisenfeld+11,coldgass}. \citet{a14_irtz} posited that the IRTZ was a manifestation of evolution, representing the stage in which a galaxy is actively shedding its ISM, and that this phase {\em follows} SF quenching (traced by the {\em u--r} colors).  In this scenario, first the galaxies move toward the elbow in Fig~\ref{fig:cmd}b, where SF is suppressed, and then they move into the IRTZ, where they quench SF and move to the optical red sequence. The H\,{\sc i} findings of \citet{serra+12} of that 40\% of field early-type galaxies still contain non-negligible reservoirs of neutral gas.  New work on poststarburst galaxies found that many poststarburst galaxies still contain non-negligible molecular reservoirs \citep{french+15,rowlands+15}, confirming the suggestion that SF quenching can take place before galaxies expel their molecular interstellar media.  It is currently unclear how often galaxies transition in this fashion, but further observations of IRTZ and poststarburst galaxies, combined with our new observations of at least some HCG galaxies (certainly those that are warm H$_2$-bright and CO-bright), have shown that this ``quenching first'' path contributes to the population of transitioning galaxies.
 
The presence of suppressed objects within the IRTZ points to a possible new method for identifying galaxies most likely to be in a phase of inefficient SF, before the exhaustion of their molecular gas.  By directly imaging the CO in molecular gas-rich galaxies that also appear in the IRTZ, we can pinpoint a population of galaxies likely to exhibit SF suppression. This result suggests that future studies might identify larger samples of suppressed galaxies by selecting galaxies based on their optical and IR colors. A larger sample of suppressed systems thus allows us to study how the injection of turbulence not only impacts the energy balance that dictates our SF laws \citep{krumholz+12}, but also allows us to study how the neutral ISM is exhausted as a galaxy transforms from a spiral into an early-type galaxy.

\section{Summary}
\label{sec:summary}
\begin{itemize}
\item We have used CARMA to map CO(1--0) of galaxies within 12 HCGs, many with elevated warm H$_2$ emission, detecting molecular gas in 14 galaxies, and unresolved 3mm radio continuum in 7 (consistent with the presence of AGNs).  Figures~\ref{fig:hcg25}--\ref{fig:hcg100} show the molecular gas data for each galaxy, including the moment0 map, moment1 map, channel maps, PVDs, and integrated spectra.  A comparison to the single-dish data from \citet{lisenfeld+14} shows that our observations have not resolved out large fractions of the flux.

\item The molecular gas morphologies of our HCG galaxies (using the metric set for the ATLAS$^{\rm 3D}$ galaxies in \citealt{alatalo+13}) indicate that HCGs are consistent with the distribution of gas morphologies found in early-type and spiral galaxies rather than ULIRGs \citep{wilson+08,ueda+14}.

\item We have shown that a large proportion of our CO-imaged HCG galaxies exhibit SF suppression (\supp) when plotted relative to both the K-S relation \citep{ken98} as well as the universal SF law from \citet{krumholz+12}.  The mean SF suppression for this sample is $\langle$\supp$\rangle$\,$\approx$\,10$\pm$5, and exhibits a bimodality.  The most extreme objects (HCG\,25b, 40c, 57a, 79a \& 82b) exhibit \supp\,$\gtrsim$\,10, and have molecular gas depletion timescales $t_{\rm dep}$\,$\geq$\,10\,Gyr.

\item The mean gas-to-dust ratio for the CO-imaged HCGs is around 170, within the range found in normal galaxies \citep{sandstrom+13,remy-ruyer+14}. We do not believe that the observed \supp\ is due to an incorrect $L_{\rm CO}$--$M$(H$_2$) conversion factor (which would appear as a highly discrepant gas-to-dust ratio in suppressed galaxies).

\item A non-negligible fraction of our CO-imaged HCG galaxies contained substantial warm H$_2$ emission \citep{cluver+13}, consistent with there being shocks injecting substantial turbulence into these systems, and the turbulent energy required to stabilize the molecular gas against collapse appears to agree within an order of magnitude with the warm H$_2$ luminosity.  As has been seen in the Milky Way \citep{kauffmann+13}, NGC\,1266 \citep{a15_sfsupp}, and 3C\,326N \citep{guillard+15}, the additional turbulence could upset the energy balance that dictates the rate of SF \citep{krumholz+12}, thereby suppressing SF.

\item HCGs with the most SF suppression are usually located within the transition regions of optical and IR color space, independent of the mass of the molecular reservoir.  This ties in well with work that indicates that galaxies are able to transition in colors and quench SF before they have shed their ISMs \citep{a14_irtz,french+15,rowlands+15}, showing how galaxies might render their molecular reservoirs infertile before expelling them. This could play an important role in understanding the blue to red galaxy transition.  

\item The {\em u--r} and {\em WISE} IRTZ colors, combined with a CO detection, are also able to select the objects most likely to exhibit SF suppression, providing an ideal sample selection criterion with which to study this phenomenon.

\end{itemize}

%%%%%%%%%%%%%%%%%%%%%%%%%%%%%%%%%%%%%%%%%%%%%%%
\acknowledgments
We thank Laure Ciesla for lending her advice and {\sc cigale} expertise in our investigation of the SF uncertainties, as well as Philip Chang for prescient advice on the theoretical grounding of the energy budget. We also thank the anonymous referee for a useful and insightful report. 
Partial support was provided  to KA, TB and PA by NASA observations through a contract issued by the Jet Propulsion Laboratory, California Institute of Technology under a contract with NASA.  Additional support for KA is provided by NASA through Hubble Fellowship grant \hbox{\#HST-HF2-51352.001} awarded by the Space Telescope Science Institute, which is operated by the Association of Universities for Research in Astronomy, Inc., for NASA, under contract NAS5-26555. UL acknowledges  support by the research projects.   AYA2011-24728 from the Spanish Ministerio de Ciencia y Educaci\'on and the Junta de Andaluc\'\i a (Spain) grants FQM108.  TB and VC would like to acknowledge partial support from the EU FP7 Grant PIRSES-GA-2012-316788.  TB acknowledges support from DGAPA-UNAM postdoctoral fellowships.  LVM work has been supported by grant AYA2011-30491-C02-01 co-financed by MICINN and FEDER funds, and the Junta de Andaluc\'ia (Spain) grants P08-FQM-4205 and TIC-114.

Support for CARMA construction was derived from the Gordon and Betty Moore Foundation, the Kenneth T. and Eileen L. Norris Foundation, the James S. McDonnell Foundation, the Associates of the California Institute of Technology, the University of Chicago, the states of California, Illinois, and Maryland, and the National Science Foundation.  This publication makes use of data products from the Wide-field Infrared Survey Explorer, which is a joint project of the University of California, Los Angeles, and the Jet Propulsion Laboratory/California Institute of Technology, funded by the National Aeronautics and Space Administration. The work is also based, in part, on observations made with {\it Herschel}, a European Space Agency Cornerstone Mission with significant participation by NASA. This research has made use of the NASA/IPAC Extragalactic Database (NED) which is operated by the Jet Propulsion Laboratory, California Institute of Technology, under contract with the National Aeronautics and Space Administration.  We acknowledge the usage of the HyperLeda database (http://leda.univ-lyon1.fr).\\

\noindent{\em Facilities:} \facility{CARMA}, \facility{{\em Herschel}}, \facility{{\it WISE}}
%%%%%%%%%%%%%%%%%%%%%%%%%%%%%%%%%%%%%%%%%%%%%%%
% BIBLIO

%\begin{thebibliography}{99}
%\end{thebibliography}
\bibliographystyle{apj}
\bibliography{ms}

%%%%%%%%%%%%%%%%%%%%%%%%%%%%%%%%%%%%%%%%%%%%%%%
% APPENDIX?
\begin{appendix}
\section{Determining the AGN contributions within CO(1--0) imaged HCG galaxies}
\label{app:decompir}
In Table~\ref{tab:gal_params}, we presented the AGN classifications of the galaxies in our sample, based both in optical emission line diagnostics as well as radio continuum. Such sources can significantly contribute to the infrared emission of the galaxies, leading to an overestimation of their star formation rates, if they are not accounted for. To disentangle the fraction of AGN contribution in the total infrared luminosity we have fitted the galaxy observed infrared (8--500$\mu$m; purely dust emission) spectral energy distributions with {\sc DecompIR} \citep{decompir}. This code simply fits the observed fluxes with sets of host-galaxy + AGN component templates, and estimates the contribution of the AGN to the total infrared luminosity. The AGN templates are described by broken power-laws at  around 40$\mu$m that fall steeply above that. From this analysis we find that only three of our sources have significant AGN contribution at the infrared bands, HCG\,68a with 9\%, and HCG\,91a with 35\%. 

\section{Comments on individual galaxies}
\label{app:individuals}

\noindent {\bf HCG\,25b}: Figure~\ref{fig:hcg25} shows that HCG\,25b is an edge-on galaxy. Deep optical imaging by \citet{eigenthaler+15} also showed that HCG\,25b is interacting strongly with HCG\,25f. A tidal tail connects the two, and the polyaromatic hydrocarbon (PAH) emission shows a small tidal feature. HCG\,25b is also a MOHEG \citep{cluver+13} and is a transitioning galaxy in the IR (i.e., it lies in the optical red sequence as well as the {\em Spitzer} IR gap and the {\em WISE} IRTZ). HCG\,25b also likely contains an AGN, classified through optical emission line diagnostics \citep{martinez+10}, and the presence of 1.4\,GHz nuclear emission\footnote{Nuclear 1.4\,GHz emission can also be due to star formation \citep{condon92}. Thus, detecting 1.4\,GHz emission in these objects does not confirm the presence of an AGN without morphological confirmation \citep{best+05} and thus are just suggestive of their presence.}. The molecular gas is morphologically classed as a disk.\\

\noindent{\bf HCG\,40c}: Figure~\ref{fig:hcg40} shows that HCG\,40c is an edge-on galaxy strongly detected with CARMA.  While HCG\,40c lies in the red part of the {\em WISE} IRTZ, it is found in the optical green valley, though near the red sequence. It also has a radio core and optically identified AGN \citep{martinez+10}.  Two Micron All-Sky Survey (2MASS) imaging also shows that HCG\,40c is tidally interacting with HCG\,40e \citep{2mass}.  The PVD of HCG\,40c (Fig~\ref{fig:hcg40}) also shows a significant bar (seen as the large velocity structure with very little position shift), similar to what was seen by \citet{alatalo+13} in several \atlas\ early-type galaxies. This is likely what is responsible for the slight appearance of broad wings in the CO spectrum.  Given the frequency with which bars arise during gravitational encounters \citep{athanassoula96,ath+bureau99}, it is unsurprising that HCG\,40c is morphologically classed as a bar+ring.\\

\noindent{\bf HCG\,47a}: Figure~\ref{fig:hcg47} shows that HCG\,47a is an oblong spiral (in both PAH and optical images from SDSS; \citealt{sdss}) that is tidally interacting with HCG\,47b. HCG\,47a is also on the cusp of being considered a MOHEG ($L_{\rm H_2,warm}/L_{7.7\mu{\rm m~PAH}}$\,=\,0.035; \citealt{cluver+13}). The center, while void of molecular gas, contains a spectrally classified AGN \citep{stern+12} as well as 1.4\,GHz emission. The CO(1--0) emission traces the oblong spiral, and so the molecular gas is morphologically classified as {\em both} a ring and a spiral.\\

\noindent{\bf HCG\,55c} (Fig~\ref{fig:hcg55}) is part of the chain of galaxies VV\,172/Arp\,239\footnote{It contains a galaxy with a highly discrepant redshift (HCG\,55e, $v$\,=\,36880\,km~s$^{-1}$; \citealt{sargent68}).}. HCG\,55 is the most distant HCG that we have imaged with CARMA. The emission in this galaxy is resolved and spans approximately 3 beam widths. HCG\,55c was not detected in 1.4\,GHz or 3mm continuum, and does not have a published optical spectrum (to determine the presence of an AGN).  However, HCG\,55c is at the edge of the {\em Spitzer} gap, and is in both the optical green valley and the {\em WISE} IRTZ.  The molecular gas in HCG\,55c is morphologically classified as a disk. \\

\noindent{\bf HCG\,57a} is found to have some of the most complex molecular gas dynamics in the sample, including three distinct kinematic components (Fig~ref{fig:hcg57}), and is interacting with HCG\,57d. HCG\,57a is also a MOHEG \citep{cluver+13}, and an in-depth discussion of both HCG\,57a and HCG\,57d can be found in \citet{a14_hcg57}.  The molecular gas in HCG\,57a is morphologically classified as mildly disrupted. \\

\noindent {\bf HCG\,57d} (Fig~\ref{fig:hcg57}) was not within the {\em Spitzer} IRS footprint, and thus we do not have warm H$_2$ information on this source.  An in-depth discussion of both HCG\,57a and HCG\,57d can be found in \citet{a14_hcg57}.  The molecular gas in HCG\,57d is morphologically classified as a ring.\\

\noindent{\bf HCG\,68a} (Fig~\ref{fig:hcg68}) was previously un-detected by the IRAM\,30m in the \atlas\ survey \citep{young+11}, but was later detected by \citet{lisenfeld+14}.  The CARMA observation helps shed light on why this was the case.  The molecular gas in HCG\,68a is not only compact (unresolved by the CARMA beam), but also very broad ($\Delta v$\,$\approx$\,$2000$ \kms).  HCG\,68a also appears to have rotation velocities in its CO emission that makes it an outlier on the M$_{\rm gal}$--$v_{\rm rot,CO}$ relation \citep{davis+11a}.  HCG\,68a and HCG\,68b were also both detected in 3mm continuum emission.  {\sc DecompIR} \citep{decompir} suggests that about 9\% of the far-IR emission \citep{bitsakis+14} originates from the AGN in this system.  The AGN in HCG\,68a has an 2--10 keV X-ray luminosity of $L_X = 1.6\times10^{40}$\,ergs~s$^{-1}$ \citep{evans+10}, which could account for the majority of the far-IR emission if the obscuring column of molecular gas is sufficiently high (which the CARMA observations suggest might be the case). HCG\,68a is also one of the strongest MOHEGs, with $L_{\rm H_2,warm}/L_{7.7\mu{\rm m~PAH}}$\,=\,0.741, and is an outlier in IR color space amongst the CO-imaged HCG galaxies.  HCG\,68a has {\em Spitzer} colors corresponding to an early-type galaxy, though these colors might also be due to a buried AGN with very few intermediate aged stars (Alatalo et al. 2015c, in prep).  HCG\,68a is also the only galaxy that is found with {\em WISE} colors completely consistent with the elliptical sample \citep{a14_irtz}, but sits within the optical green valley. The molecular gas in HCG\,68a is morphologically classified as a disk. \\

\noindent{\bf HCG\,79a} (Fig~\ref{fig:hcg79}), a member of Seyfert's Sextet \citep{seyfert51}\footnote{although HCG\,79c later identified to be a background galaxy}, is a near edge-on red early-type galaxy with a prominent dust lane. HCG\,79a is optically classified as having an AGN and is found in the {\em WISE} IRTZ. The molecular gas in HCG\,79a is morphologically classified as a disk. \\

\noindent{\bf HCG\,82b} (Fig~\ref{fig:hcg82}) is the only CO-imaged HCG that was successfully detected in 3mm continuum but not in 1.4\,GHz continuum. Despite this, there is an optically identified AGN \citep{martinez+10}, so the 3mm continuum is most likely due to an AGN. HCG\,82b is a MOHEG \citep{cluver+13} located in the red sequence, as well as the {\em Spitzer} gap and the {\em WISE} IRTZ. Figure~\ref{fig:hcg82} shows HCG\,82b contains a stellar bar, where the molecular gas does not appear aligned with the bar. The lip seen at the edge of the molecular disk seems consistent with a warp, and thus the molecular gas in this galaxy is morphologically classified as mildly disrupted.\\

\noindent{\bf HCG\,91a} (Fig~\ref{fig:hcg91}) is a nearly face-on spiral galaxy that is blue in optical and IR colors, with the CO(1--0) being brightest in the northeast quadrant of the galaxy.  Deep H$\alpha$ imaging from \citet{eigenthaler+15} seems to indicate that this might be the location of a bow shock.   It is the only galaxy in this survey classified as a Seyfert 1 \citep{cluver+13}. The deep optical imaging of HCG\,91a also indicates that it is undergoing a significant interaction with HCG\,91c \citep{eigenthaler+15}, showing multiple tidal tails, including one connecting the two interacting galaxies. \citet{vogt+15} also showed that the CO and optical line velocities were offset from that of the H\,{\sc i}.  HCG\,91a has many different kinematic components but appears to have a blue-shifted line wing that runs directly north--to--south from the nucleus of the galaxy.  {\sc DecompIR} \citep{decompir} suggests that up to 35\% of the far-IR \citep{bitsakis+14} from the center is due to an AGN.  The molecular gas in HCG\,91a is morphologically classified as a spiral.\\

\noindent{\bf HCG\,95c} (Fig~\ref{fig:hcg95}) mainly shows regular rotation, but the CO(1--0) emission extends into the tidal tail between HCG\,95a and 95c, which is also seen prominently in the 8$\mu$m non-stellar emission in the channel map of Fig~\ref{fig:hcg95}.  Shells are also present in the optical light.  {\em Spitzer} IRAC colors place this galaxy on the edge of the {\em Spitzer} IR gap closer to those of dusty spirals, but its H$_2$/PAH ratio indicates that it is a MOHEG \citep{cluver+13}. Unlike other transition galaxies, this one has a high specific star formation rate. HCG\,95c contains radio emission, and is spectrally classified as an AGN \citep{martinez+10}.  The molecular gas in HCG\,95c is morphologically classified as mildly disrupted. \\

\noindent{\bf HCG\,96a} (Fig~\ref{fig:hcg96}) is the only CO-imaged HCG galaxy that could be identified as containing an AGN solely from its mid-IR spectrum \citep{cluver+13}, that is also classified as an AGN by optical spectroscopy \citep{martinez+10}.  Deep optical imaging \citep{eigenthaler+15} shows that HCG\,96a is undergoing an interaction with HCG\,96c, with multiple tidal tails present, including stellar light connecting HCG\,96a \& 96c \citep{verdes-montenegro+97}. Optical and {\em WISE} IR colors are all consistent with a star-forming galaxy, though this object is located in the AGN wedge in {\em Spitzer} colors \citep{lacy+04}. Figure~\ref{fig:hcg96} shows that the molecular gas contains multiple components, including spiral structure, a bar component and a ring. Thus the molecular gas in HCG\,96a is morphologically classified as both a spiral and a bar+ring.\\

\noindent{\bf HCG\,96c} (Fig~\ref{fig:hcg96}) is interacting with HCG\,96a, as mentioned above, and appears in the optical green valley, and has {\em Spitzer} and {\em WISE} colors consistent with star-forming galaxies. The 1.4\,GHz emission indicates that HCG\,96c might contain a radio-bright AGN. HCG\,96c also has a broad CO(1--0) spectrum, but that is possibly due to its edge-on orientation.  The molecular gas in HCG\,96c is morphologically classed as a disk. \\

\noindent{\bf HCG\,100a} (Fig~\ref{fig:hcg100}) exhibits prominent spiral structure, with prolific SF activity in its center (seen in optical light and PAH emission in Fig~\ref{fig:hcg100}). HCG\,100a does not exhibit any outward signs of interaction, such as tidal tails, and is identified as star-forming in both optical and IR colors.  HCG\,100a is spectrally identified as containing an AGN \citep{martinez+10}, consistent with its detection in 3mm continuum and 1.4\,GHz emission. The majority of the molecular gas in HCG\,100a seems to be undergoing regular rotation, but there also seems to be a minor-axis component that includes higher velocities (seen as wings in the CO(1--0) spectrum), possibly due to a bar or an outflow. Despite the presence of a putative minor axis component of the molecular gas in HCG\,100a, we morphologically classify this molecular gas as a disk. \\

\section{Complete figures for each HCG}
\label{app:figures}
\begin{figure*}[b!]
\centering
\subfigure{\includegraphics[width=\textwidth]{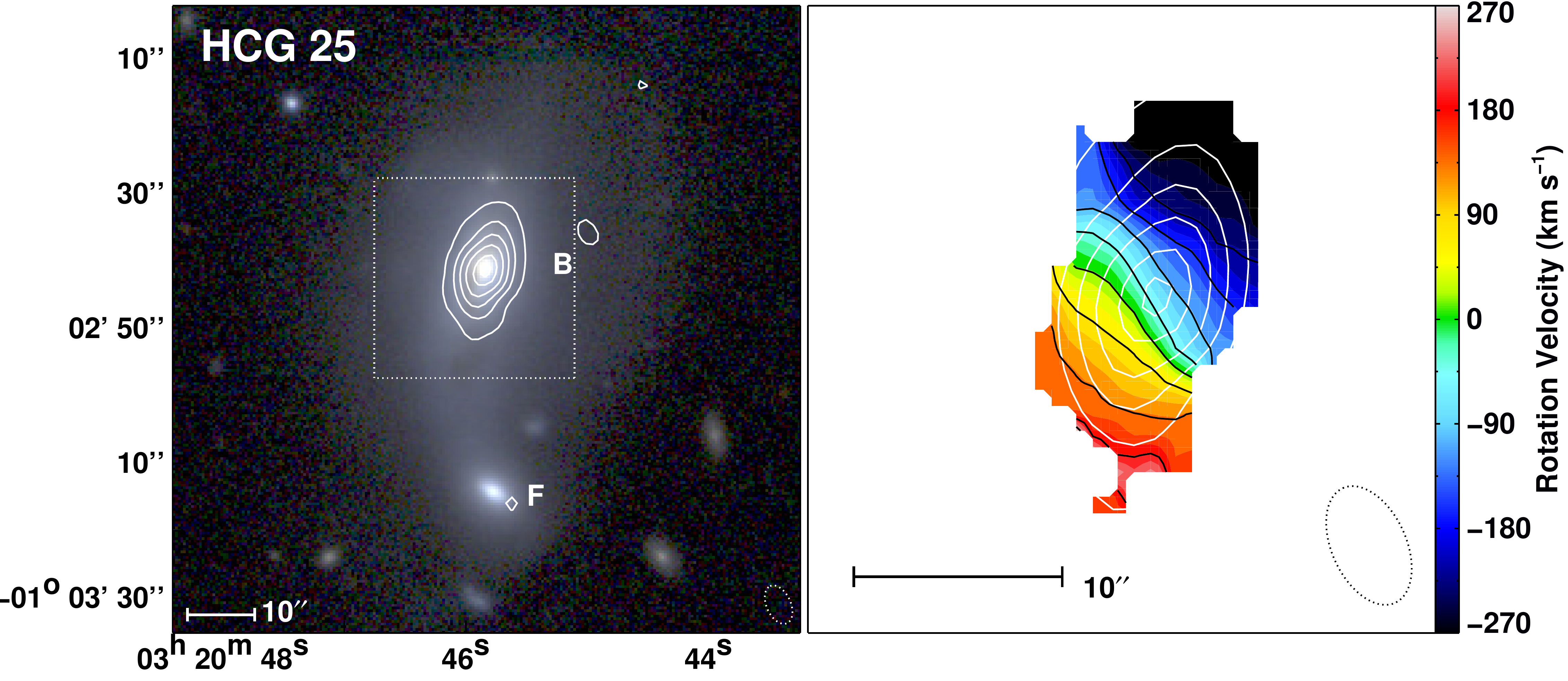}}
\subfigure{\includegraphics[height=5.1cm]{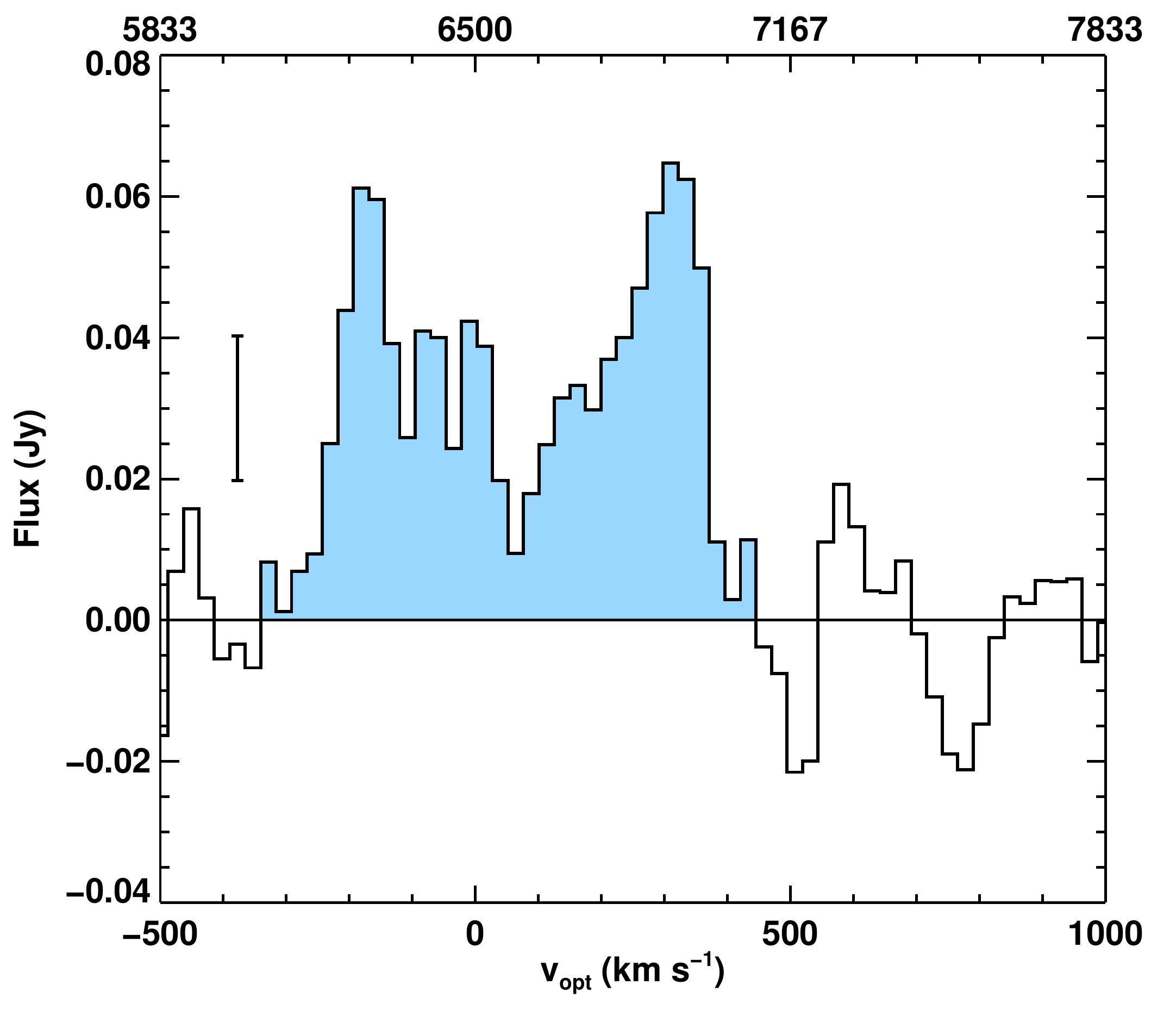}}
\subfigure{\includegraphics[height=5.1cm]{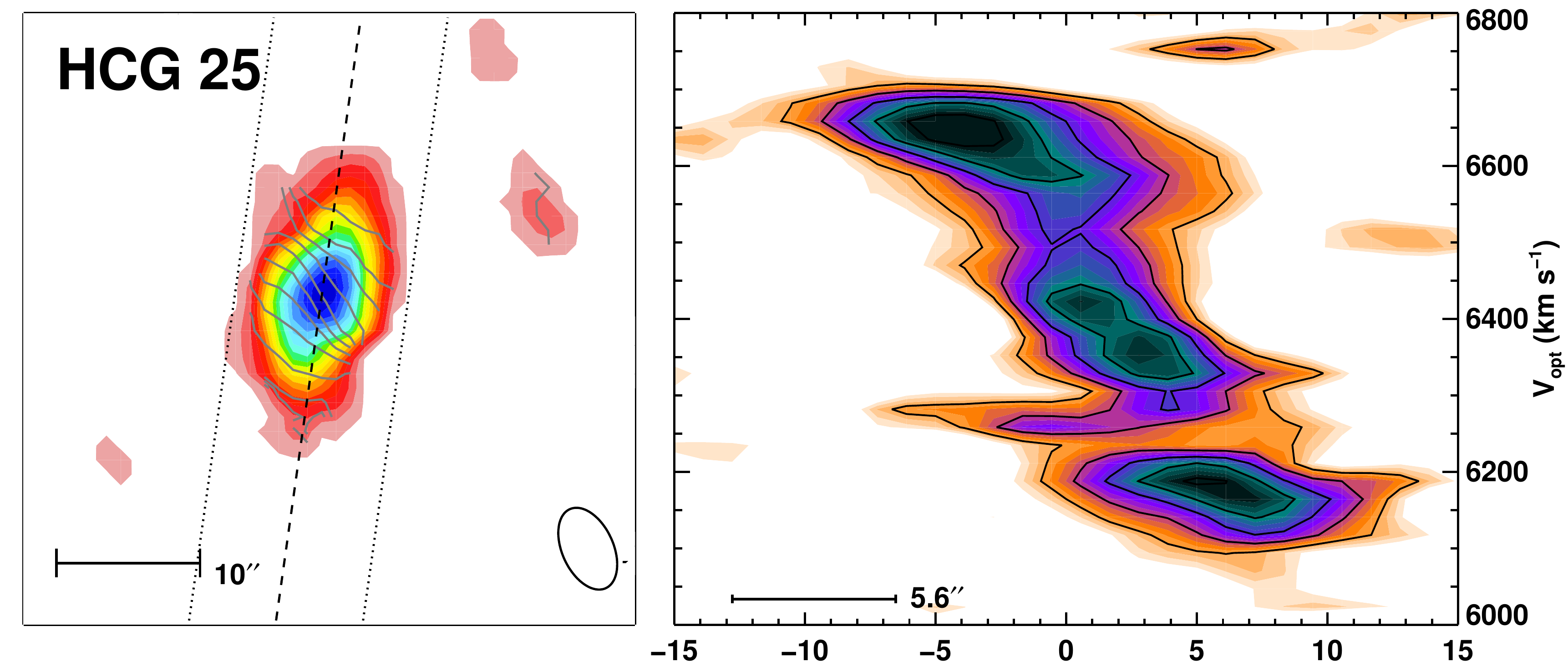}}
\subfigure{\includegraphics[width=0.77\textwidth]{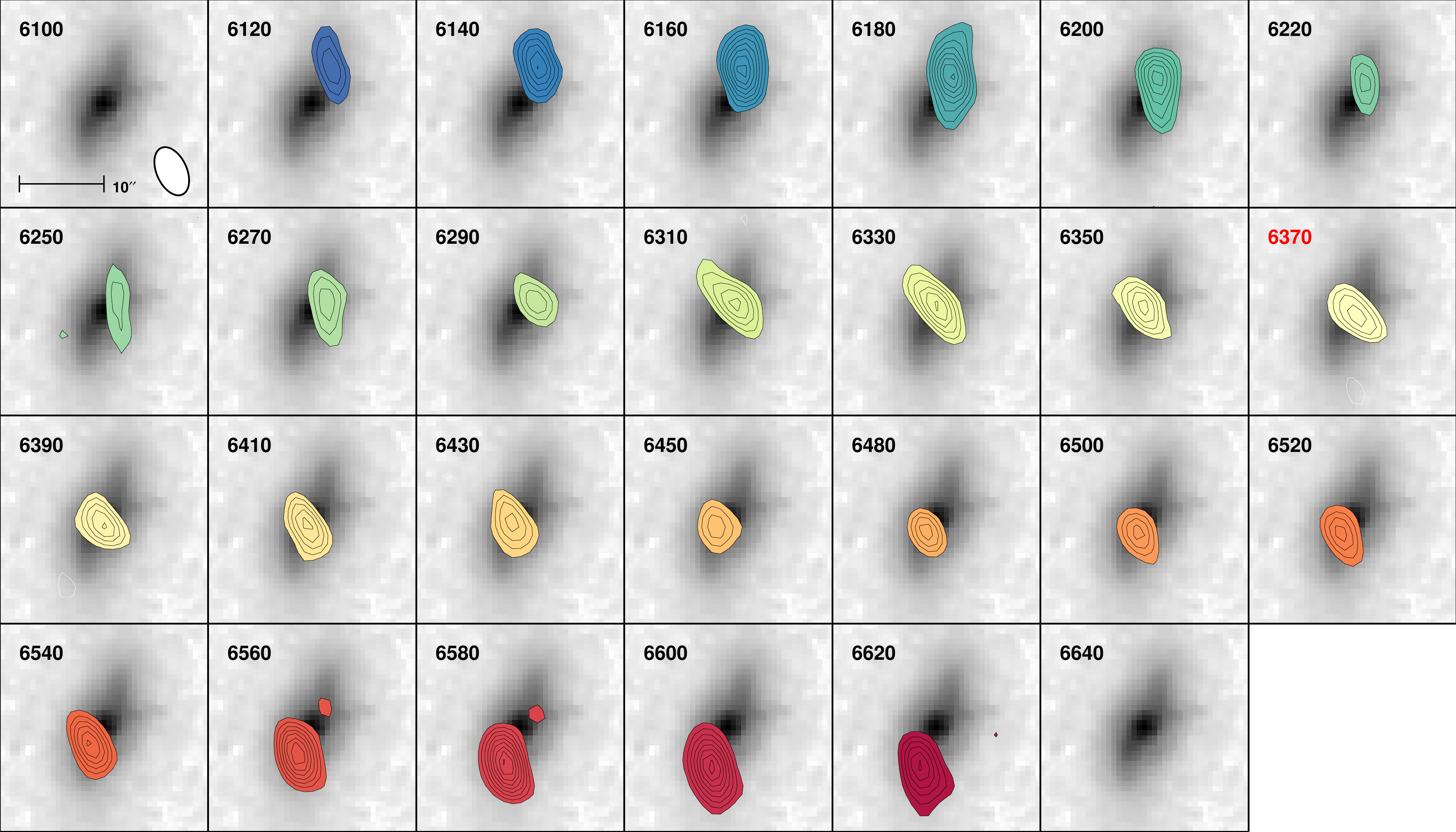}}
\caption{HCG\,25. Channel map contours are in 1$\sigma$ steps. Elements of this figure are described in \S\ref{app:figures}.}
 \label{fig:hcg25}
 \end{figure*}

\begin{figure*}[h!]
\subfigure{\includegraphics[width=\textwidth]{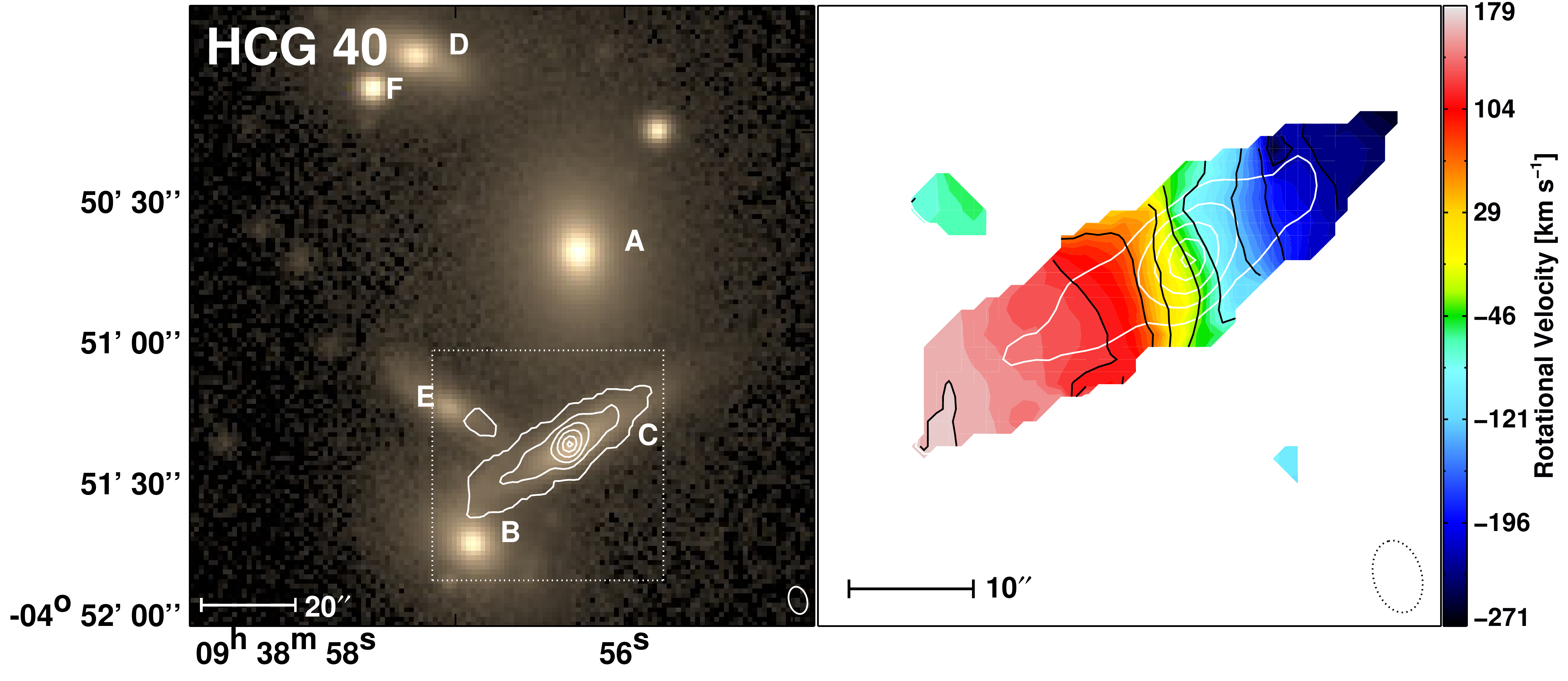}}
\subfigure{\includegraphics[height=4.9cm,clip,trim=0.9cm 1.4cm 0.1cm 1.5cm]{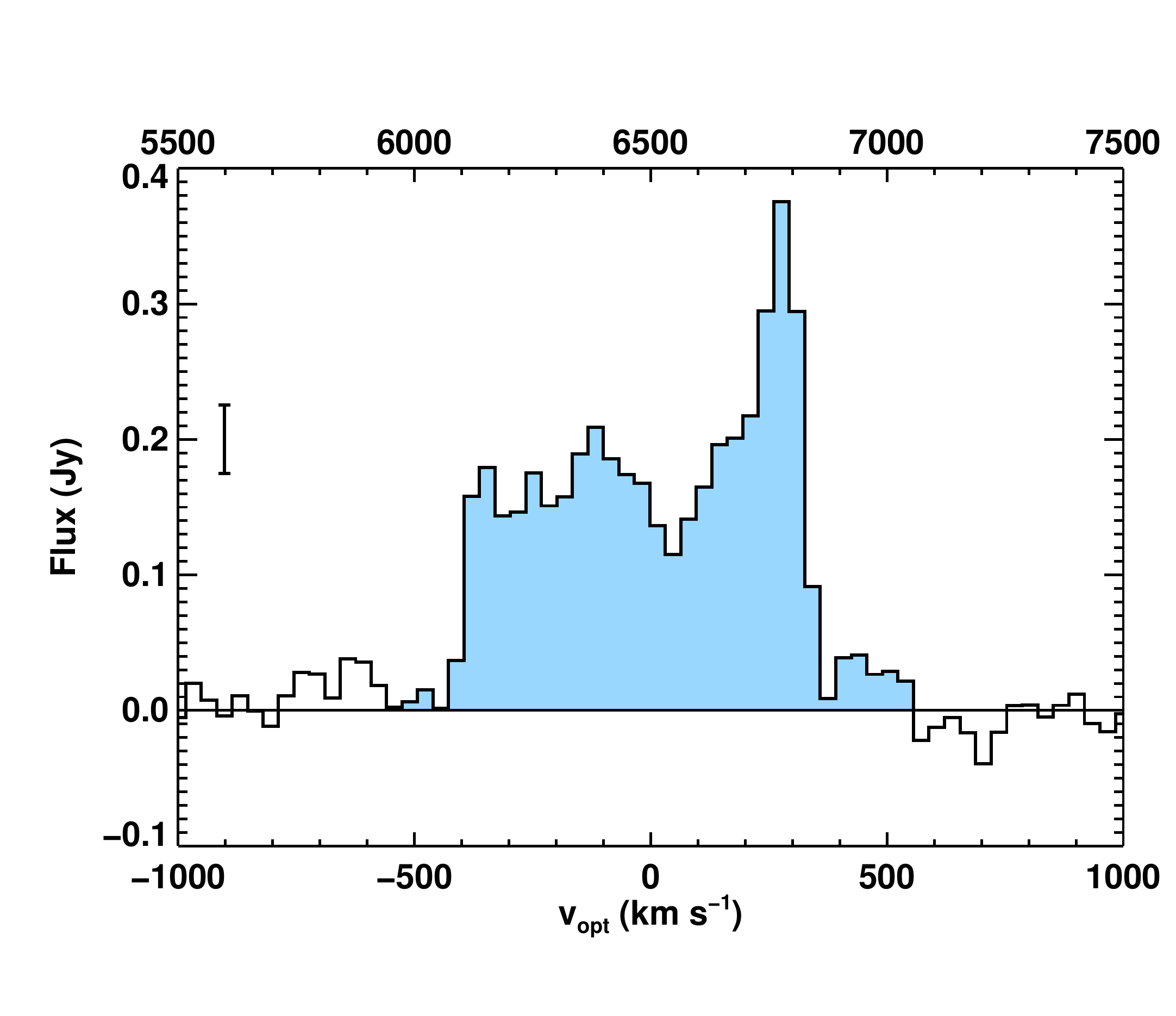}}
\subfigure{\includegraphics[height=4.9cm]{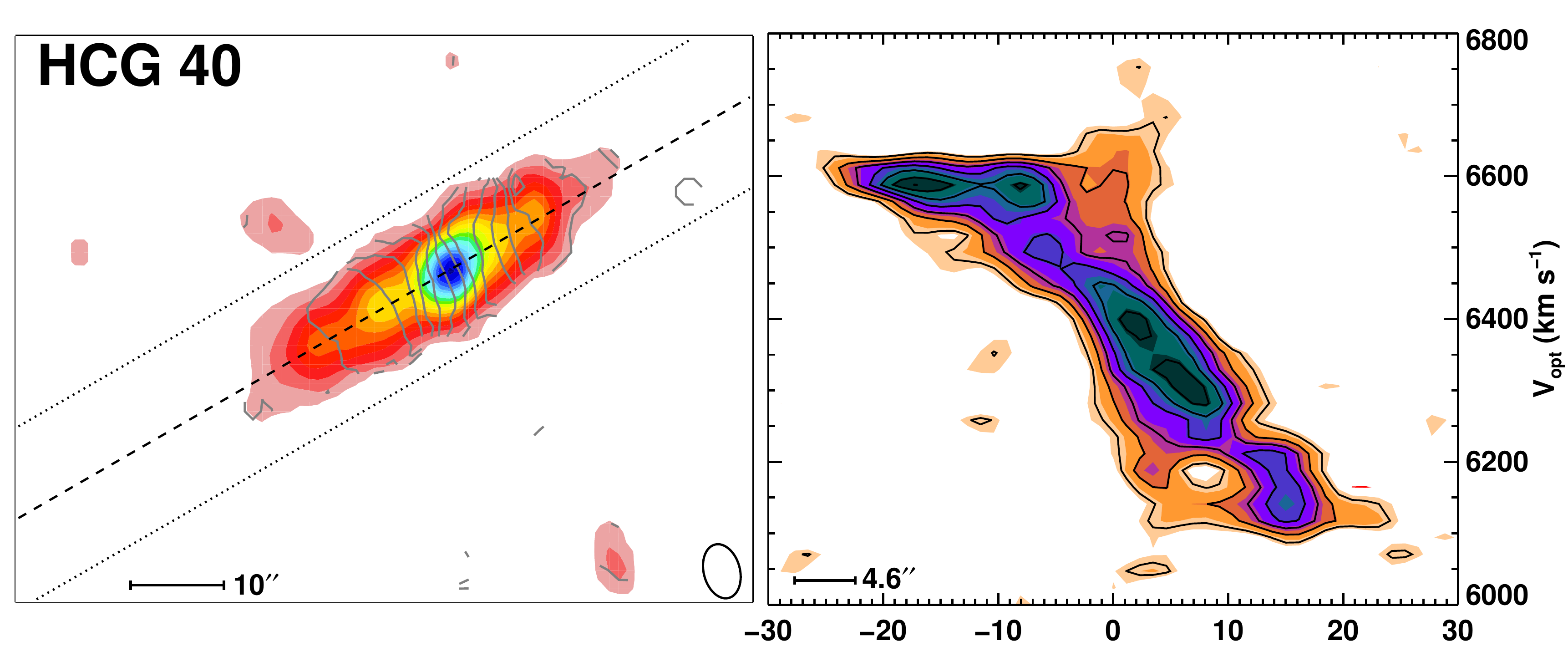}}
\subfigure{\includegraphics[width=\textwidth]{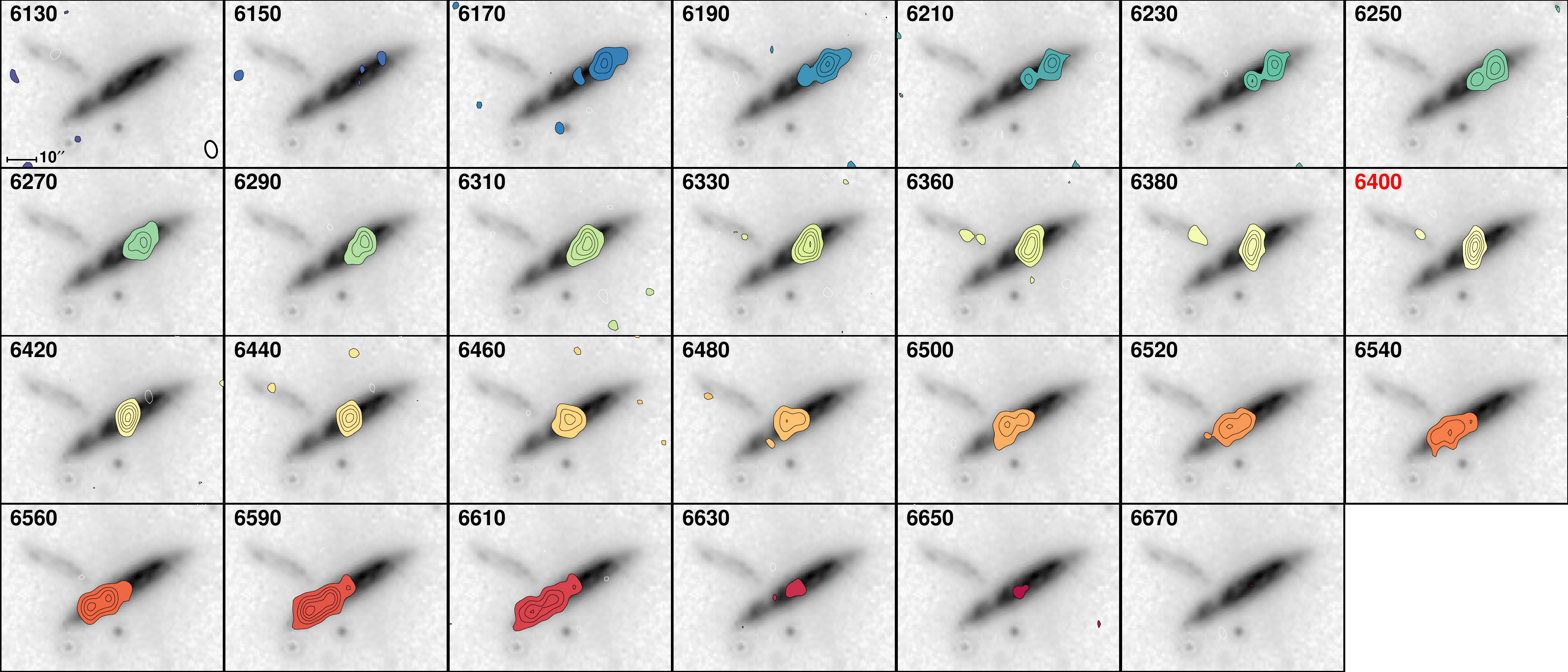}}
\caption{HCG\,40.  Channel map contours are in 3$\sigma$ steps.}
 \label{fig:hcg40}
 \end{figure*}

\begin{figure*}[h!]
\subfigure{\includegraphics[width=\textwidth,clip,trim=0cm 0cm 1cm 0cm]{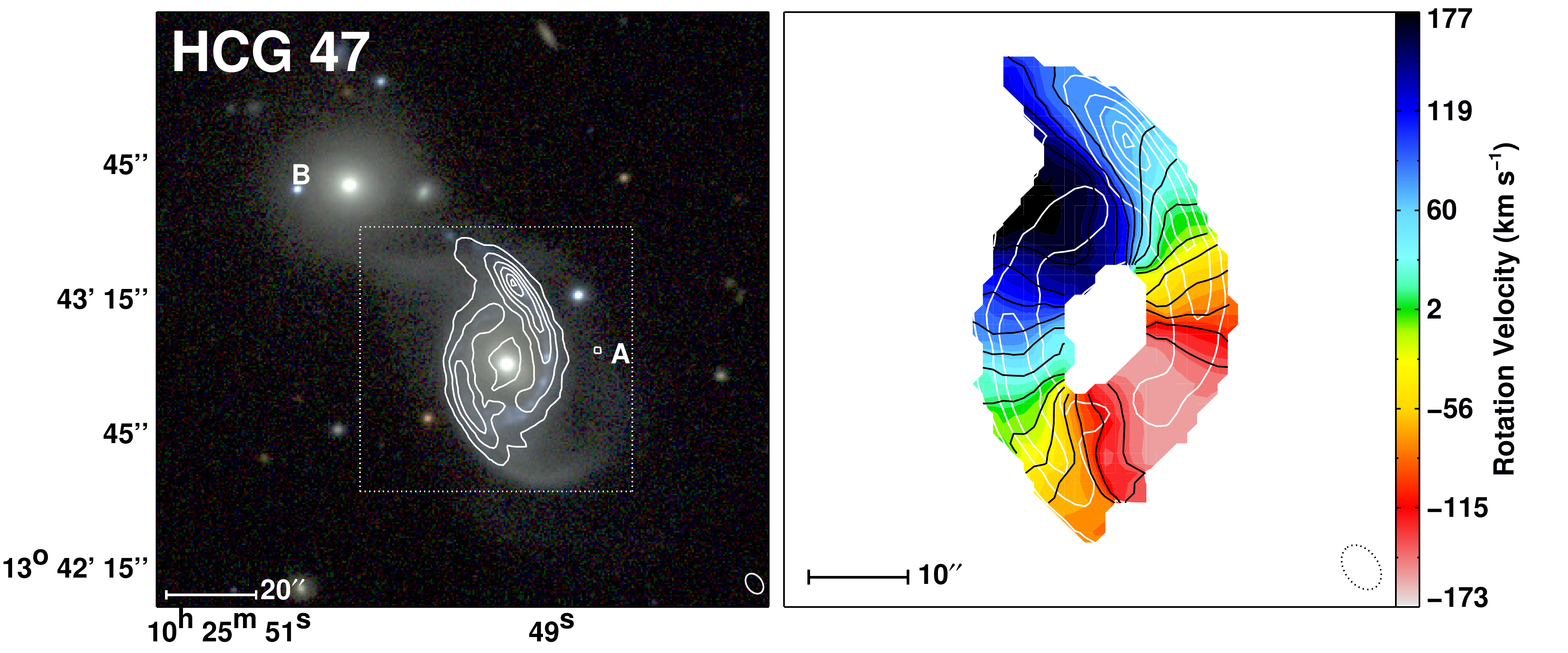}}
\subfigure{\includegraphics[height=4.9cm]{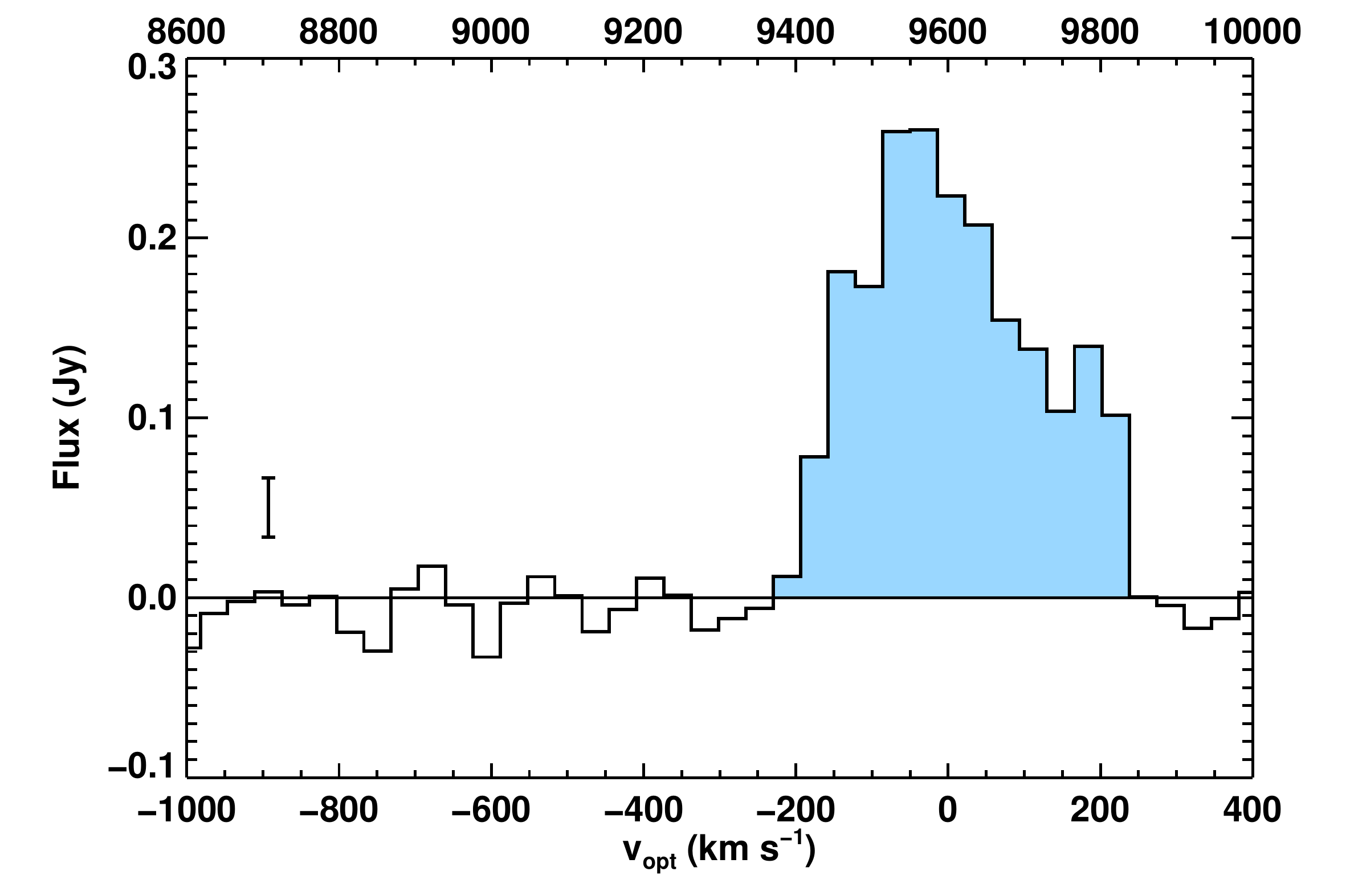}}
\subfigure{\includegraphics[height=4.9cm]{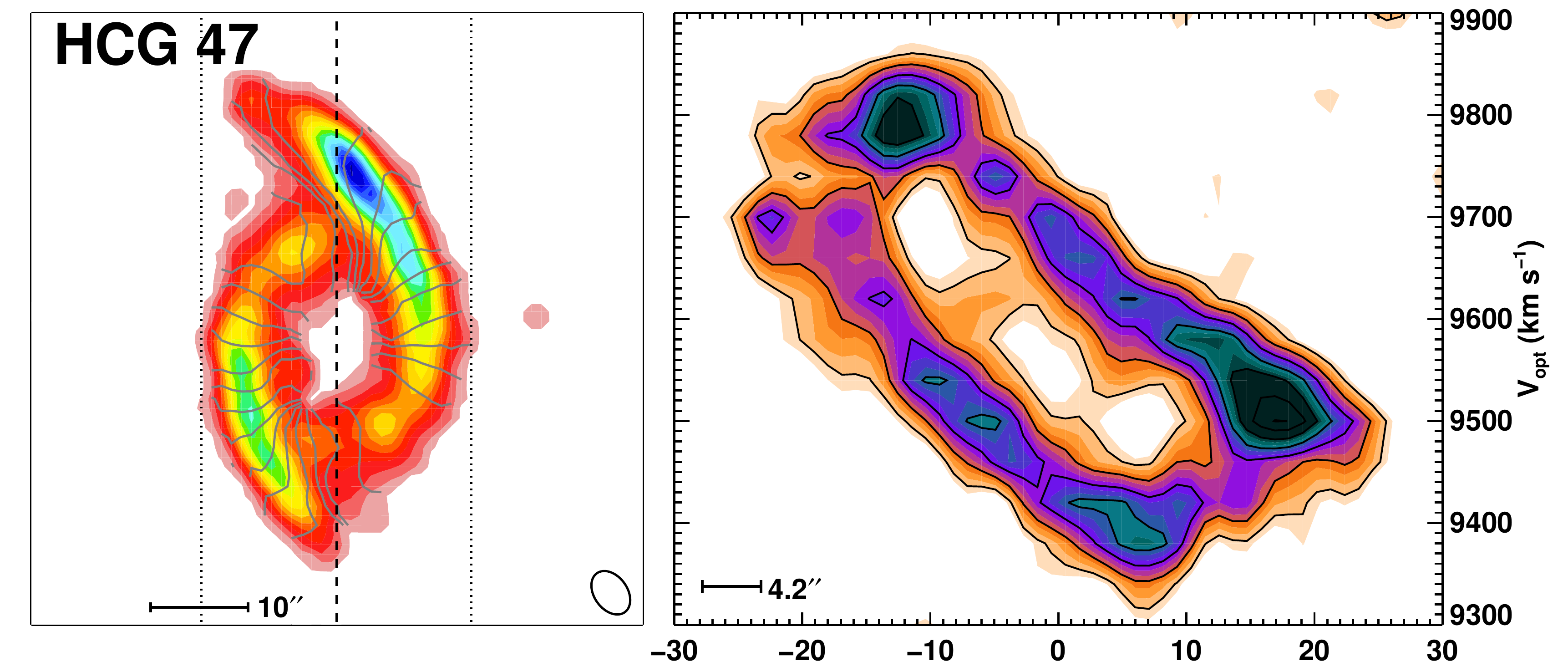}}
\subfigure{\includegraphics[width=\textwidth]{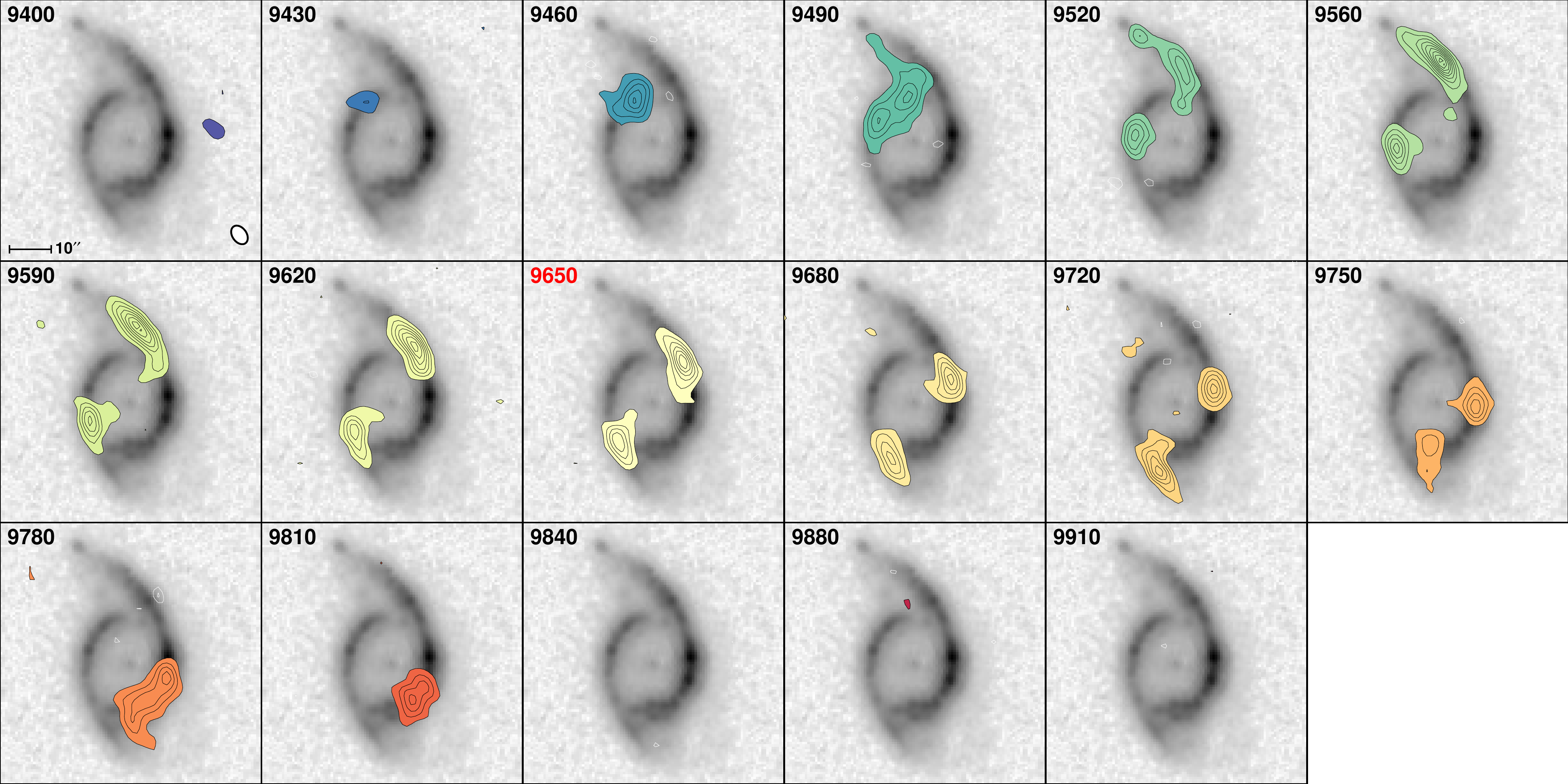}}
\caption{HCG\,47. Channel map contours are in 3$\sigma$ steps.}
 \label{fig:hcg47}
 \end{figure*}

\begin{figure*}[h!]
\subfigure{\includegraphics[width=\textwidth]{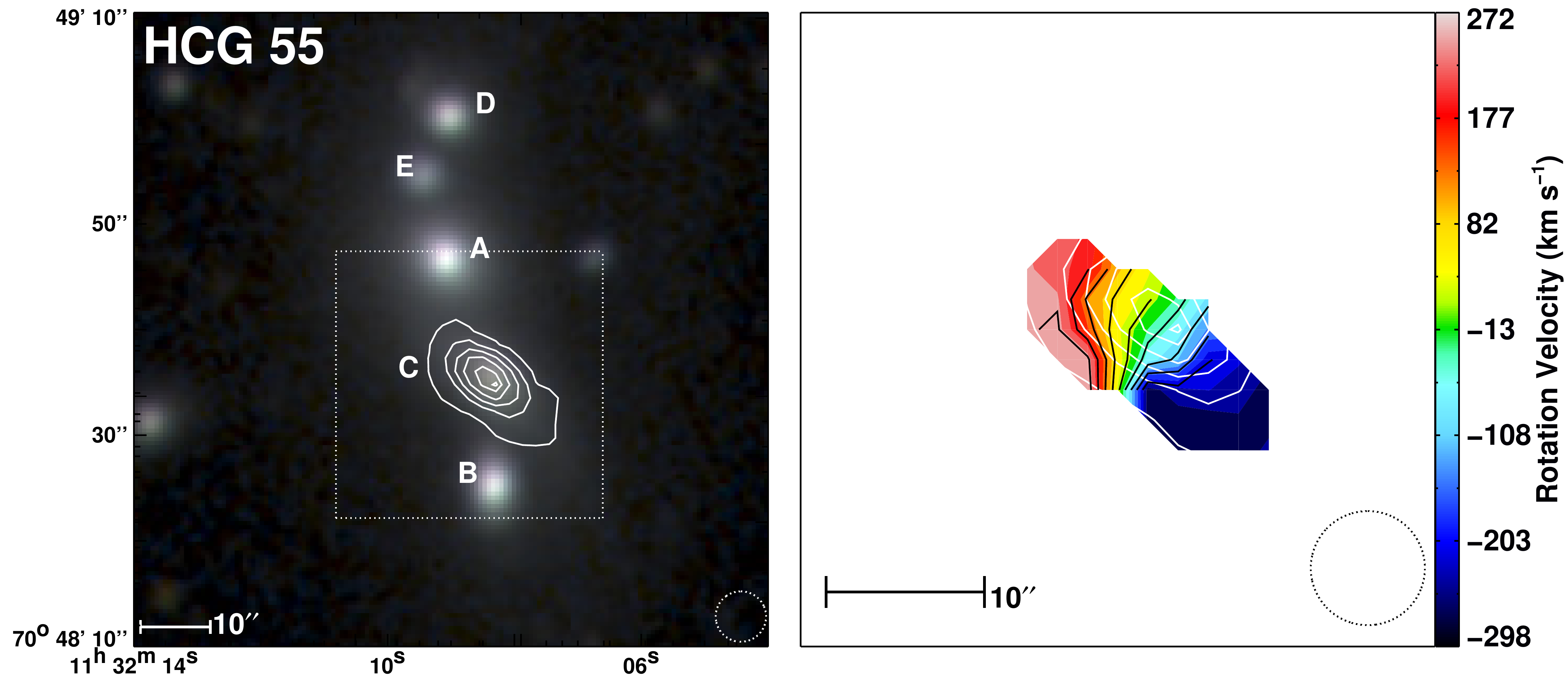}}
\subfigure{\includegraphics[height=5.2cm]{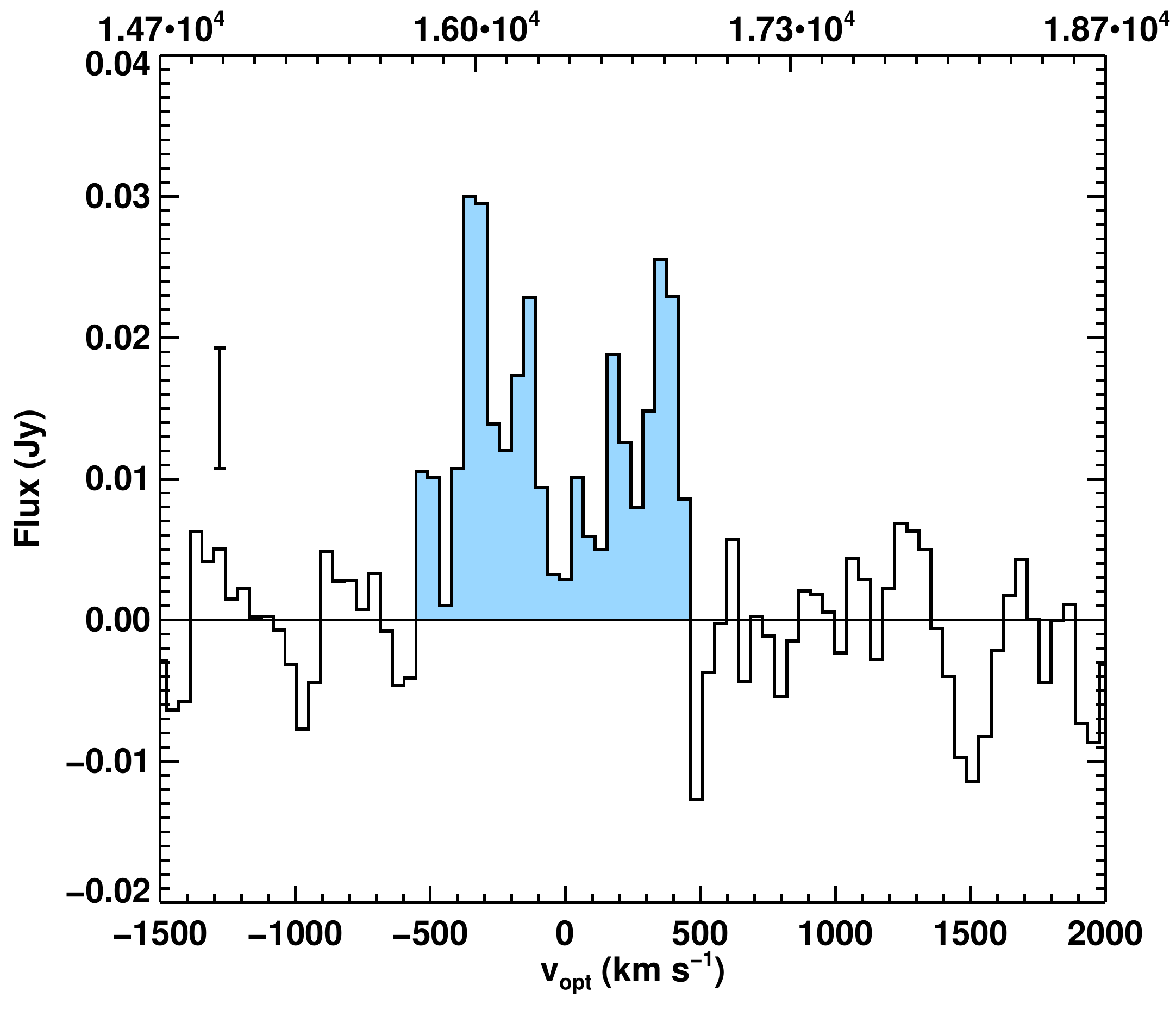}}
\subfigure{\includegraphics[height=5.2cm]{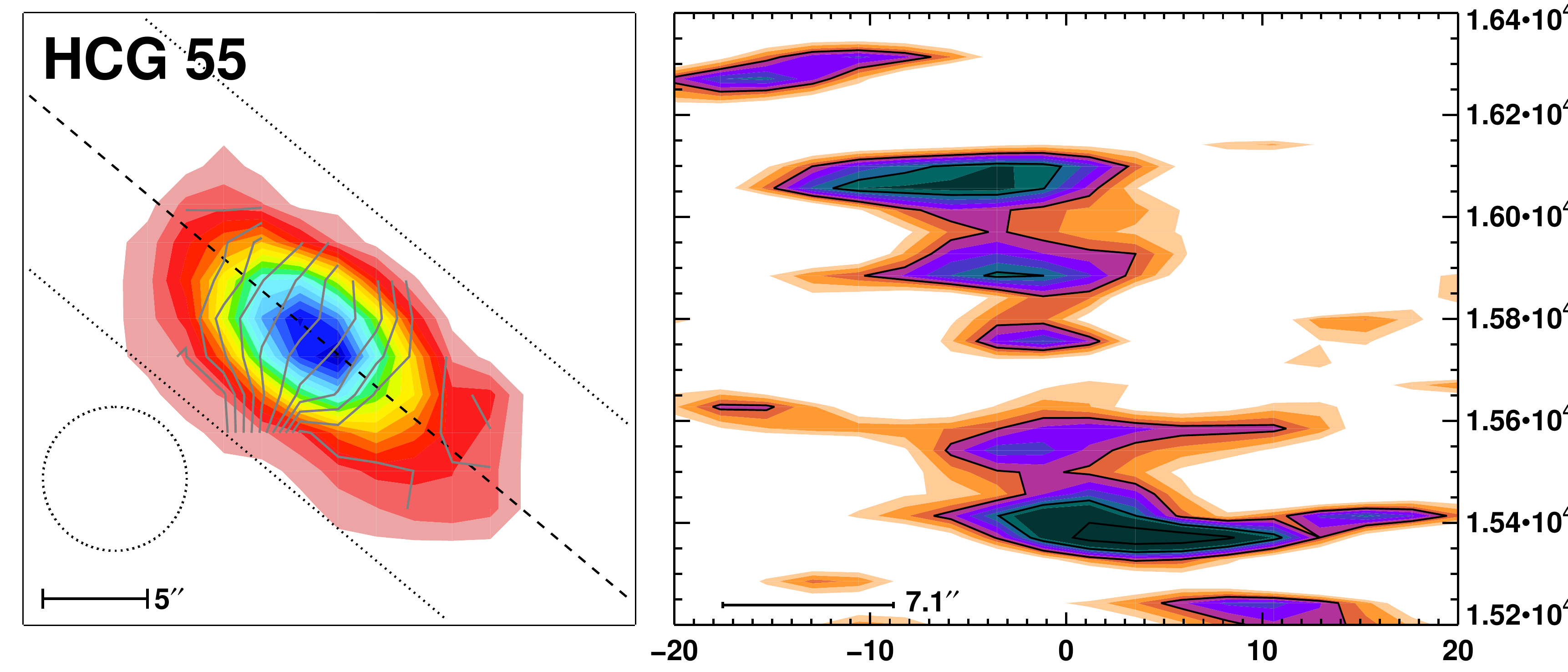}}
\subfigure{\includegraphics[width=\textwidth]{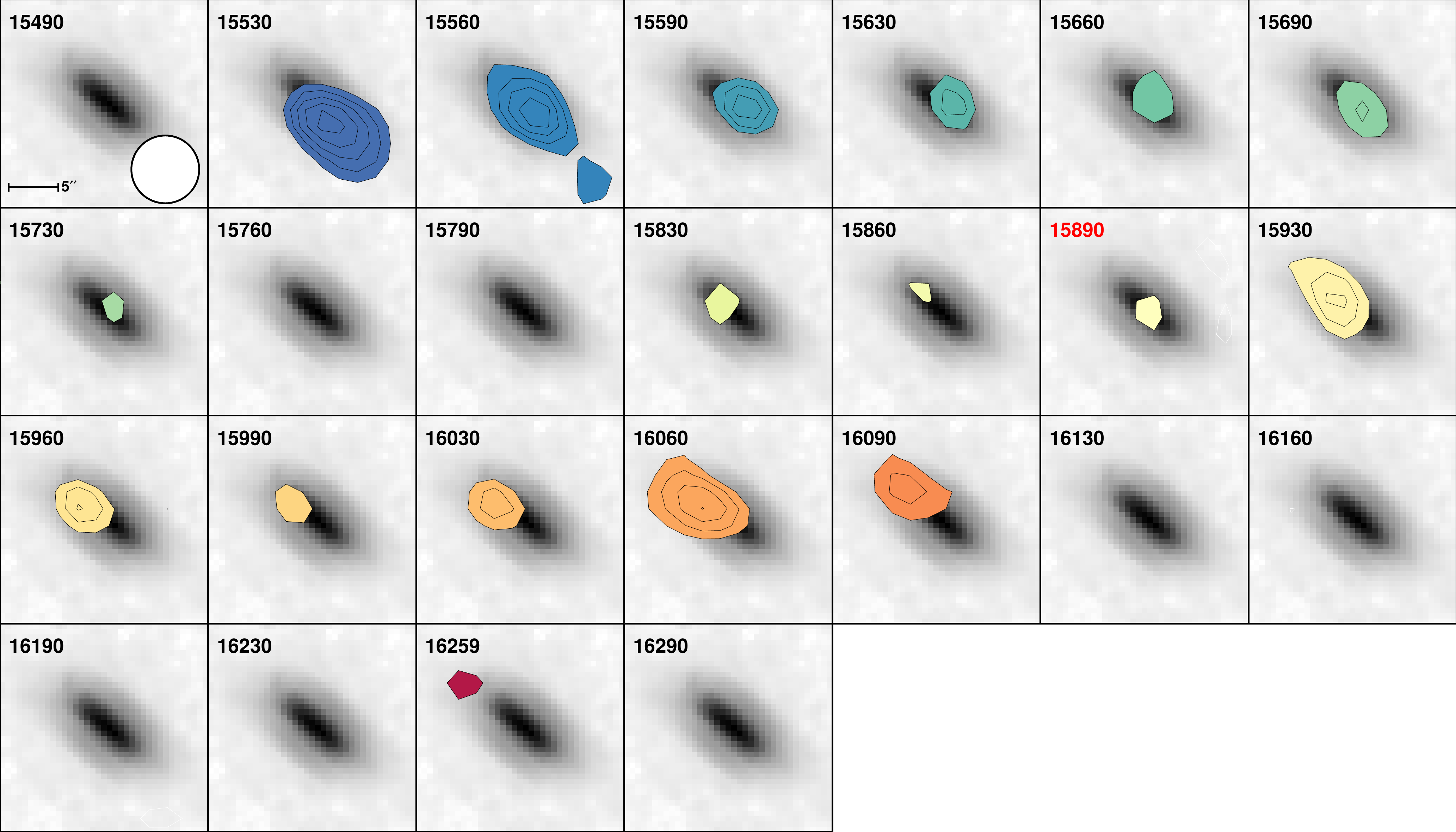}}
\caption{HCG\,55. Channel map contours are in 1$\sigma$ steps.}
 \label{fig:hcg55}
 \end{figure*}

\begin{figure*}[h!]
\subfigure{\includegraphics[width=\textwidth]{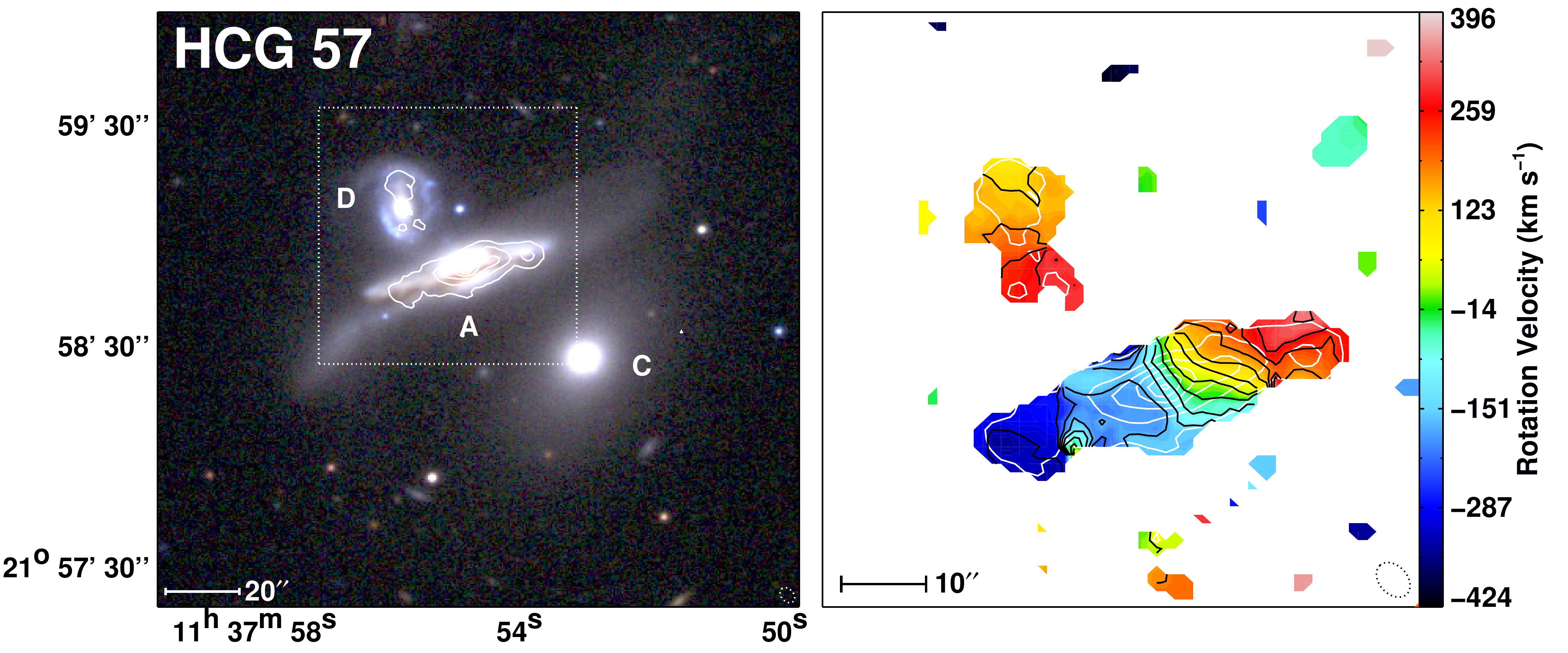}}
\subfigure{\includegraphics[height=4.8cm]{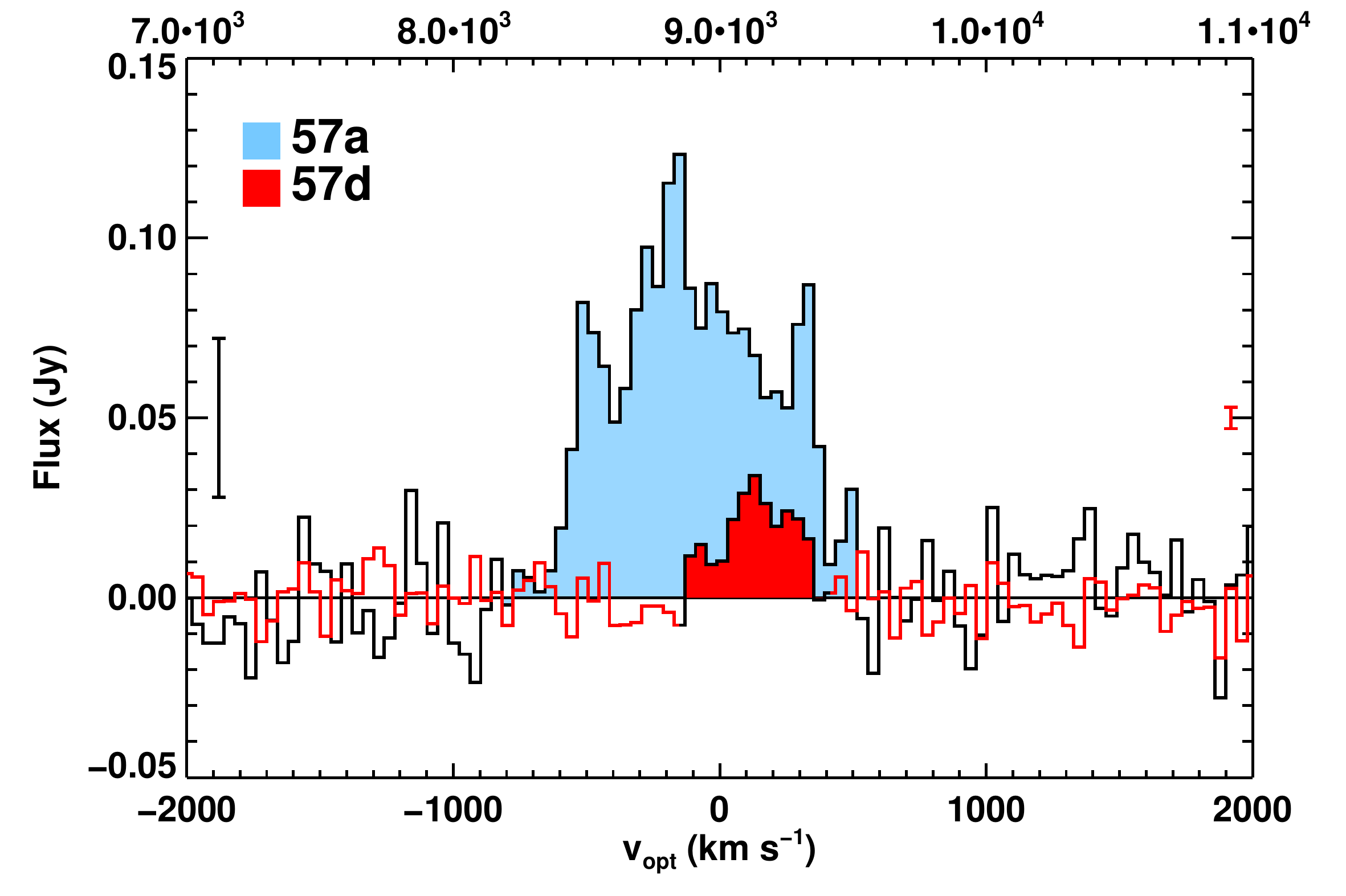}}
\subfigure{\includegraphics[height=4.8cm]{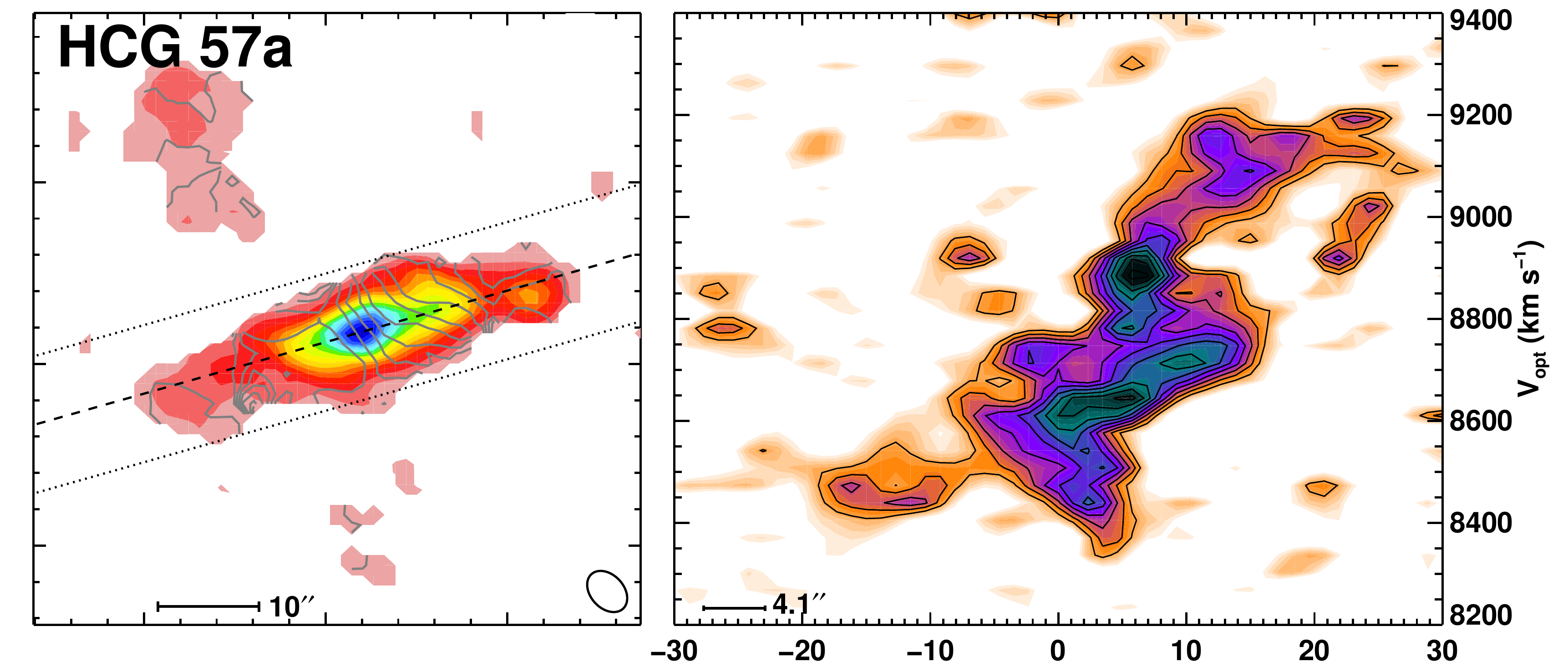}}
\subfigure{\includegraphics[width=\textwidth]{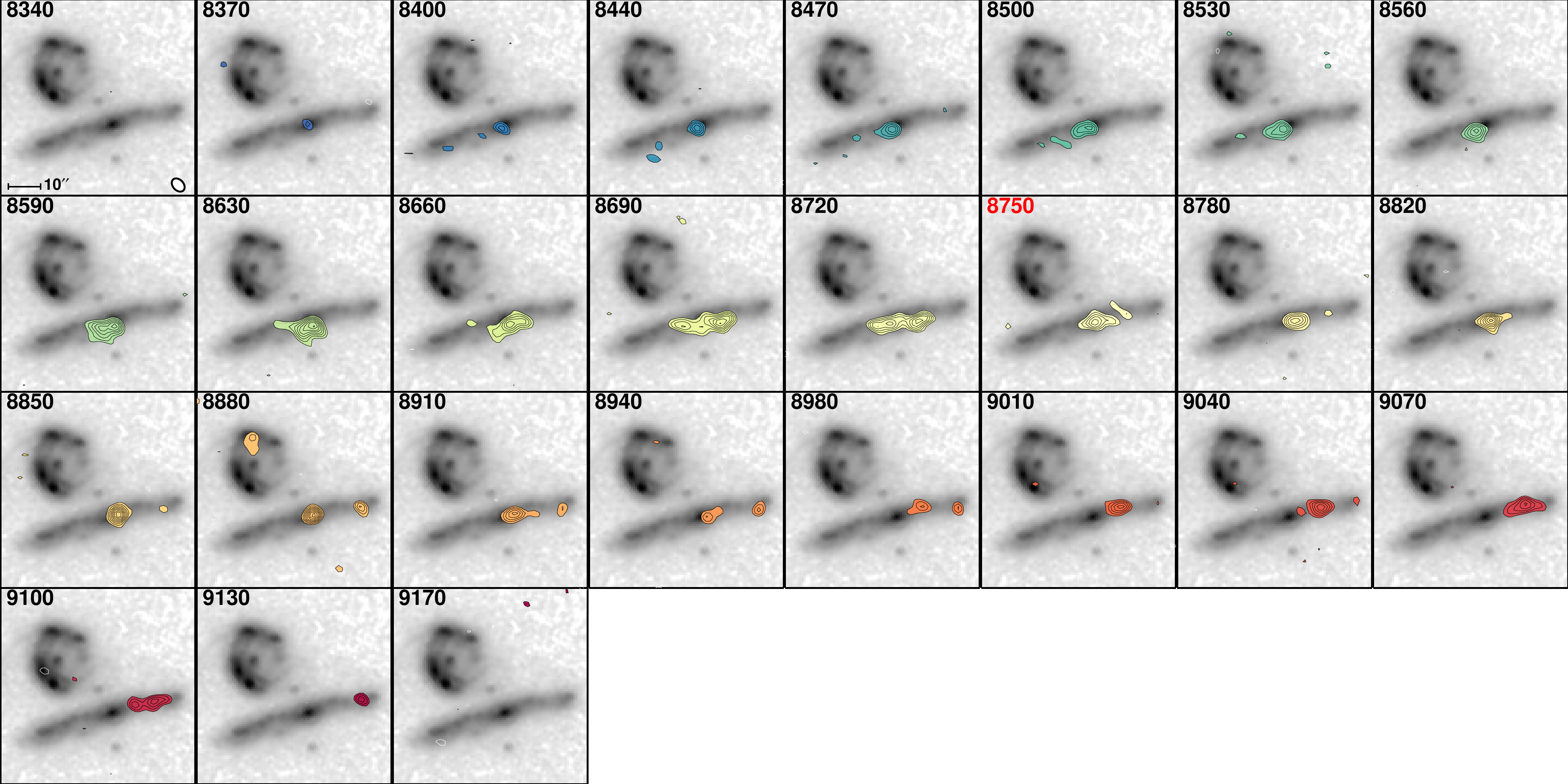}}
\caption{HCG\,57. Channel map contours start at $\pm$2.5$\sigma$ and are in 1$\sigma$ steps.  The PVD of 57d can be found in Fig~\ref{fig:PV57d+96c}}
 \label{fig:hcg57}
 \end{figure*}

\begin{figure*}[h!]
\subfigure{\includegraphics[width=\textwidth]{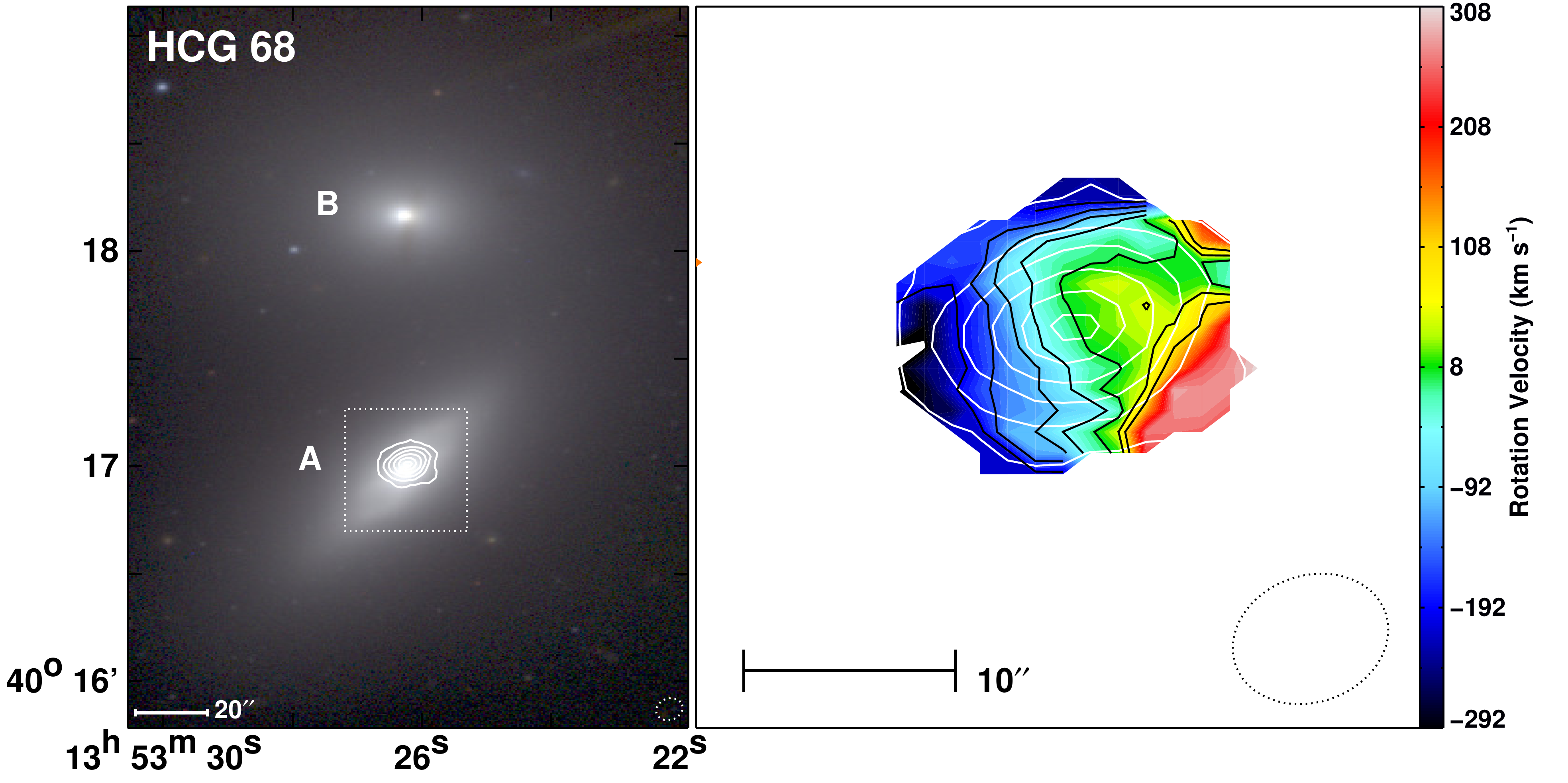}}
\subfigure{\includegraphics[height=4.9cm,clip,trim=0.5cm 1.4cm 0cm 2cm]{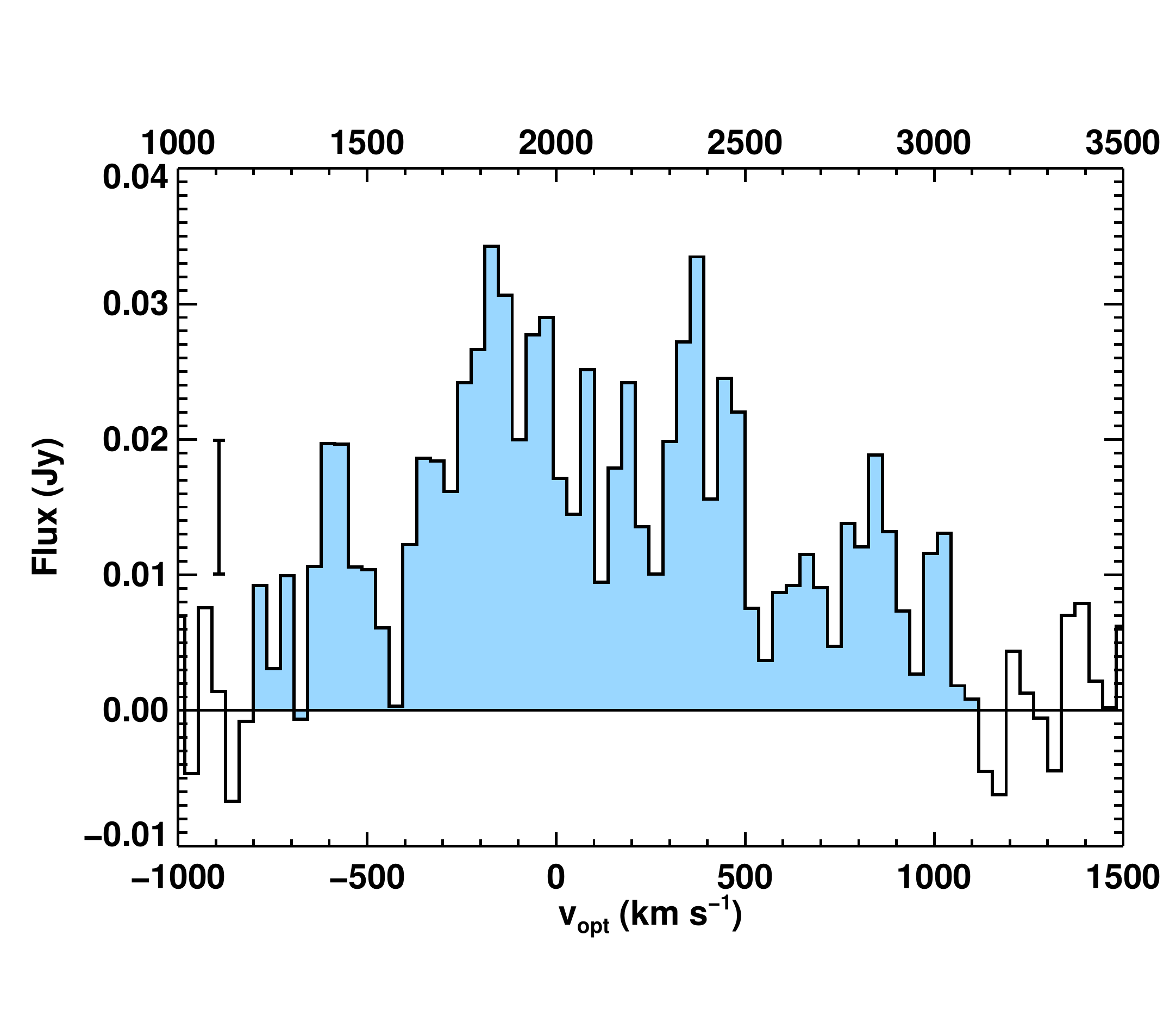}}
\subfigure{\includegraphics[height=4.9cm]{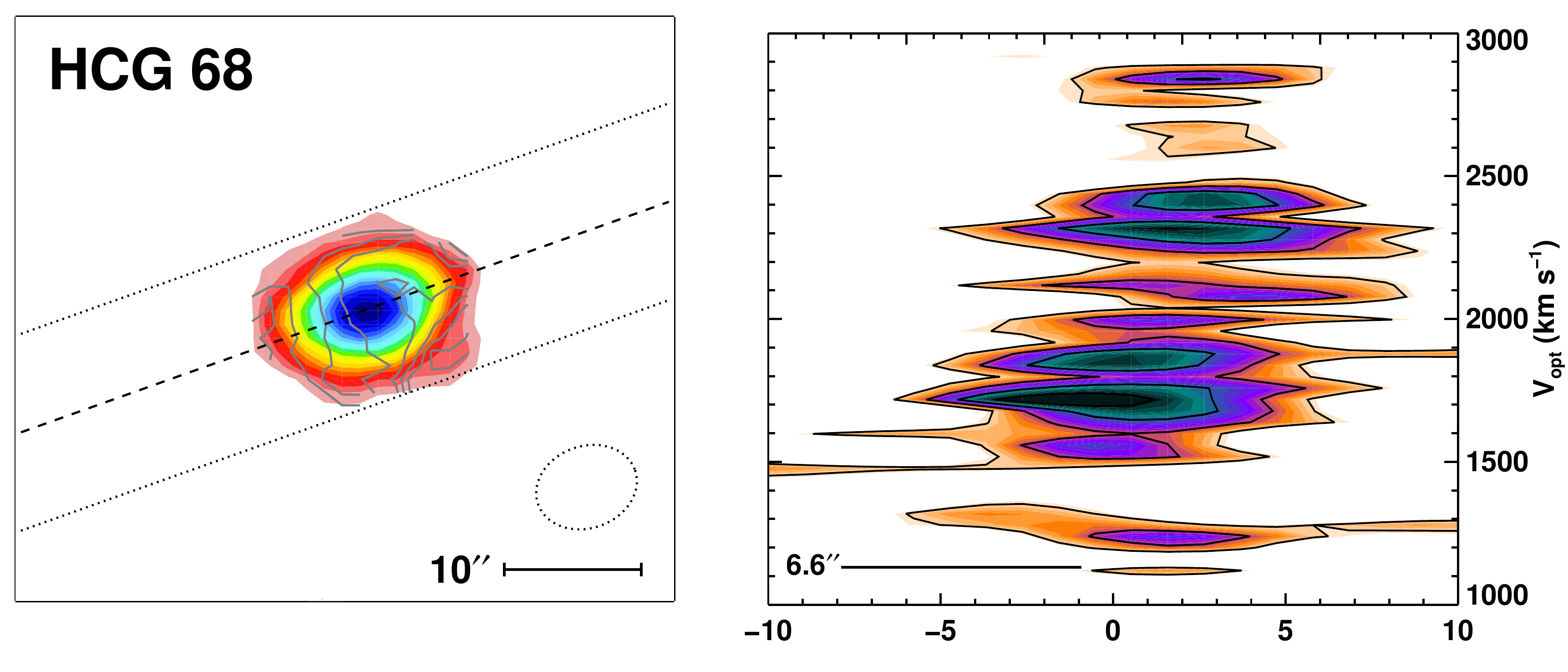}}
\subfigure{\includegraphics[width=\textwidth]{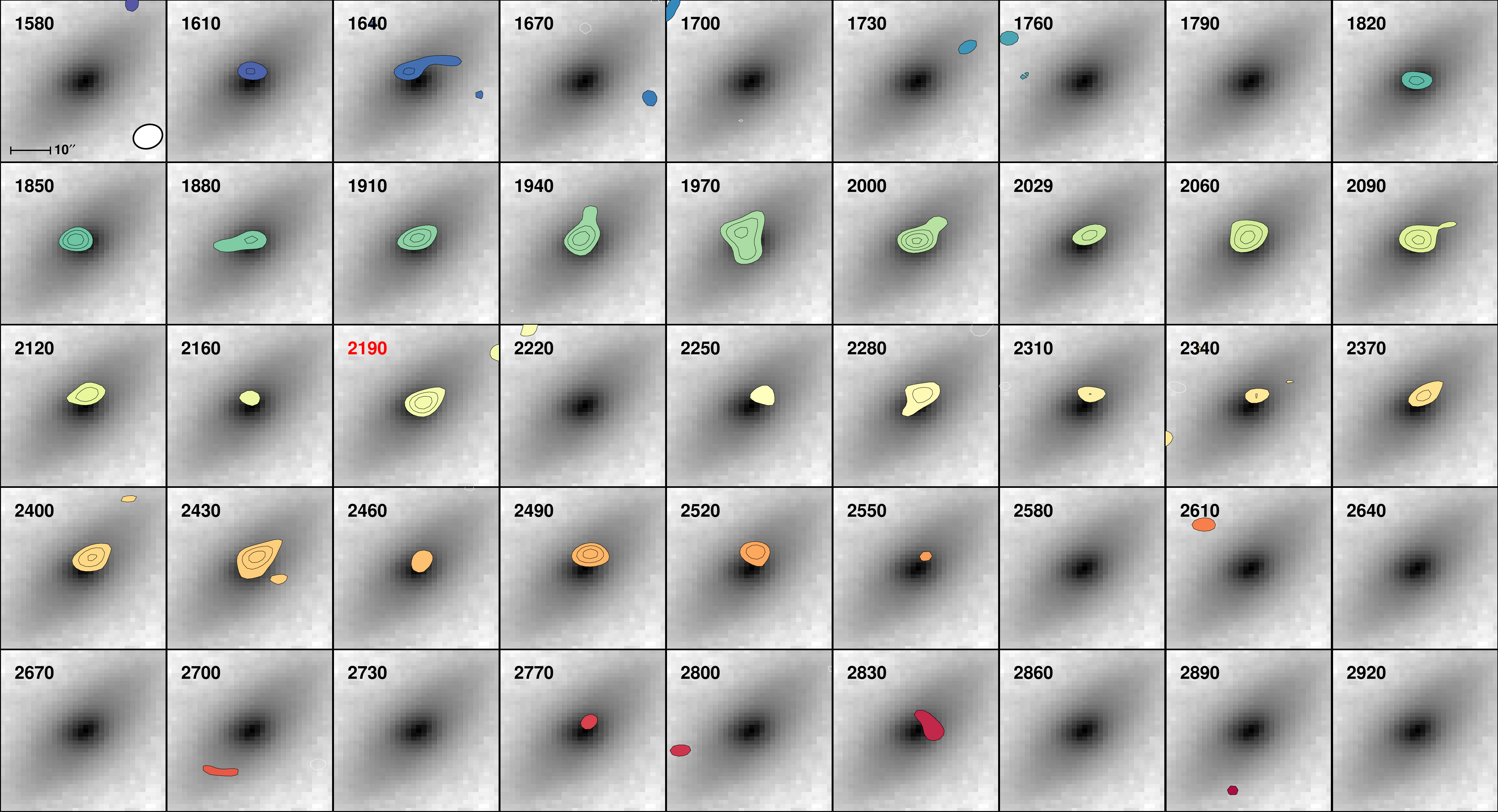}}
\caption{HCG\,68. Channel map contours are in 1$\sigma$ steps.}
 \label{fig:hcg68}
 \end{figure*}

\begin{figure*}[h!]
\subfigure{\includegraphics[width=\textwidth,clip,trim=0cm 0cm 1.4cm 0cm]{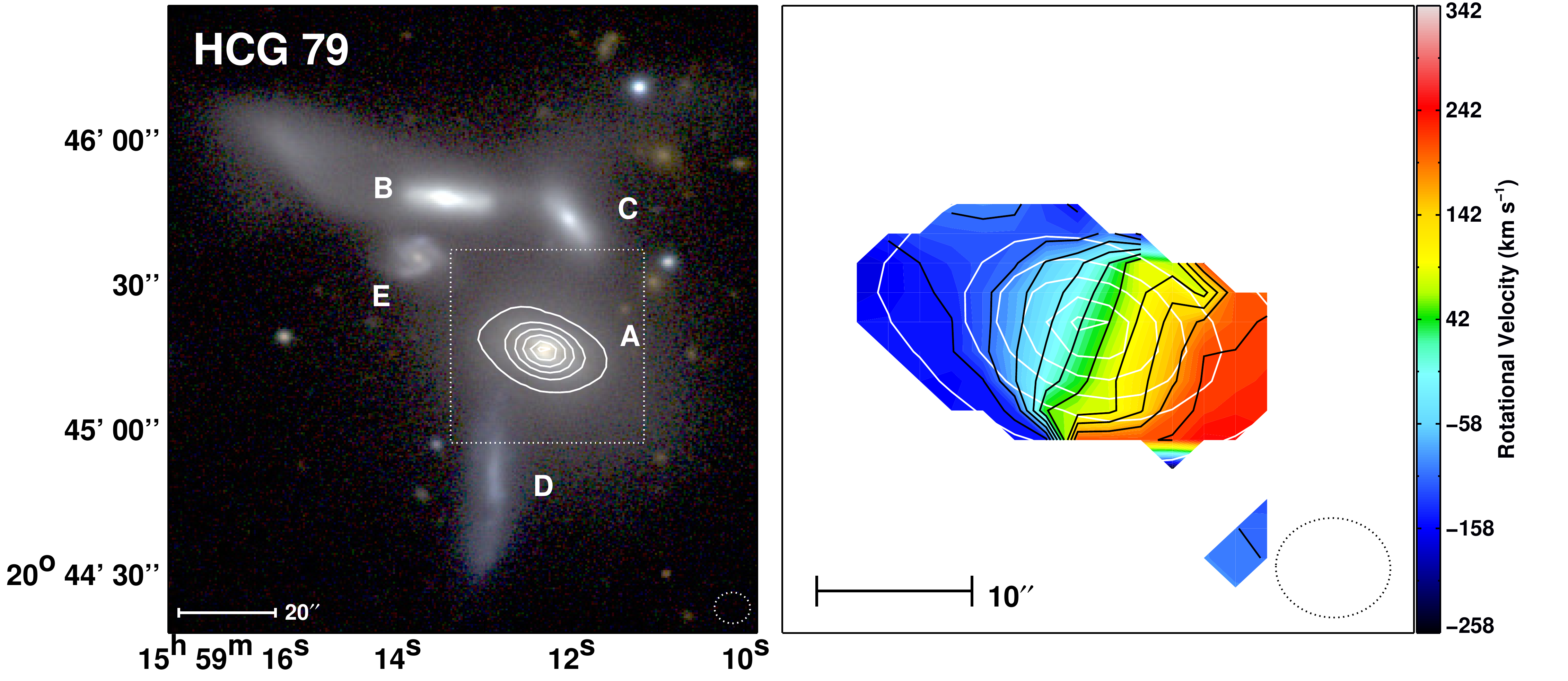}}
\subfigure{\includegraphics[height=4.9cm]{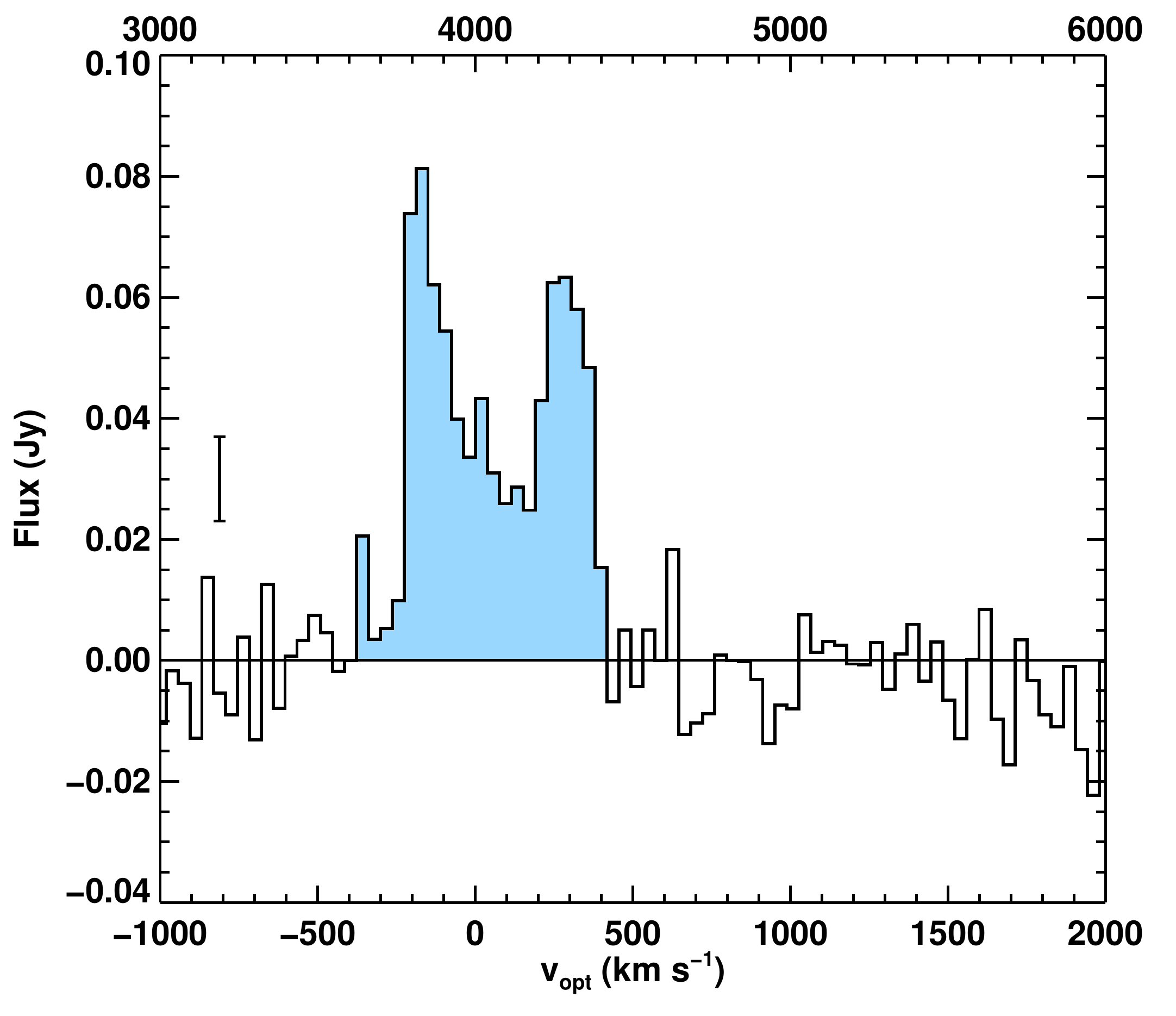}}
\subfigure{\includegraphics[height=4.9cm]{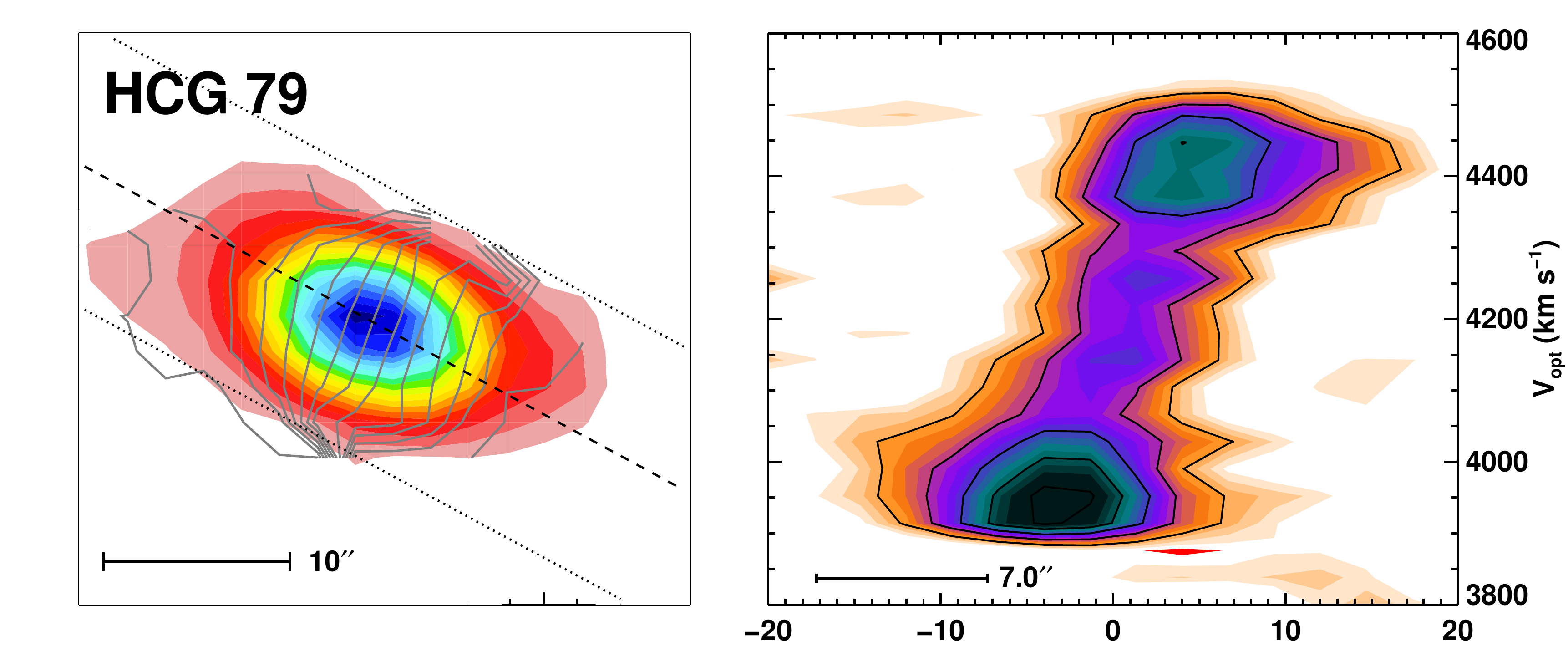}}
\subfigure{\includegraphics[width=\textwidth]{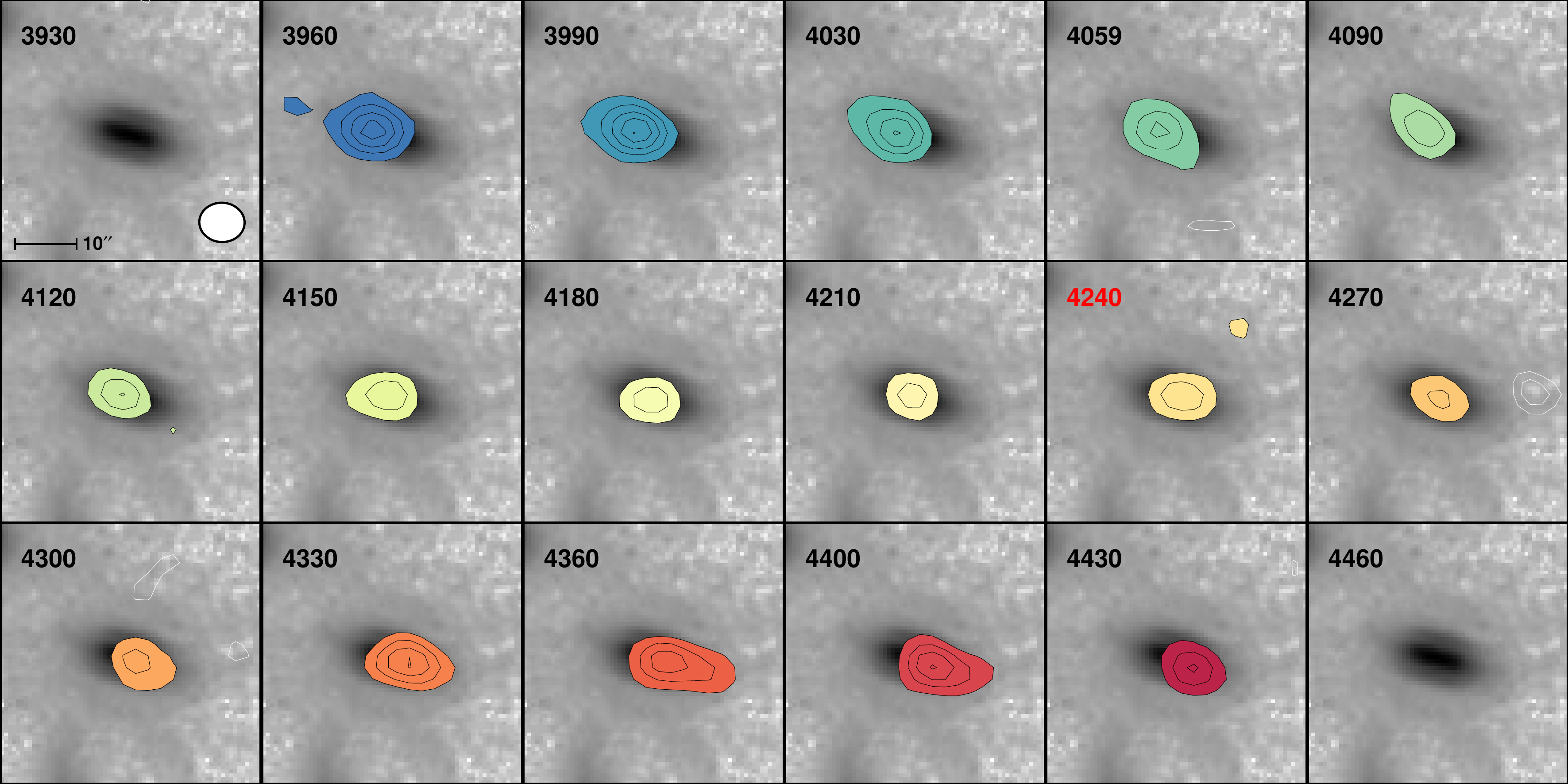}}
\caption{HCG\,79. Channel map contours are in 3$\sigma$ steps.}
 \label{fig:hcg79}
 \end{figure*}

\begin{figure*}[h!]
\subfigure{\includegraphics[width=\textwidth]{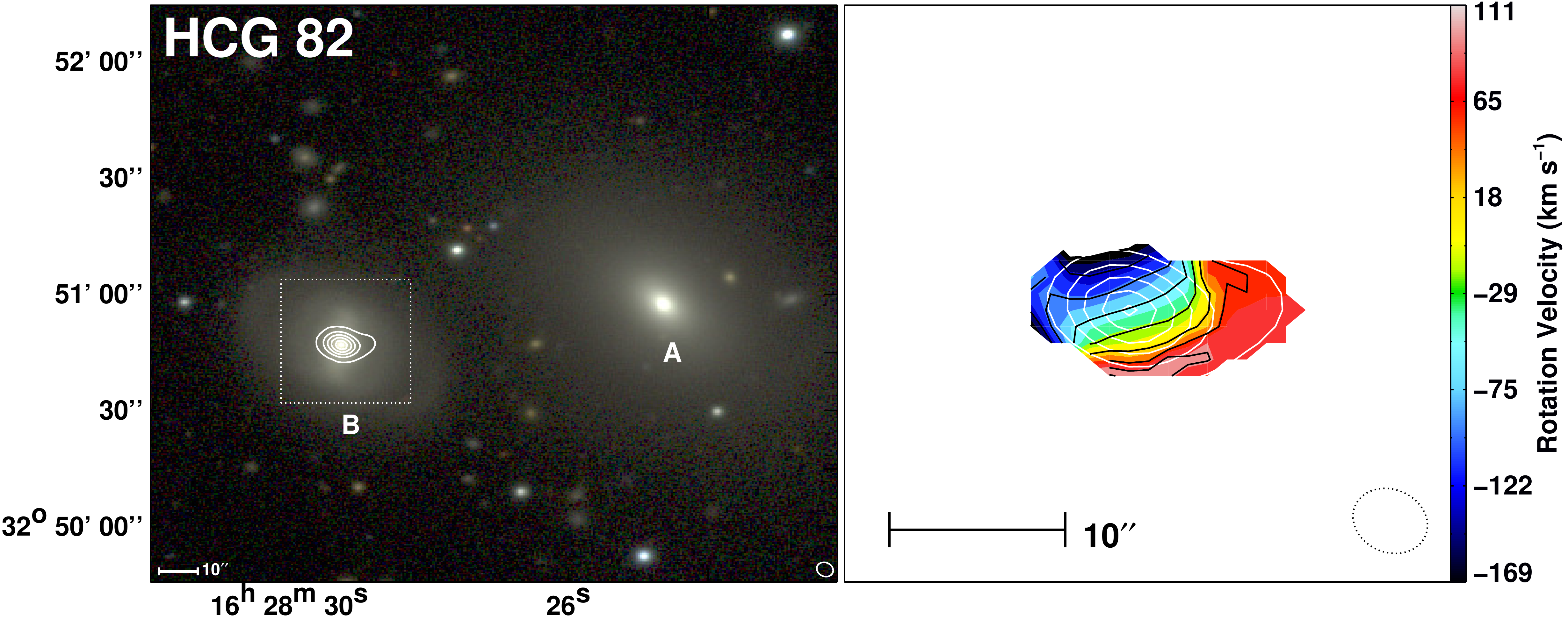}}
\subfigure{\includegraphics[height=4.8cm]{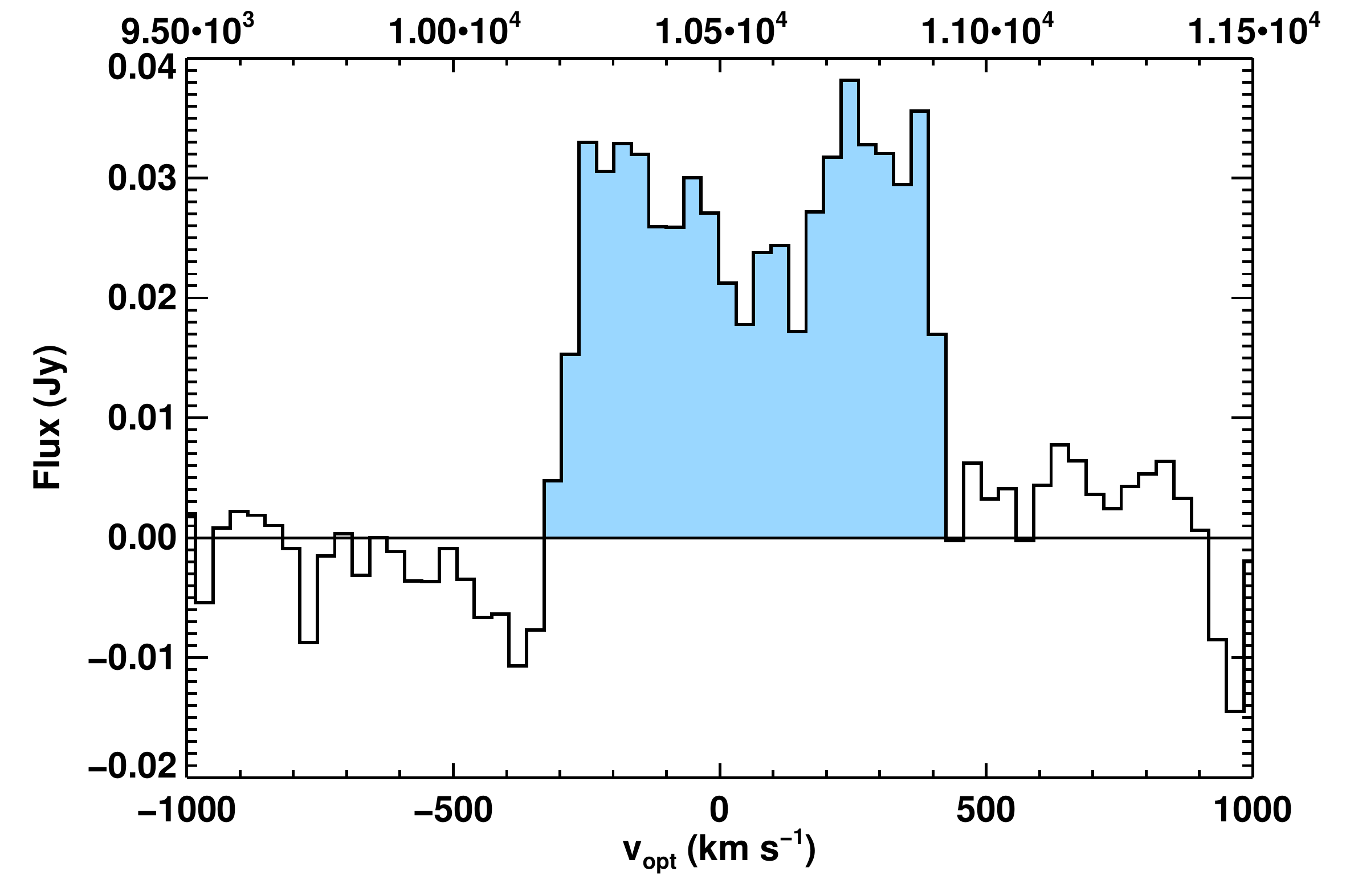}}
\subfigure{\includegraphics[height=4.8cm]{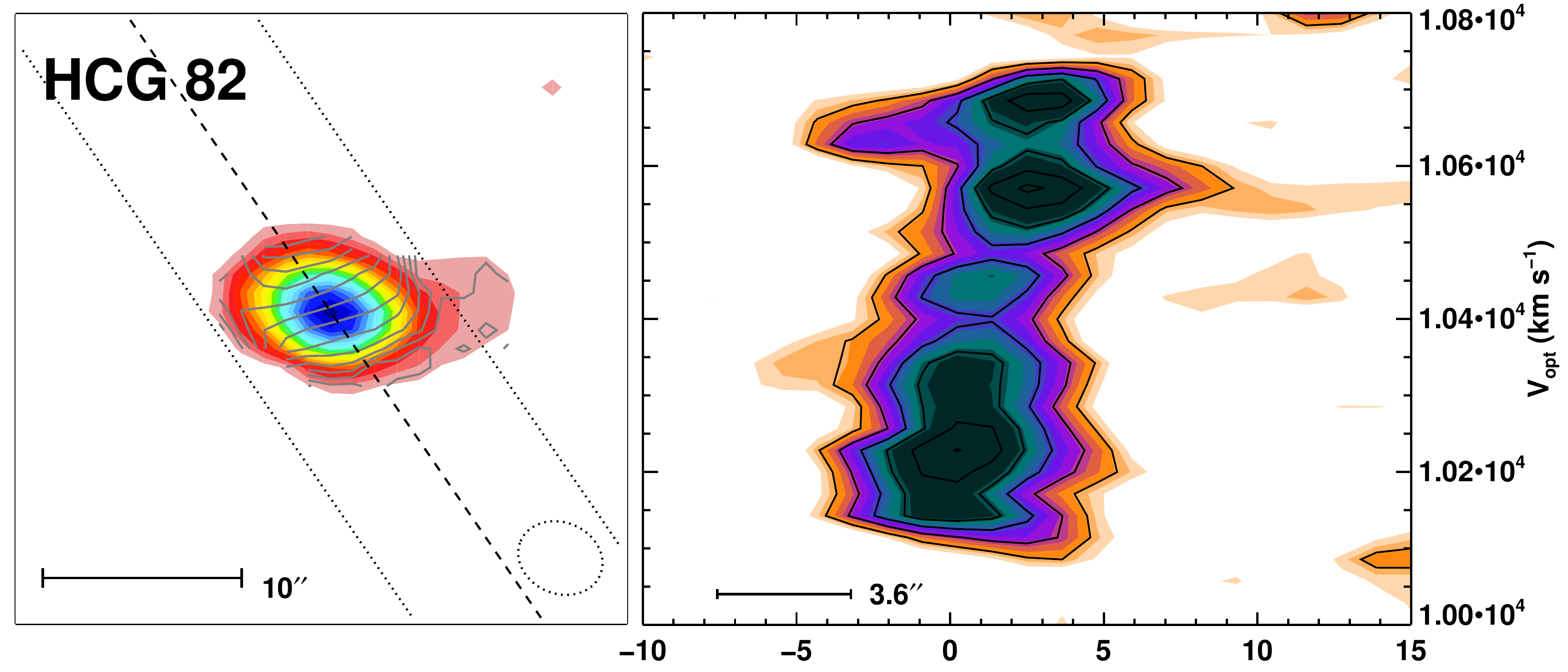}}
\subfigure{\includegraphics[width=\textwidth]{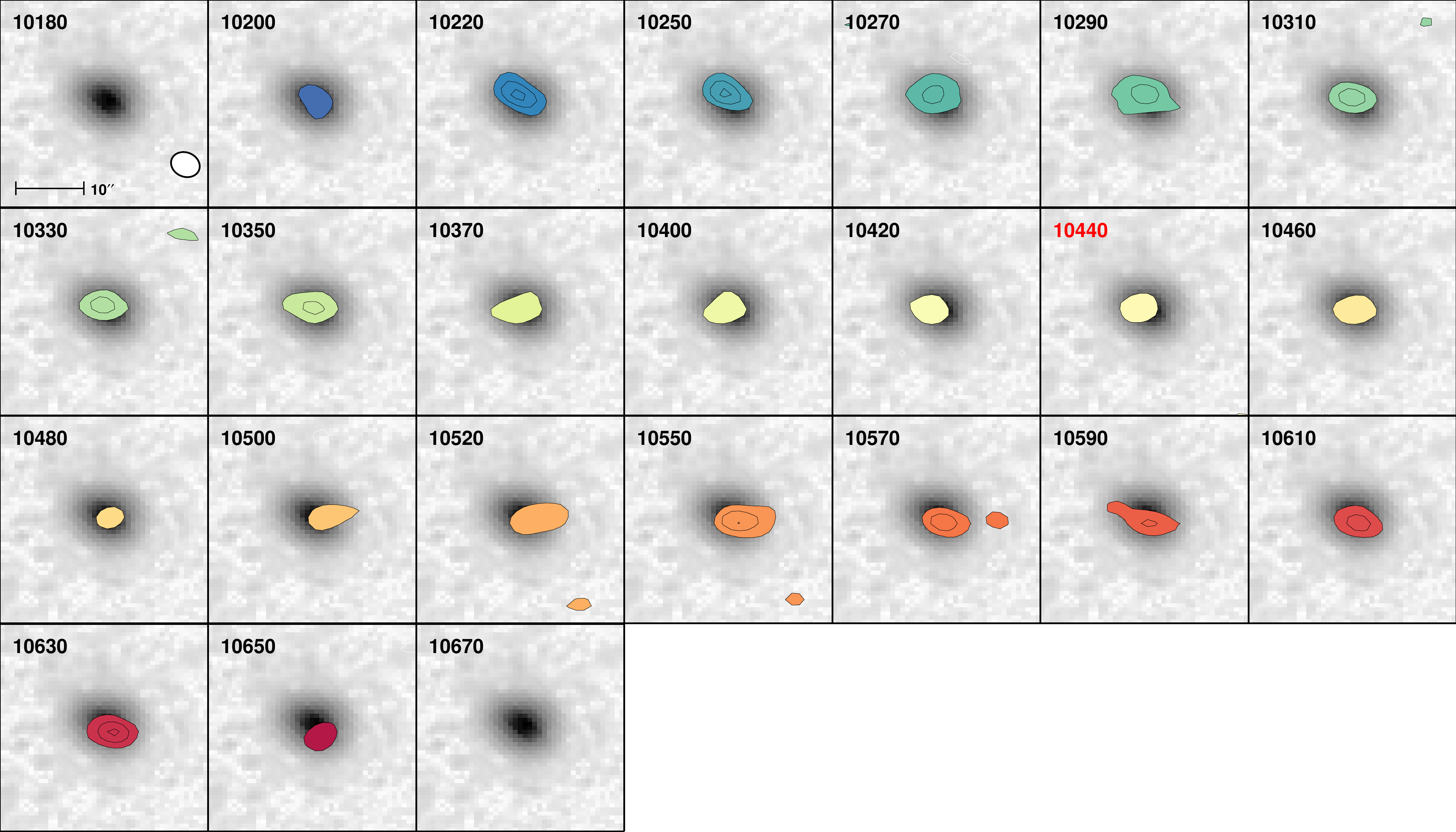}}
\caption{HCG\,82. Channel map contours are in 3$\sigma$ steps.}
 \label{fig:hcg82}
 \end{figure*}

\begin{figure*}[h!]
\subfigure{\includegraphics[width=\textwidth]{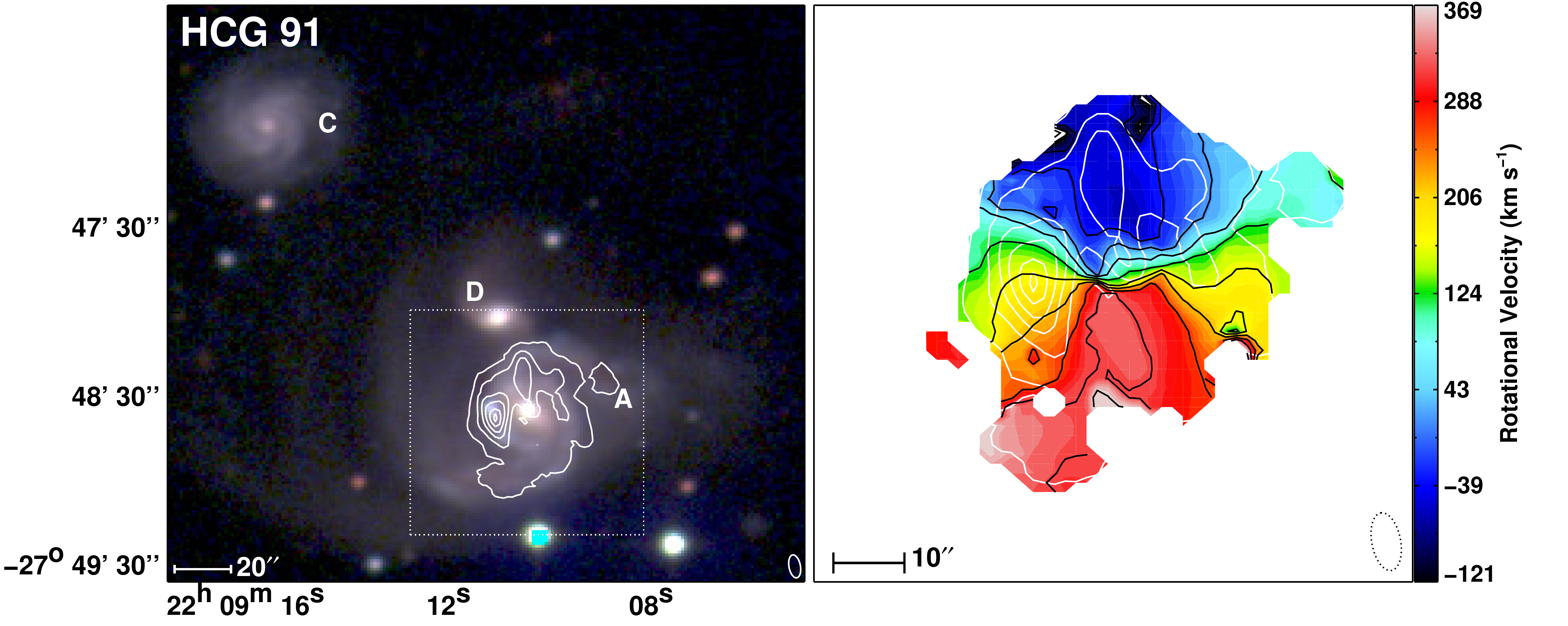}}
\subfigure{\includegraphics[height=4.9cm,clip,trim=0.9cm 1.4cm 0.1cm 2cm]{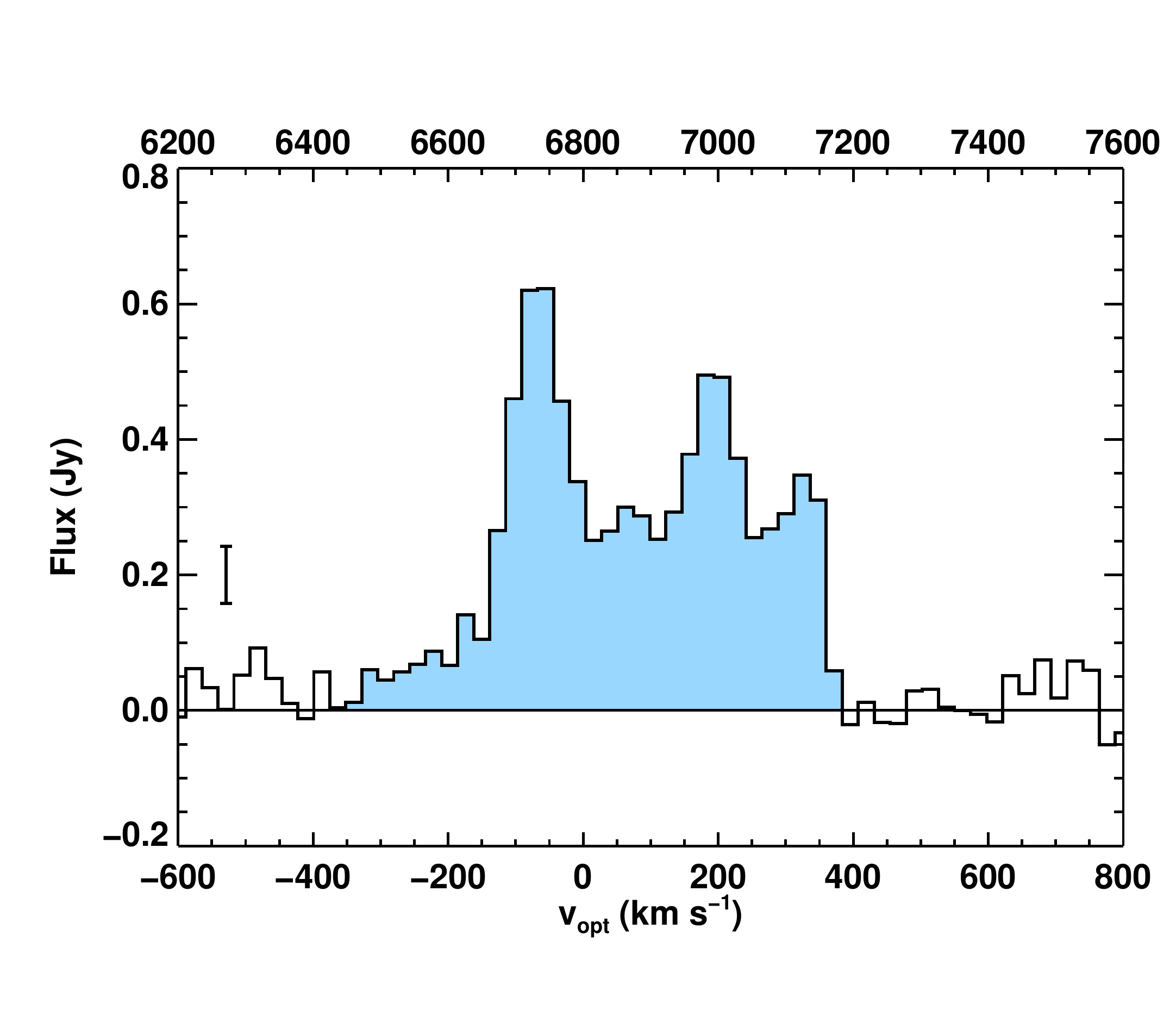}}
\subfigure{\includegraphics[height=4.9cm]{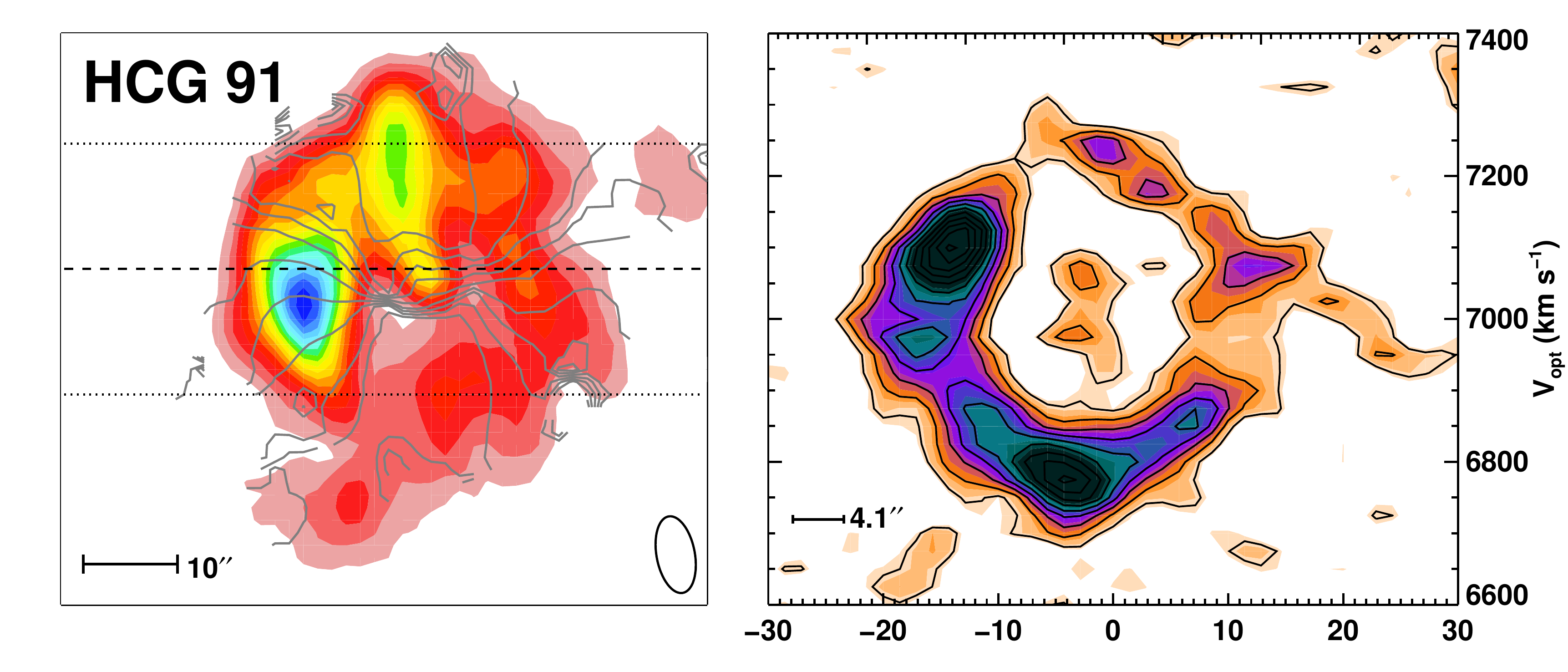}}
\subfigure{\includegraphics[width=\textwidth]{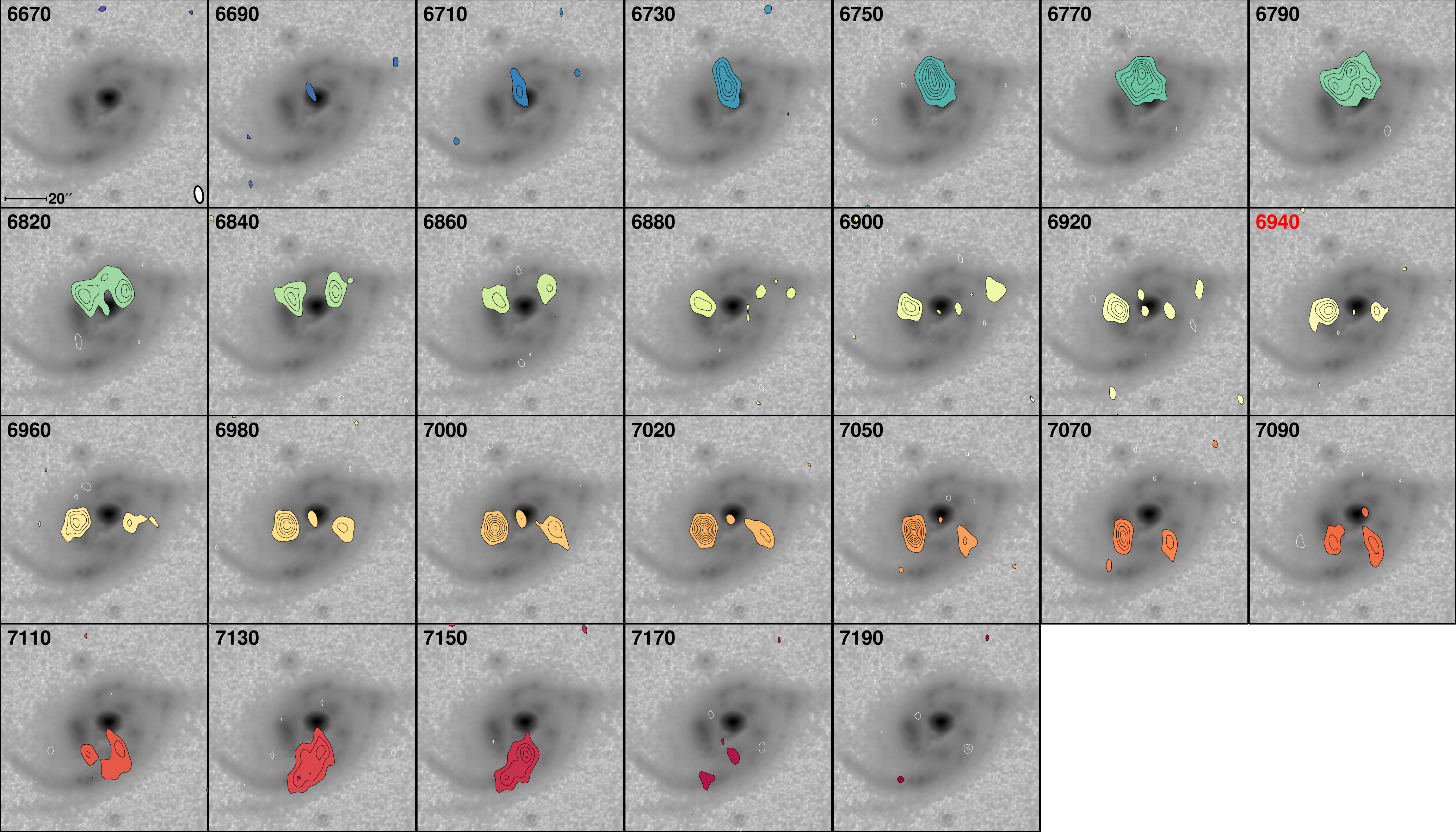}}
\caption{HCG\,91. Channel map contours are in 3$\sigma$ steps.}
 \label{fig:hcg91}
 \end{figure*}

\begin{figure*}[h!]
\subfigure{\includegraphics[width=\textwidth]{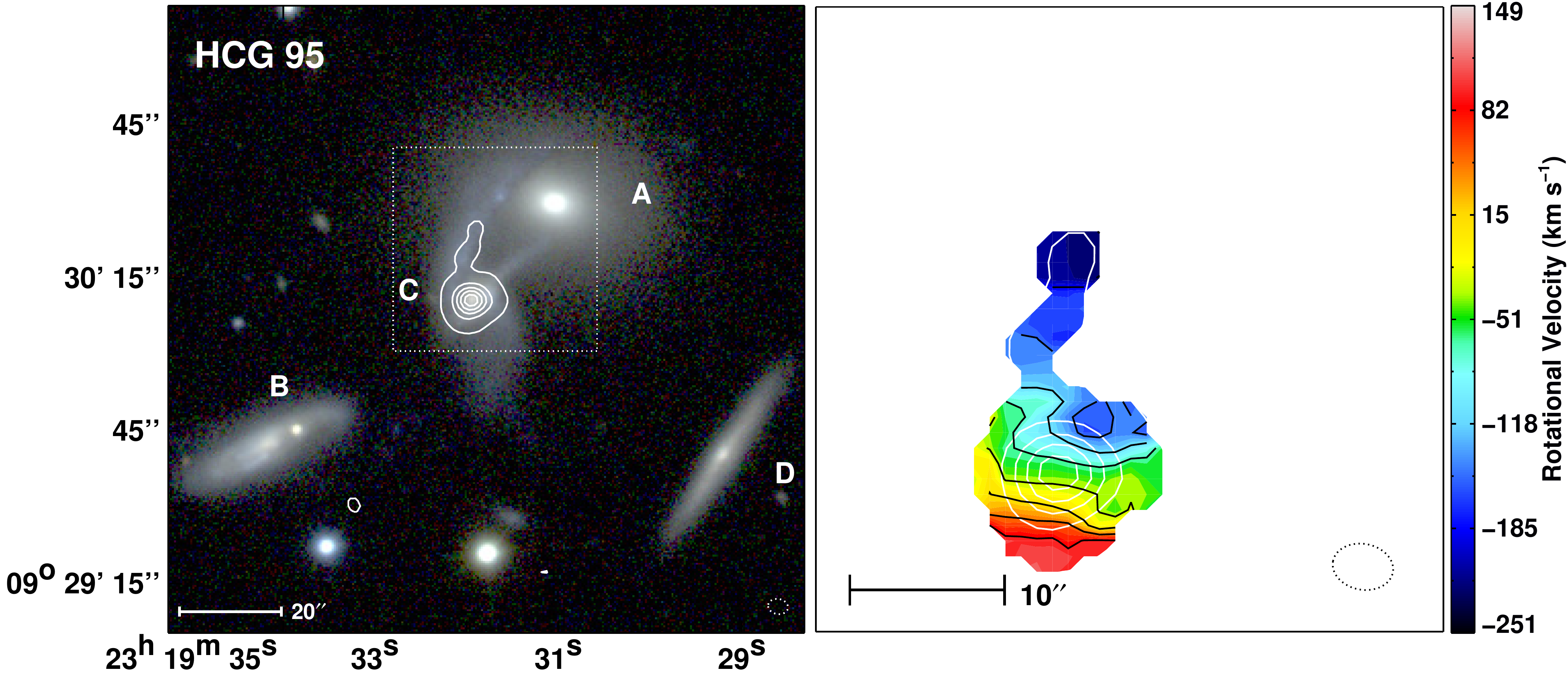}}
\subfigure{\includegraphics[height=4.9cm]{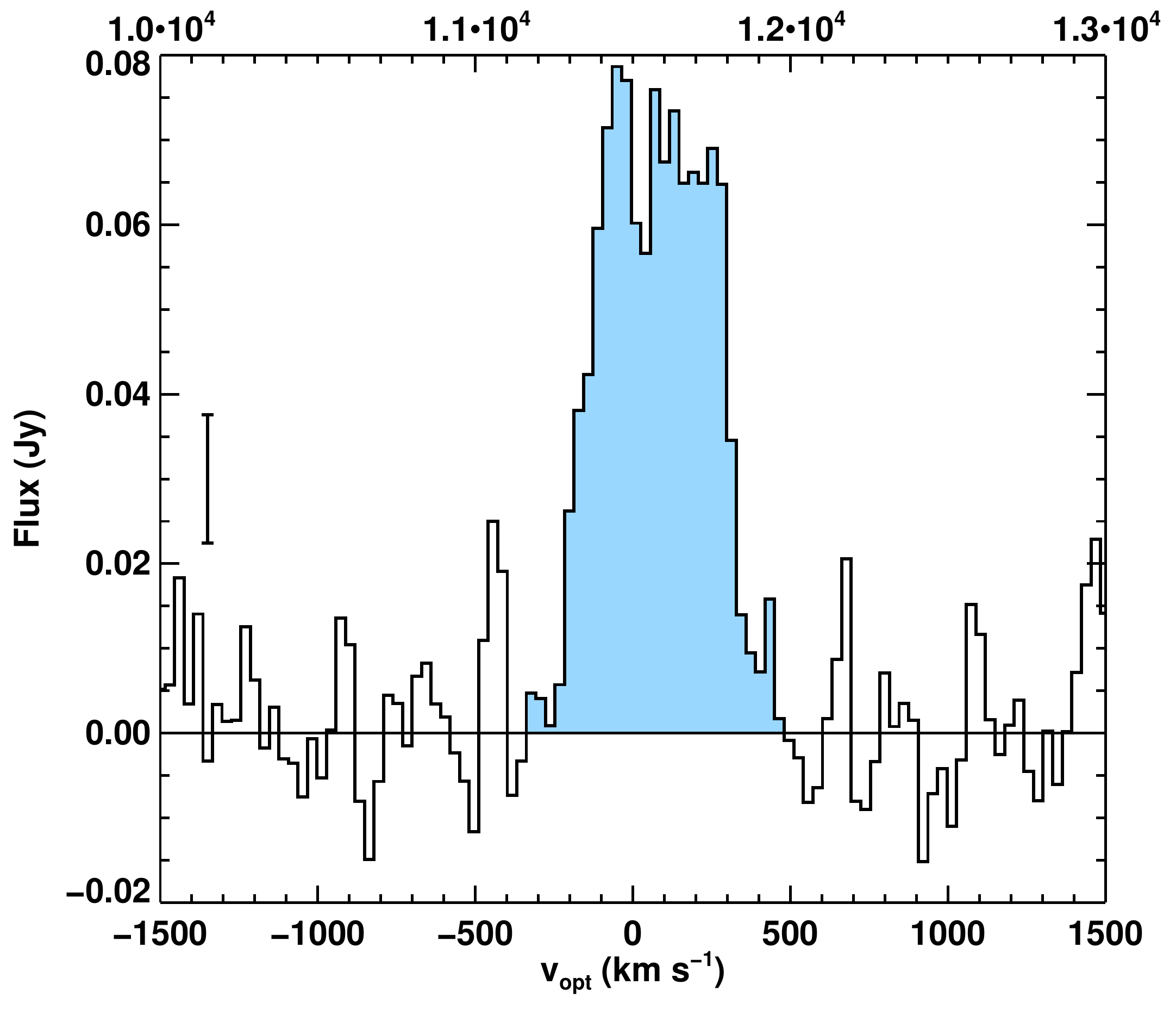}}
\subfigure{\includegraphics[height=4.9cm]{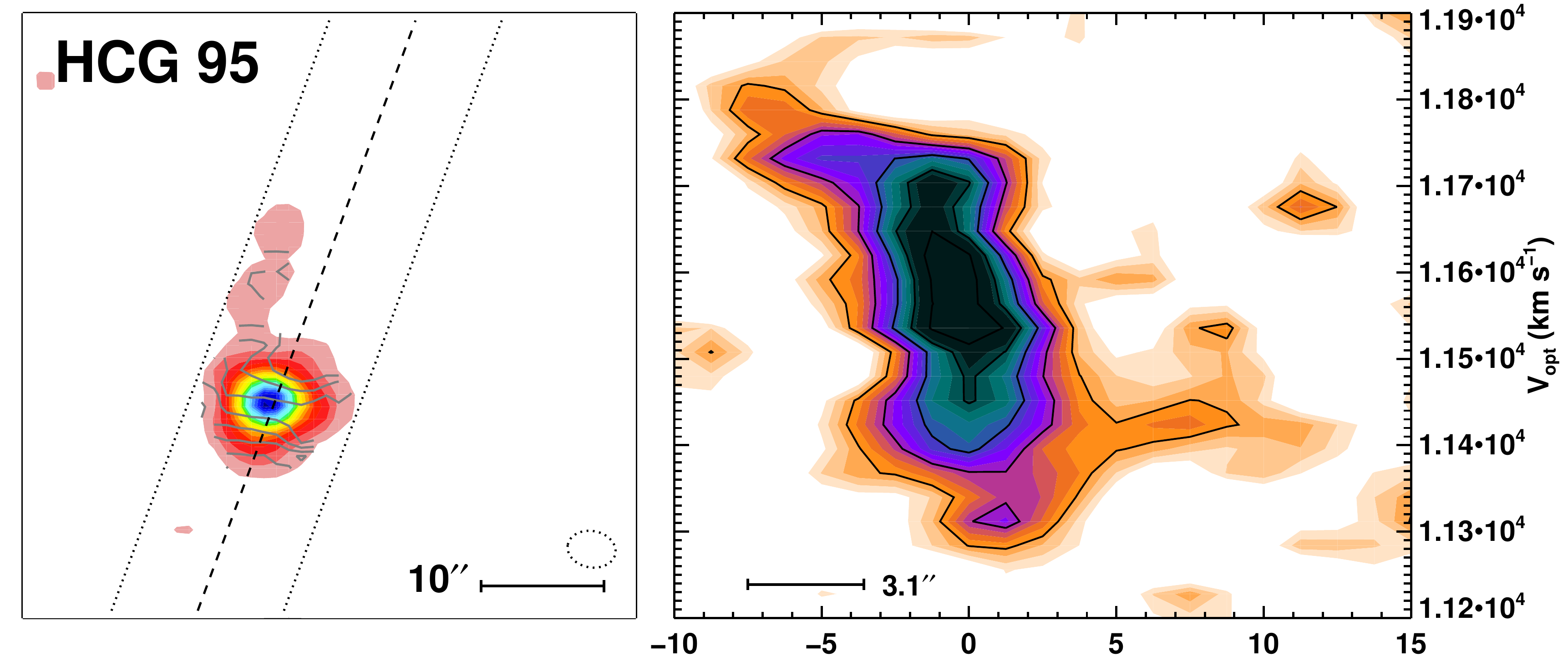}}
\subfigure{\includegraphics[width=\textwidth]{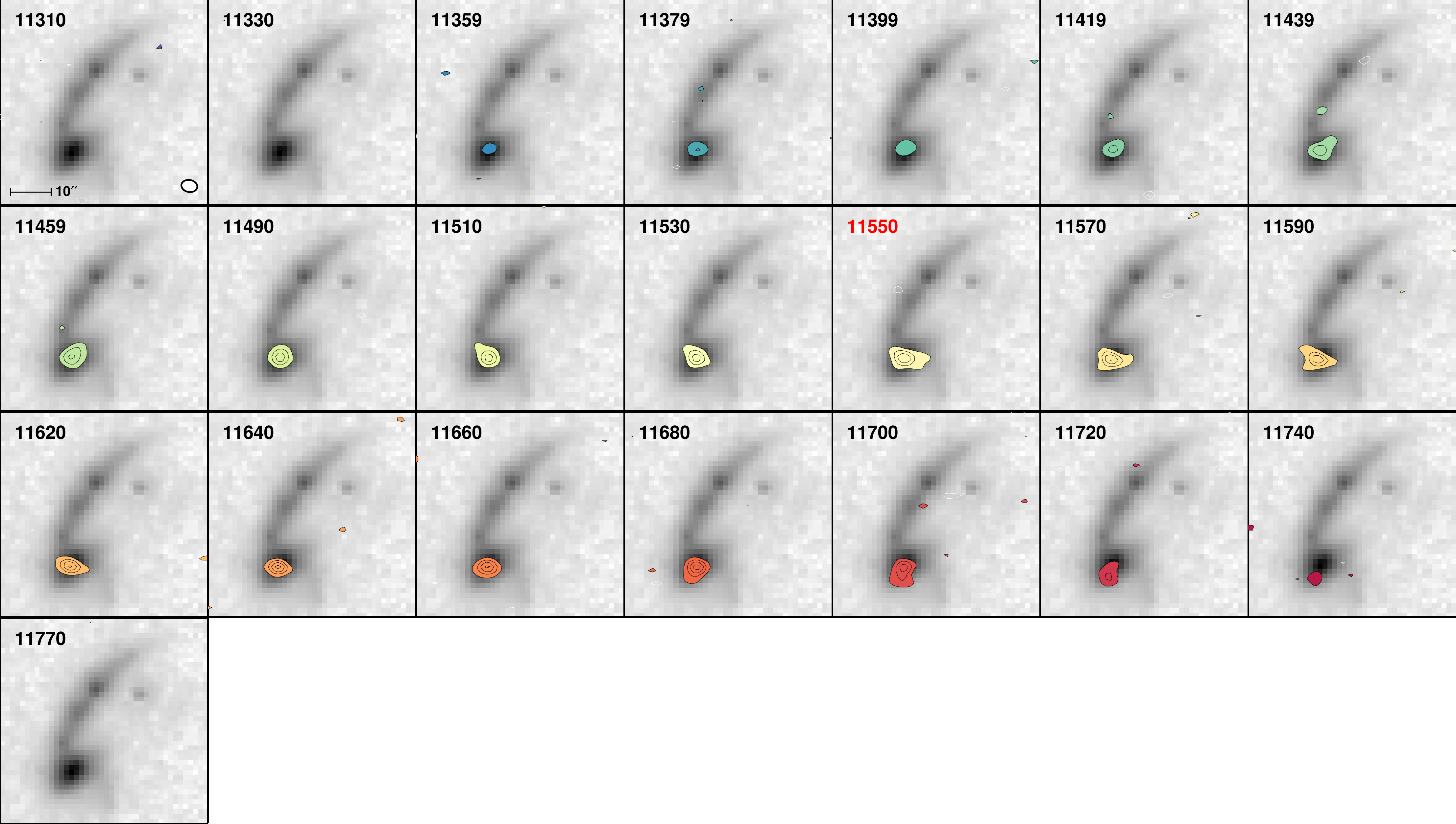}}
\caption{HCG\,95. Channel map contours are in 3$\sigma$ steps.}
 \label{fig:hcg95}
 \end{figure*}

\begin{figure*}[h!]
\subfigure{\includegraphics[width=\textwidth]{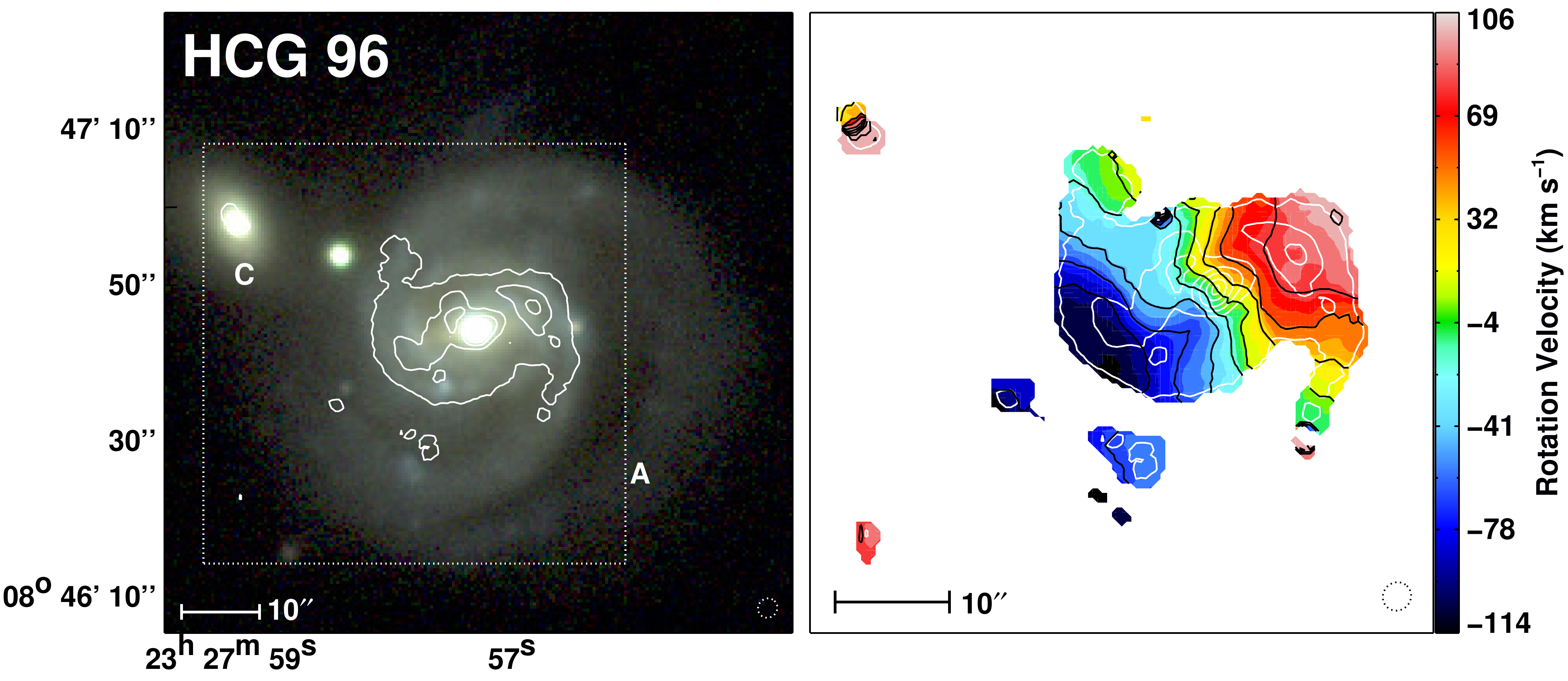}}
\subfigure{\includegraphics[height=4.9cm]{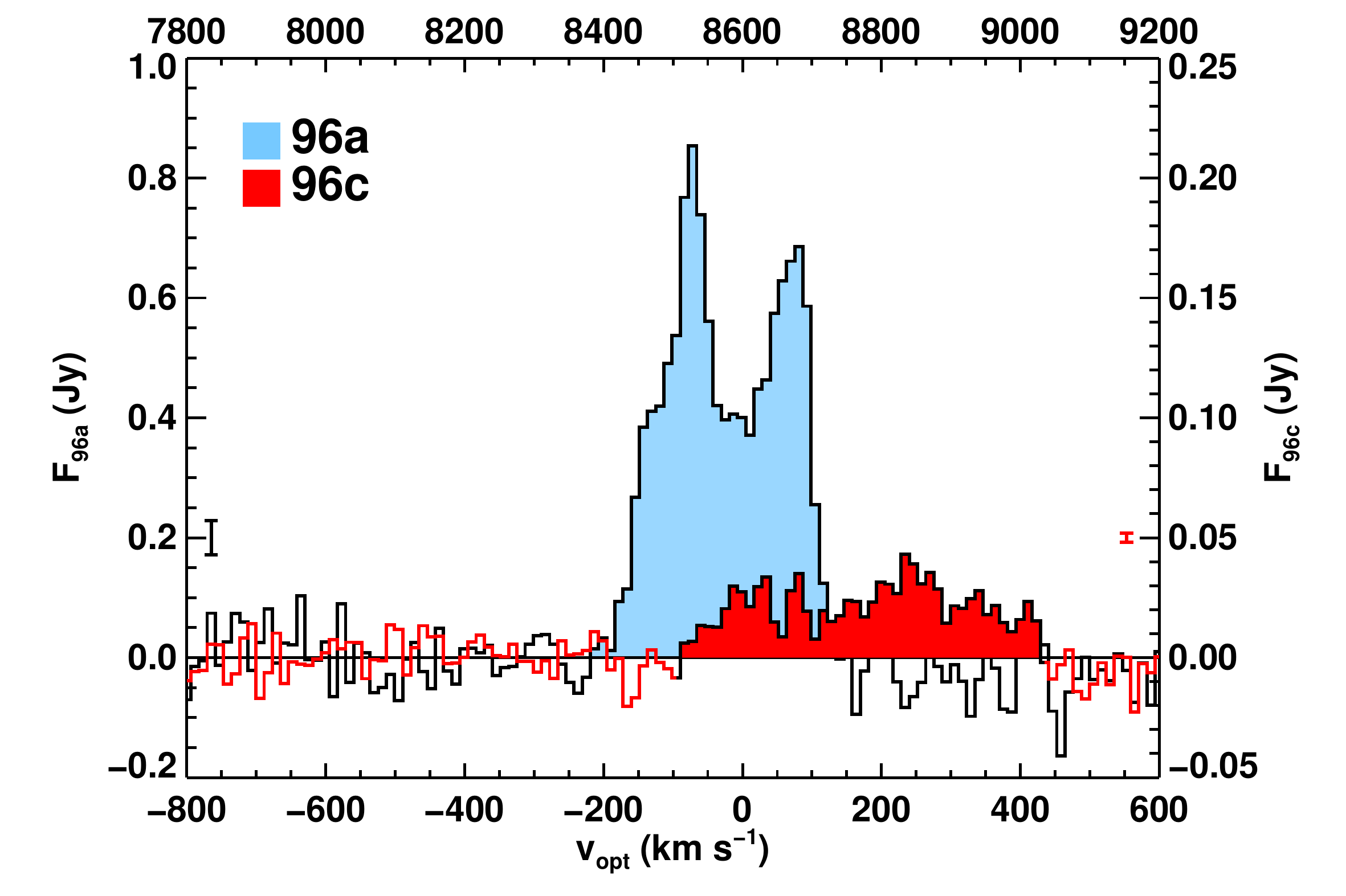}}
\subfigure{\includegraphics[height=4.9cm]{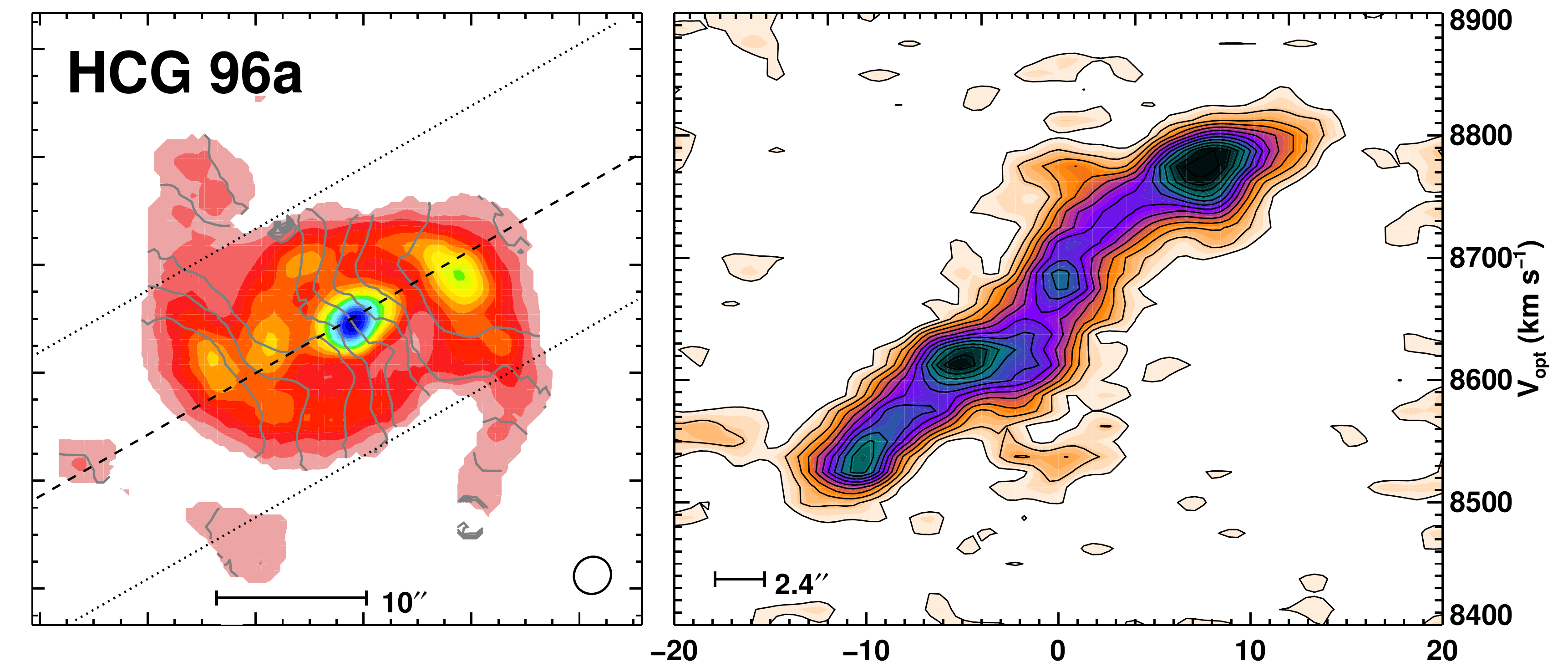}}
\subfigure{\includegraphics[width=\textwidth]{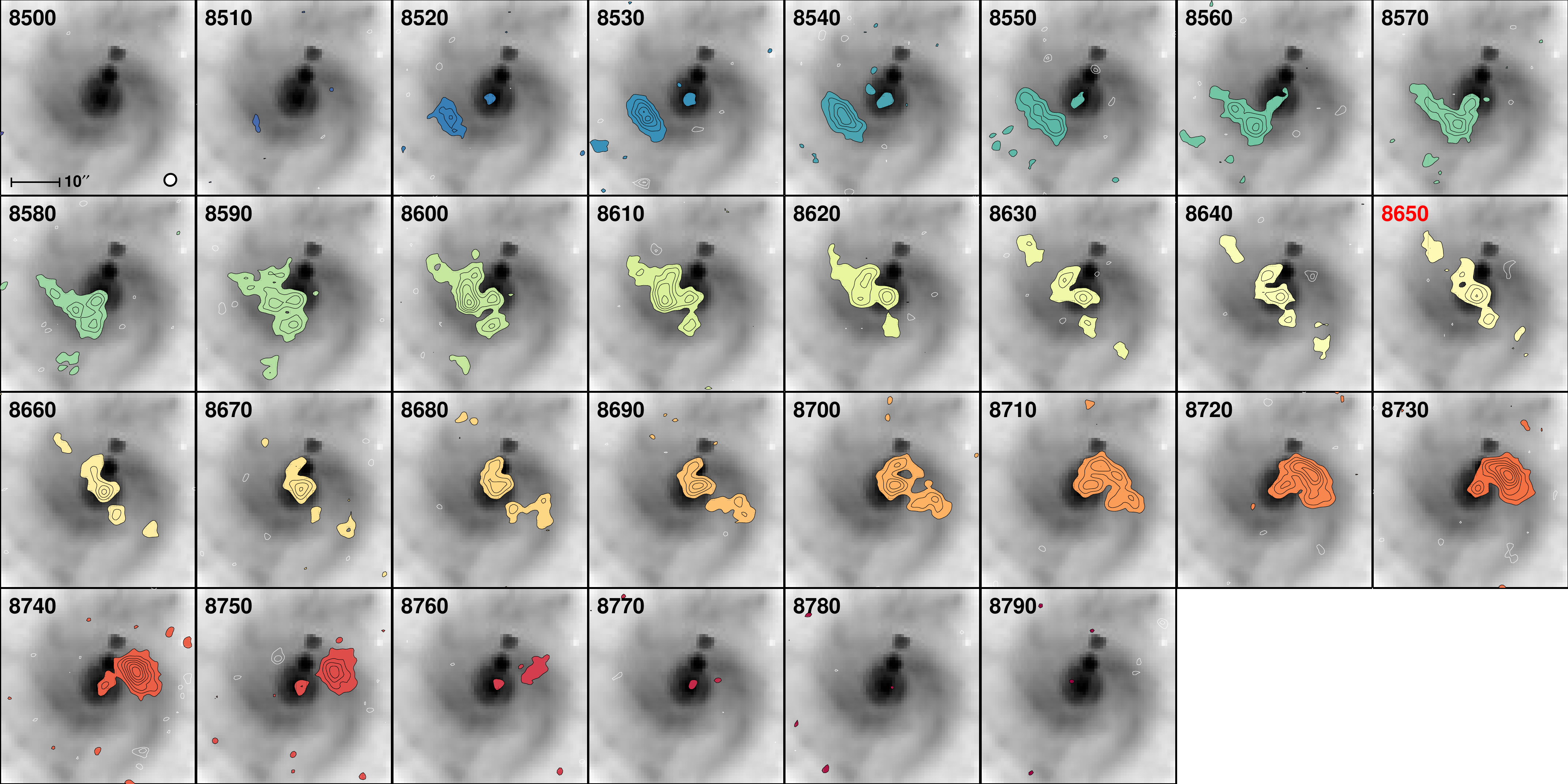}}
\caption{HCG\,96. Channel map contours are in 3$\sigma$ steps.  Channel maps and a PVD of 96c can be found in Figures\,\ref{fig:HCG96c} and \ref{fig:PV57d+96c}, respectively.}
 \label{fig:hcg96}
 \end{figure*}

\begin{figure*}[h!]
\subfigure{\includegraphics[width=\textwidth]{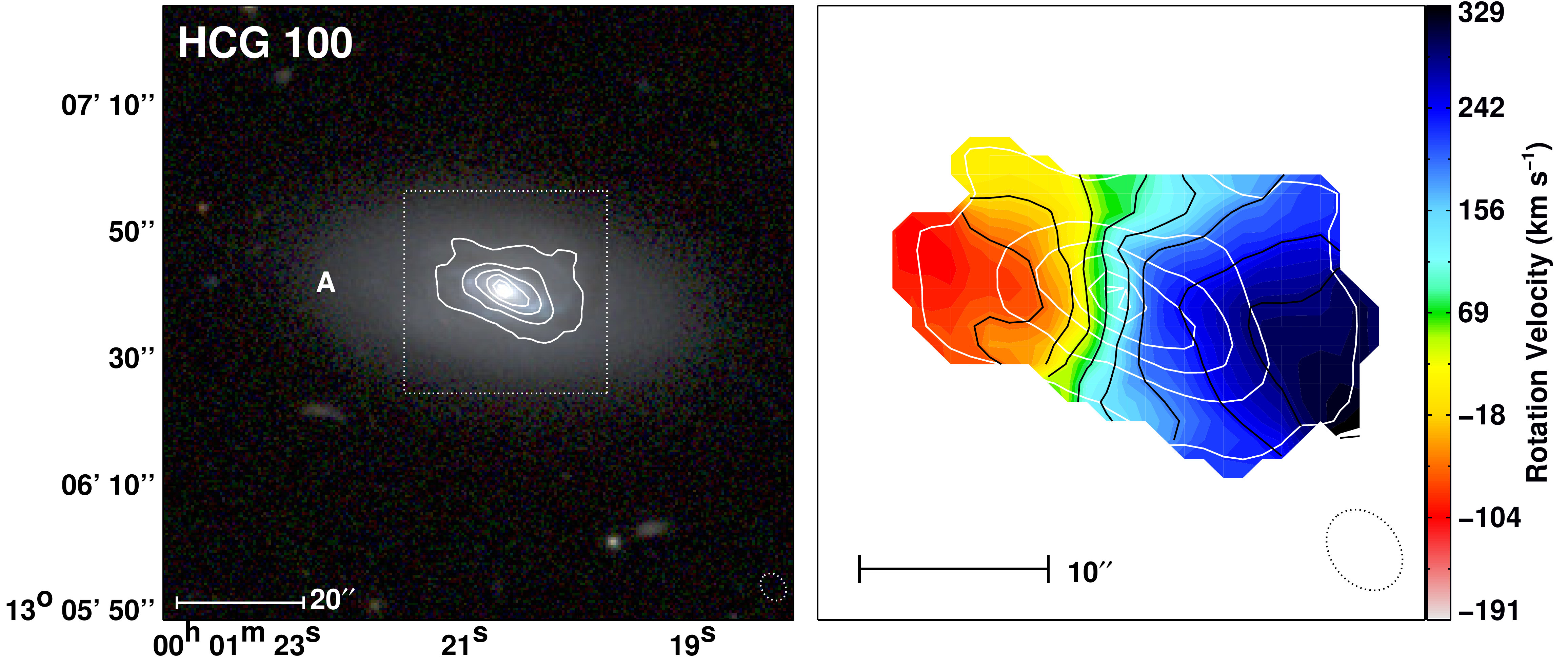}}
\subfigure{\includegraphics[height=5.1cm]{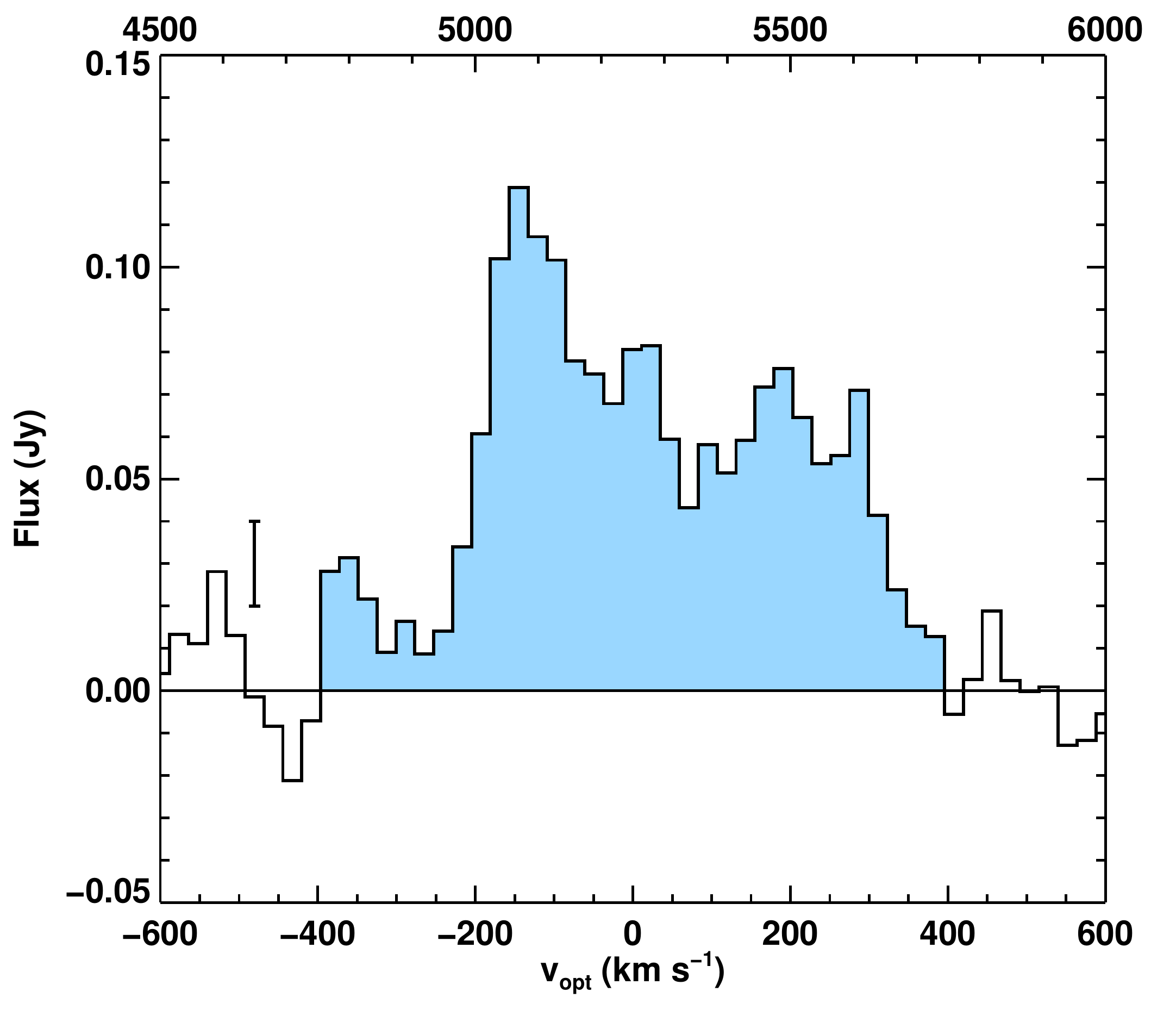}}
\subfigure{\includegraphics[height=5.1cm]{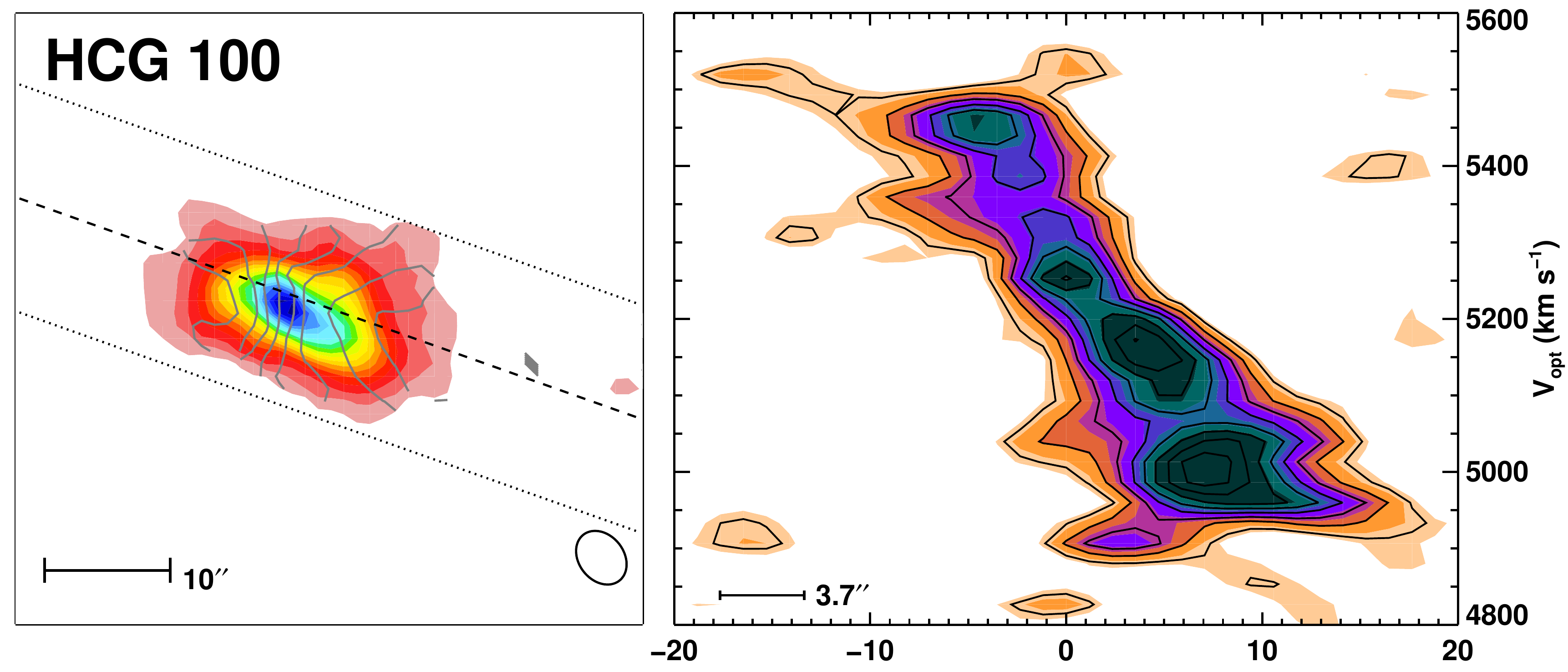}}
\subfigure{\includegraphics[width=\textwidth]{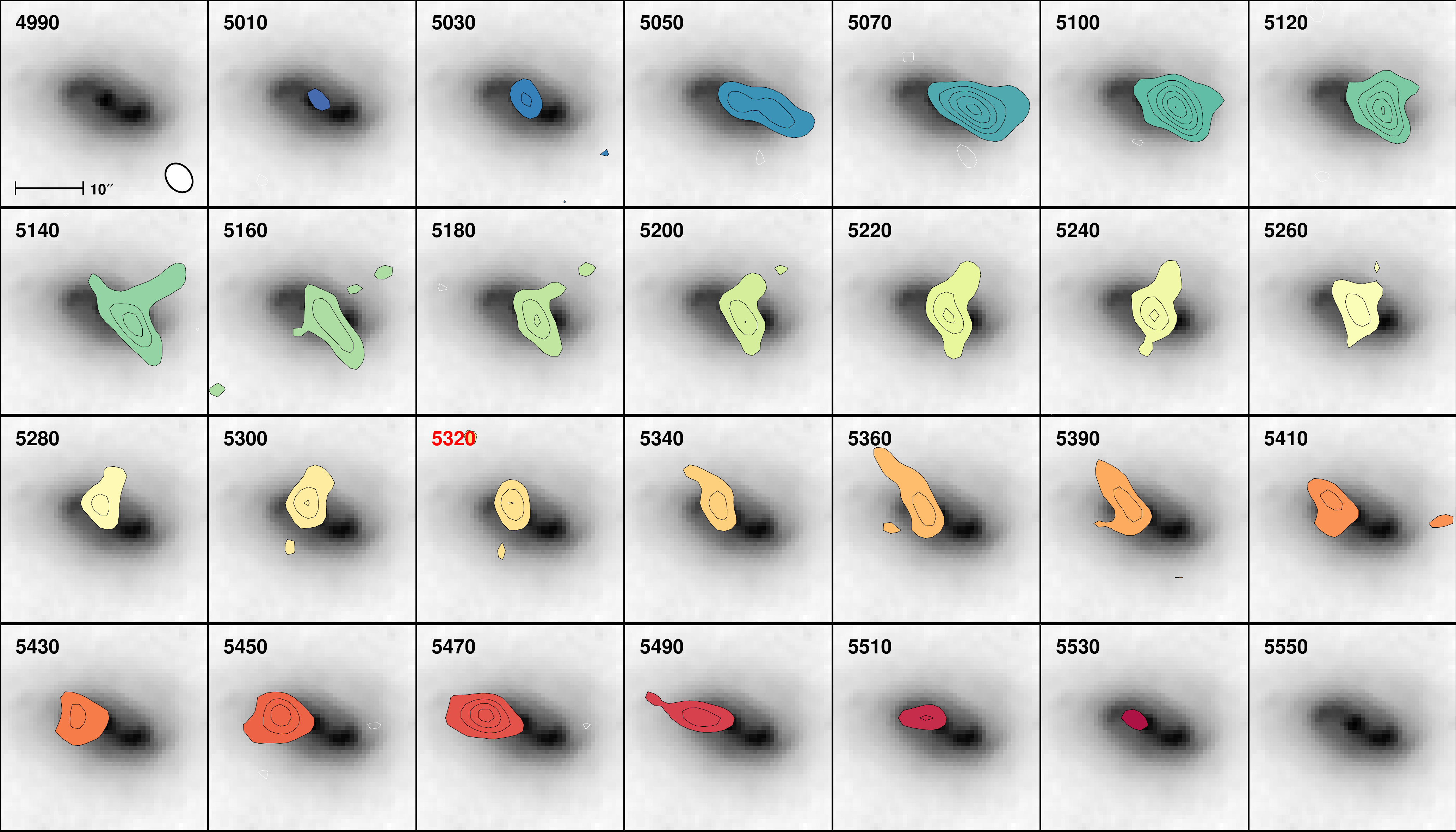}}
\caption{HCG\,100. Channel map contours are in 3$\sigma$ steps.}
 \label{fig:hcg100}
 \end{figure*}

\begin{figure*}[h!]
 \centering
 \subfigure{\includegraphics[width=0.6\textwidth]{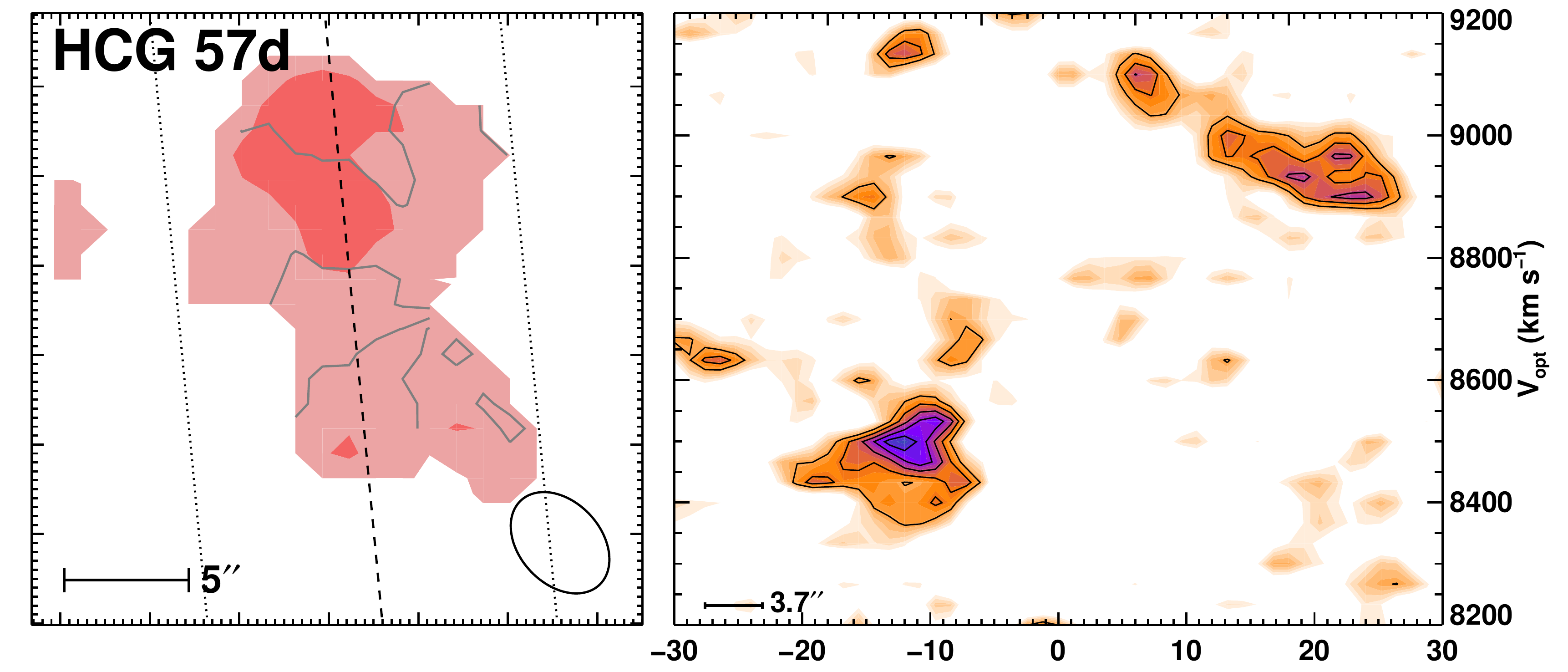}}
 \subfigure{\includegraphics[width=0.6\textwidth]{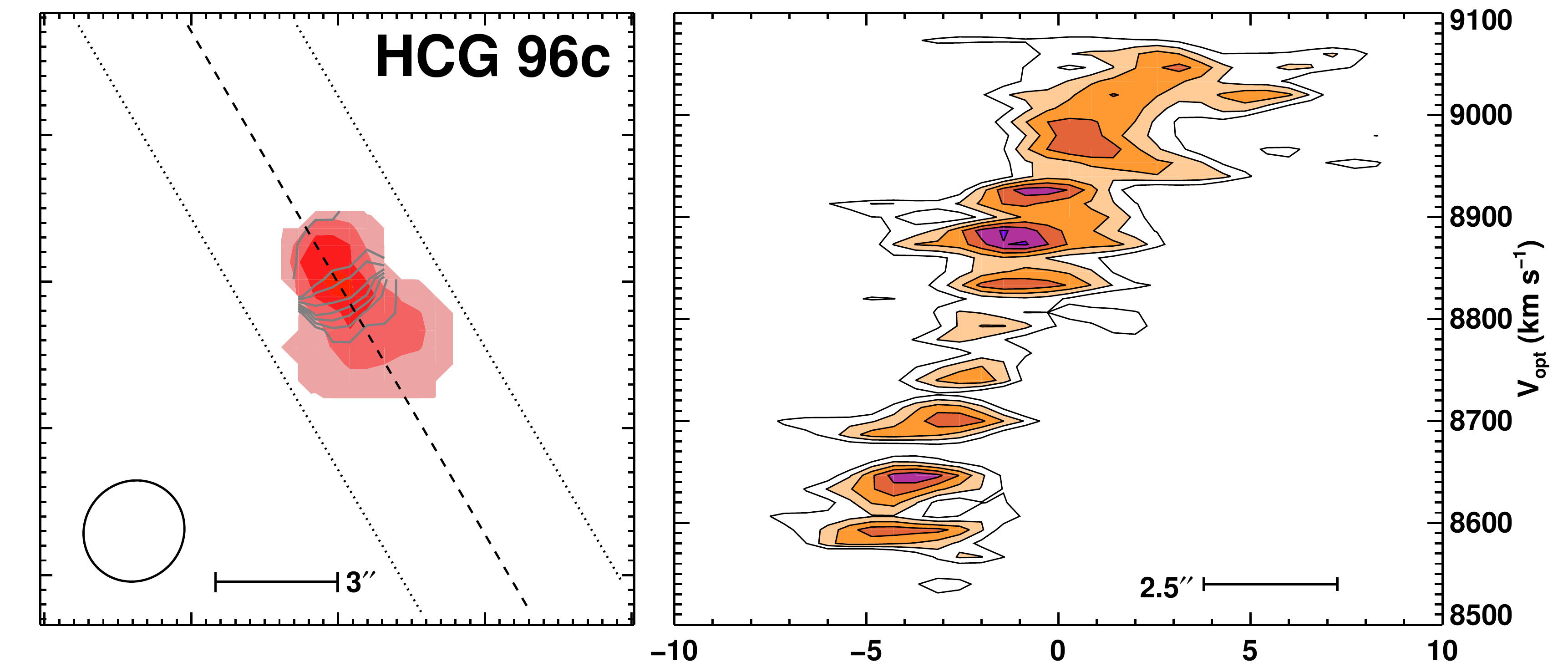}}
\caption{PVDs of HCG\,57d (top) and HCG\,96c (bottom).  The molecular gas in HCG\,57d appears to be consistent with a ring (seen in PAH emission in Figure~\ref{fig:hcg57} and described in \citealt{a14_hcg57}).  The molecular gas in HCG\,96c seems to be consistent with a rotating disk.}
 \label{fig:PV57d+96c}
 \end{figure*}

\begin{figure*}[h!]
\includegraphics[width=0.97\textwidth]{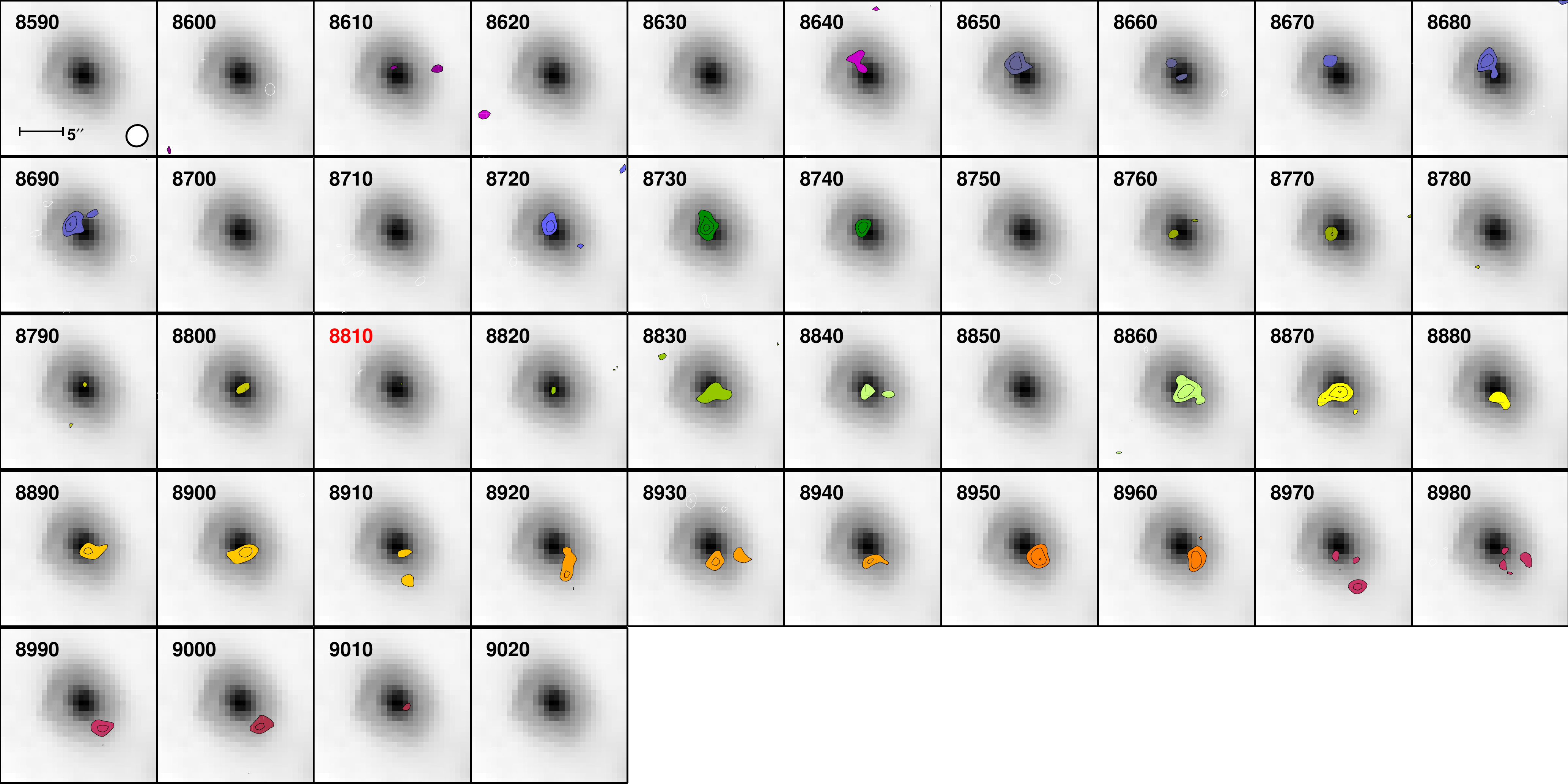}
 \caption{HCG 96c. Channel map contours are in 1$\sigma$ steps.}
 \label{fig:HCG96c}
 \end{figure*}

\noindent Figures for each individual HCG are shown in Figures \ref{fig:hcg25}--\ref{fig:hcg100}.  For each HCG, we show five figures, as well as an additional figures for HCGs where CARMA detected 2 galaxies: HCG\,57d (Fig~\ref{fig:PV57d+96c}a) and 96c (Figs. \ref{fig:PV57d+96c}b \& \ref{fig:HCG96c}). The outlined box in the image (white dotted line) represents the area of the corresponding moment1 map. North is up and East is left in all images.\\

\noindent {\bf Top left:}  The CO(1--0) integrated intensity (moment0) map (white contours) overlaid on an optical 3-color image, either {\em g,r,i} from the Sloan Digital Sky Survey (HCG\,25, HCG\,47, HCG\,57, HCG\,68, HCG\,79, HCG\,82, HCG\,95, HCG\,96, HCG\,100), the Digitized Sky Survey (HCG\,40, HCG\,55), or the {\em Swift} archive (HCG\,91). \\

\noindent {\bf Top right:} The CO(1--0) mean velocity (moment1) map, overlaid with the moment0 map (white contours).  The velocities are relative to the systemic velocity of each source.\\

\noindent {\bf Middle left:} The integrated CO(1--0) spectrum, which was constructed by summing all pixels in each channel using the moment0 mask as a clip mask.\\

\noindent {\bf Middle center \& right:} The position--velocity diagram (PVD), taken by slicing the CO(1--0) cube in a plane and summing over a specific region.  The moment0 map (middle center) outlines the PVD integration area (dashed black line for the center, dotted gray lines for the boundaries), and the corresponding PVD shows the velocity structure tangent to the velocity slice (middle right).\\

\noindent {\bf Bottom:} The CO(1--0) channel maps are overlaid on the {\em Spitzer} 8$\mu$m nonstellar emission. The {\em Spitzer} 8$\mu$m nonstellar maps were created by subtracting a scaled 3.6$\mu$m map from IRAC from the 8.0$\mu$m IRAC map using the scale factor of 0.232 for late-type galaxies \citep{helou+04}.  The shading corresponds to a signal-to-noise ratio of 3 in each channel (except where noted otherwise), with additional contours in either 1$\sigma$ or 3$\sigma$ steps.  The panel that corresponds to the systemic velocity has its velocity labeled in red.\\

%\section{Galaxies Detected by the Owens Valley Radio Observatory}
%\begin{figure*}[h!]
%\includegraphics[width=\textwidth,clip,trim=0cm 0cm 2.5cm 0cm]{figures/hcg16c_3color.pdf}
%\vskip -2mm
%\caption{HCG\,16c}
%\label{fig:HCG16c}
%\vskip -2mm
%\end{figure*}
%
%\begin{figure*}[h!]
%\includegraphics[width=\textwidth]{figures/hcg16d_3color.pdf}
%\vskip -2mm
%\caption{HCG\,16d}
%\label{fig:HCG16d}
%\vskip -2mm
%\end{figure*}
%
%
%\begin{figure*}[h!]
% \includegraphics[width=\textwidth]{figures/hcg37_3color.pdf}
% \vskip -2mm
%\caption{HCG\,37}
% \label{fig:HCG37}
% \vskip -2mm
% \end{figure*}
%
%\begin{figure*}[h!]
%\includegraphics[width=\textwidth]{figures/hcg44_3color.pdf}
%\vskip -2mm
%\caption{HCG\,44}
%\label{fig:HCG44}
%\end{figure*}

 \end{appendix}
 %%%%%%%%%%%%%%%%%%%%%%%%%%%%%%%%%%%%%%%%%%%%%%%
% END

 \end{document}